\documentclass[mnsc,nonblindrev]{informs3_nojournal}
\usepackage{endnotes}
\let\footnote=\endnote

\usepackage{natbib}
 \bibpunct[, ]{(}{)}{,}{a}{}{,}%
 \def\bibfont{\small}%
 \def\bibsep{\smallskipamount}%
 \def\bibhang{24pt}%

\usepackage{bbm}
\usepackage{xifthen,enumitem,comment}
\setlist[enumerate,1]{label=\normalfont{(\roman*)},leftmargin=2em}
\usepackage{xcolor}
\usepackage{etoolbox}
\makeatletter
\patchcmd{\env@cases}{1.2}{0.96}{}{}
\makeatother
\usepackage{algorithm,algpseudocode}
\usepackage{caption}
\usepackage{subcaption}
\usepackage[hypertexnames=false]{hyperref}
\hypersetup{colorlinks=true,linkcolor={blue!50!black},linkbordercolor=red,citecolor={blue!50!black},
}
 \usepackage{cleveref}
 \hypersetup{
 	colorlinks = true,
 	citecolor={blue!50!black},
 }
\usepackage{booktabs,tikz}
\usetikzlibrary{matrix,decorations.pathreplacing}
\ifx\argmax\undefined\DeclareMathOperator*{\argmax}{arg\,max}\fi
\ifx\argmin\undefined\DeclareMathOperator*{\argmin}{arg\,min}\fi

\newcommand*{\QED}{\leavevmode\unskip\penalty9999 \hbox{}\nobreak\hfill
    \quad\hbox{$\square$}}

\providecommand{\allone}{\boldsymbol{1}}

\renewcommand{\P}{\mathbb{P}}
\providecommand{\E}{\mathbb{E}}
\providecommand{\R}{\mathbb{R}}
\providecommand{\F}{\mathbb{F}}
\providecommand{\G}{\mathbb{G}}
\providecommand{\cB}{\mathcal{B}}
\providecommand{\cC}{\mathcal{C}}
\providecommand{\cF}{\mathcal{F}}
\providecommand{\cJ}{\mathcal{J}}
\providecommand{\cM}{\mathcal{M}}
\providecommand{\cP}{\mathcal{P}}
\providecommand{\cR}{\mathcal{R}}
\providecommand{\cS}{\mathcal{S}}
\providecommand{\cV}{\mathcal{V}}
\providecommand{\Rad}{\mathfrak{R}}
\providecommand{\Rev}{\mathrm{Rev}}
\providecommand{\OPT}{\mathsf{OPT}}
\newcommand{\bv}{\boldsymbol{v}}
\newcommand{\bq}{\boldsymbol{q}}

\newcommand{\of}[1]{\left(#1\right)}
\newcommand{\off}[1]{\left[#1\right]}
\newcommand{\offf}[1]{\left\{#1\right\}}
\newcommand{\bigof}[1]{\big(#1\big)}

\newcommand{\bigoff}[1]{\big[#1\big]}
\newcommand{\bigofff}[1]{\big\{#1\big\}}
\newcommand{\Bigof}[1]{\Big(#1\Big)}
\newcommand{\biggof}[1]{\bigg(#1\bigg)}
\newcommand{\Bigoff}[1]{\Big[#1\Big]}

\usepackage{cases}
 \usepackage{upgreek}

\OneAndAHalfSpacedXI
\TheoremsNumberedThrough     \ECRepeatTheorems

\EquationsNumberedThrough    \MANUSCRIPTNO{}

\usepackage{array}
\usepackage{delarray}
\begin{document}

\RUNTITLE{Multi-Item Screening with a Maximin-Ratio Objective}
\RUNAUTHOR{Shixin Wang}
\TITLE{\Large Multi-Item Screening with a Maximin-Ratio Objective}
\ARTICLEAUTHORS{\AUTHOR{Shixin Wang}
\AFF{H. Milton Stewart School of Industrial \& Systems Engineering, Georgia Institute of Technology 
} }

\ABSTRACT{

In multi-item screening, optimal selling mechanisms are challenging to characterize and implement, even with full knowledge of valuation distributions.
In this paper, we aim to develop tractable, interpretable, and implementable mechanisms with strong performance guarantees in the absence of precise distributional knowledge. In particular, we study robust screening with a maximin ratio objective. We show that given the marginal support of valuations, the optimal mechanism is separable: each item’s allocation probability and payment depend only on its own valuation and not on other items’ valuations. However, we design the allocation and payment rules by leveraging the available joint support information. This enhanced separable mechanism can be efficiently implemented through randomized pricing for individual products, which is easy to interpret and implement. Moreover, our framework extends naturally to scenarios where the seller possesses marginal support information on aggregate valuations for any product bundle partition, for which we characterize a bundle-wise separable mechanism and its guarantee. Beyond rectangular-support ambiguity sets, we further establish the optimality of randomized grand bundling mechanisms within a broad class of ambiguity sets, which we term ``$\boldsymbol{\rho}-$scaled invariant ambiguity set".

 }

\maketitle    
\section{Introduction}
The multi-item screening problem is important and difficult to solve in general. 
For one item, charging a posted price is proven optimal \citep{myerson1981optimal,riley1983optimal}.
However, with $n \geq 2$ items, even when a seller has complete knowledge of the buyer's joint valuation distribution, the optimal mechanism becomes significantly more challenging to characterize or implement \citep{manelli2007multidimensional, daskalakis2013mechanism, daskalakis2014complexity}.  
In this regard, the robust mechanism design literature addresses these difficulties by focusing on mechanisms that perform well under limited distributional information, which offers a promising approach to deriving tractable and interpretable robustly optimal solutions. 
A significant finding in multi-item robust screening shows that if the seller only knows marginal distribution but has no knowledge of how item valuations are correlated, then separate selling is robustly optimal under maximin revenue objectives \citep{carroll2017robustness, gravin2018separation}.  That is, selling each product at its monopoly price guarantees the highest possible worst-case revenue across all possible correlation structures compatible with the given marginals. 
This finding is significant because the separate selling mechanism is straightforward to implement and interpret. However, although separate selling maximizes revenue in the worst case, it can perform significantly worse than the hindsight optimal mechanism in less adversarial distributions, leading to a low performance ratio.
Moreover, obtaining the precise marginal distribution may not always be feasible in practice. The sellers typically possess only coarse information, such as the support of valuations (i.e., the range of plausible values for each item). 

Motivated by these challenges, we revisit the robust screening problem by adopting the maximin ratio as the objective — a metric that quantifies the worst-case ratio of a mechanism’s revenue to the optimal revenue achievable under full distributional knowledge. Moreover, we relax the informational requirement to scenarios where the seller knows only the support of the buyer’s valuations, with no additional constraints on marginal distributions or correlations. 
By shifting the objective from worst-case revenue to worst-case performance ratio relative to the optimum, we develop mechanisms that are not only robust but also competitive across all possible valuation distributions compatible with the support. We aim to deliver tractable, interpretable, and robustly optimal solutions, offering insights into how limited information shapes the trade-offs between robustness and revenue efficiency.

We show that when the seller knows only the marginal support information of each product, a separable mechanism is still robustly optimal. In contrast to the case under maximin revenue objectives, where each item’s mechanism design is independent of other items’ distributional information, the optimal robust mechanism under the maximin ratio objective can and should exploit the joint support information across items. Concretely, this optimal separable mechanism has the following properties: 
(i) there exist feasible one-dimensional selling mechanisms for all items, such that the original allocation function for each item in the multi-item screening problem is the same as the allocation function for the same item in the one-dimensional mechanisms, and the payment function in the multi-item mechanism is the sum of the payment functions in the one-dimensional mechanisms for all the items; (ii) each item’s allocation probability depends only on its own valuation (and not on other items’ valuations); (iii)  the design of each item’s allocation and payment rules can leverage the joint support information for all items, unlike in traditional decomposed separable mechanisms where these functions are independent of other items' distributional information.

The advantage of this separable mechanism is twofold. First, despite integrating joint support information into its design, the mechanism retains the operational simplicity of separate selling: the seller can post randomized prices for each item independently, with price distributions tailored using the joint support information. This avoids the complexity of probabilistic mixed-bundling mechanisms. Second, by exploiting joint support information—even without distributional details—the mechanism achieves strictly stronger performance guarantees than robust mechanisms that ignore the support information of other items. Our results demonstrate that leveraging joint support information can enhance robustness without sacrificing simplicity, bridging tractability and performance in multi-item screening under ambiguity.

The main challenge in proving the optimality of the separate selling mechanism is characterizing the worst-case valuation distribution and the hindsight optimal selling strategy for the adversary. 
As highlighted by previous literature \citep{daskalakis2014complexity,babaioff2020simple}, it is generally challenging to characterize the optimal selling mechanism under a given joint distribution, since the optimal mechanism can be non-monotonic, randomized, unrepresentable, or computationally intractable. 
Accordingly, the maximin ratio objective creates challenges that are distinct from those in the maximin-revenue objective \citep{carroll2017robustness}. 
In our study, we address these challenges by constructing a comonotonic family of distributions for which nature’s optimal response is degenerate—i.e., supported on a single valuation that simultaneously determines the buyer’s type and the hindsight benchmark.  Under this adversary's decision, the seller's problem is reduced from a multi-dimensional mechanism design problem to a one-dimensional functional optimization. Next, by carefully constructing the subset of comonotonic distributions, we can reduce the seller's problem from functional optimization to scalar optimization. This key restriction on the adversary's strategy ensures that a simple posted price selling mechanism is the optimal mechanism for the seller. 
With this construction, we can explicitly determine the hindsight optimal mechanism and calculate its corresponding performance ratio. 
As long as this minimum performance ratio in this dual problem coincides with the feasible performance ratio achieved by the separable mechanism in the primal problem, we are able to prove the optimality of the separable mechanism.

Our analysis further reveals that the robust optimality of separable mechanisms extends beyond marginal support information to general ambiguity sets, provided these sets include the adversary’s optimal strategy.
Moreover, we demonstrate that our framework adapts naturally to settings where the seller has coarse information about aggregate valuations for specific product bundles. For instance, if the seller knows the support of buyers’ total valuations for any partition of items into bundles, the robustly optimal mechanism is to sell each bundle separately. This result highlights a separation principle: optimality is preserved by treating each predefined bundle as an independent ``item'', even when the supports of products within a bundle are not independent. Notably, we identify a sufficient condition for the optimality of randomized grand bundling—a mechanism where all items are sold as a single bundle with a randomized price.  This condition applies to a wide range of commonly used ambiguity sets, including those containing all distributions supported on triangular, ellipsoidal, or $\ell_1$-ball-shaped regions.
Taken together, these findings offer valuable insights for mechanism design under a maximin ratio objective, addressing diverse types of ambiguity sets in multi-item screening.

\subsection{Literature Review}
The simple and elegant optimal selling mechanism found in the single-item screening problem \citep{myerson1981optimal} does not extend to the multi-item settings, even for the two-item cases.
Even with independent values, the optimal mechanism may require an infinite menu of lotteries \citep{daskalakis2013mechanism,daskalakis2015strong}.
With correlations, \cite{hart2013menu} and \cite{briest2015pricing} show that simple mechanisms, such as separate selling, bundling, or any deterministic mechanism, can not guarantee any positive fraction of the optimal revenue. 
\cite{hart2017approximate} demonstrate that separate selling achieves a performance ratio of $\mathcal{O}(\frac{1}{(\ln k)^2})$ for $k$ independent goods, and bundling achieves $\mathcal{O}(\frac{1}{\ln k})$ when values are independently and identically distributed (i.i.d). Then \cite{babaioff2020simple} show that the better of separation and bundling achieves a constant performance ratio of $1/6$ for independent goods. Existing works provide \emph{approximations} for given simple mechanisms under \emph{known}, \emph{independent} valuation distributions. Our work contributes to the multi-item mechanism design literature by providing robustly \emph{optimal} mechanisms that operate \emph{without} precise knowledge of the valuation distributions or their correlations.

Multi-item robust screening solves for the optimal mechanism when the seller only has partial information about the buyer's valuation. \cite{carroll2017robustness} shows that with known marginal valuation distributions, selling each item separately at its own monopoly price maximizes worst‑case revenue.
\cite{gravin2018separation} adopt a different approach to show the optimality of separate selling under budget constraints.  \cite{che2021robustly} find that partition-wise separation is robustly optimal when the seller knows the marginal mean and aggregated moments for different partitions of products.
Besides the maximin revenue objective, \cite{koccyiugit2022robust} show that if the seller only knows the marginal support of the buyer's valuations, then separation is also robustly optimal when adopting the minimax absolute regret objective. 
Our work complements the multi-dimensional robust screening literature by incorporating the performance ratio as the robust metric. 

    Beyond the mechanism design literature, our paper relates closely to research on product bundling, which traces back to the seminal works of \cite{stigler1963united,adams1976commodity,schmalensee1984gaussian,mcafee1989multiproduct}. Recent research has explored more practical and implementable mechanisms: \cite{chu2011bundle} show numerically that bundle‑size pricing closely approximates optimal mixed bundling. \cite{ma2021reaping} introduce “pure bundling with disposal for cost’’ (PBDC), allowing customers to return unwanted items at the production cost. \cite{li2022convex} develop a tractable convex approximation that uses moment information on valuations. \cite{chen2023component} analyze ``component pricing with a bundle‑size discount,'' prove its asymptotic optimality, and provide a mixed‑integer linear program for price optimization. \cite{sun2025partition} propose the ``single bundle with the rest'' (SBR) framework and solve it efficiently under multivariate normal valuations. Previous studies have yielded valuable insights within particular practical and implementable mechanism families. Our analysis complements these works by characterizing the optimal mechanism among all feasible mechanisms without predetermining a mechanism type.

In the single-item screening problem, the robust mechanism is studied under the maximin revenue objective for ambiguity sets based on Prohorov metric \citep{bergemann2011robust}, moments information \citep{pinar2017robust,carrasco2018optimal,chen2022distribution,chen2023screening}, mean preserving contraction \citep{du2018robust,chen2023screening},
Wasserstein metric \citep{li2019revenue,chen2023screening}, mean absolute deviation \cite{chen2023screening}, and shape of demand curve \citep{cohen2021simple}. 
\cite{chen2022distribution} also identify conditions under which grand bundling outperforms separate selling for the multi-product setting.
Under the minimax absolute regret metric, \cite{bergemann2008pricing} provides the static optimal robust mechanism, and \cite{caldentey2017intertemporal} study the dynamic pricing under support information. For the performance ratio objective, optimal or near-optimal mechanism is studied under support \citep{eren2010monopoly,wang2024minimax}, mean-support \citep{wang2024minimax}, moments \citep{giannakopoulos2023robust, wang2024minimax}, quantile with regular/MHR \citep{allouah2023optimal}, quantile-support ambiguity sets \citep{wang2025power} and samples \citep{huang2018making,allouah2022pricing}. However, the analysis for the single-item screening with performance ratio metric can not be extended to multi-item screening directly, due to the complexity of characterizing the optimal hindsight policy in the multi-item screening problem. 

In addition to the single-buyer problem, robust mechanism design is also studied when there are multiple buyers and the seller only knows partial information of buyers' joint value profiles, under maximin revenue \citep{bandi2014optimal,carrasco2015robust, koccyiugit2020distributionally,he2022correlation,suzdaltsev2020optimal,suzdaltsev2022distributionally},  minimax regret \citep{anunrojwong2022robustness,koccyiugit2024regret}, or performance ratio \cite{azar2013parametric,azar2013optimal,bei2019correlation,allouah2020prior,anunrojwong2023robust}. 

Competitive ratio is also important in helping quantify pricing efficacy. For instance, 
\cite{besbes2019static} study the performance guarantee of different metrics for static pricing.
\cite{elmachtoub2021value} bound the gap between the best deterministic price and personalized pricing.
\cite{wang2025power} investigates the performance of optimal finite-menu robust screening. \cite{bei2019correlation,jin2020tight} evaluate simple mechanisms such as anonymous posted pricing, second-price auction, and sequential posted pricing relative to the optimal auction. Methodologically, our work draws on the tools of distributionally robust optimization \citep{chen2007robust,delage2010distributionally, goh2010distributionally,wiesemann2014distributionally, mohajerin2018data, gao2023distributionally}.

\subsection*{Notation}
$\R_+^J$ denotes the non-negative orthant of $J$-dimensional Euclidean space. We denote $\Delta(\cV)$ as the set of all probability distributions supported on $\cV$. 
The subscription $-j$ represents the components $\cJ\setminus j$. For instance, $\bv_{-j} = \of{v_{j'}}_{j'\in \cJ, j'\neq j}$. We use $\Sigma(\cJ)$ to represent the set of all permutations on set $\cJ$; that is, each $\sigma\in \Sigma(\cJ)$ is a bijective mapping $\sigma: \cJ\to \cJ$.
The subscription $k_1:k_2$ represents the $k_1$th to $k_2$th components in a vector.
For instance, $\boldsymbol{\omega}_{1:k}$ denotes the first to the $k$th components in vector $\boldsymbol{\omega}$.
We use $t^{M}(\cdot)$ to denote the payment rule under mechanism $M$, and we omit the superscript $M$ when there is no confusion. We denote $e_j$ the unit vector with 1 at position $j$ and 0 elsewhere. We refer to ``decreasing/increasing'' and ``positive/negative'' in the weak sense.  \section{Model Formulation}
\label{sec:basic formulation}

We consider a monopolist selling $J$ products denoted by $\cJ:=\{1,\dots, J \}$ to a buyer, who values the products at a valuation vector $\bv\in \R_+^J$. The buyer's valuations $\bv$ is unknown to the seller and thus modeled as a random vector drawn from an unknown distribution $\F$. 
Valuations are additive, which means the buyer values a subset $\cS\subseteq \cJ$ of products at $\sum_{j\in \cS} v_j$.
The seller knows neither the buyer's valuation $\bv$ nor its distribution $\F$, but knows that $\bv$ lies within $\cV:=[\underline{v}_1,\overline{v}_1]\times [\underline{v}_2,\overline{v}_2]\times\dots\times [\underline{v}_J,\overline{v}_J]$, where $\underline{v}_j$ and $\overline{v}_j$ represent the minimum and maximum valuations for product $j$, respectively.   
We assume the upper bound of the valuation for each product is finite and at least one product has a positive lower bound, i.e., $\overline{v}_j<\infty, \forall j\in \cJ$ and $\max_{j\in\cJ}\underline{v}_j>0$.
Let $\Delta(\cV)$ denote all probability distributions defined on $\cV$. 
The seller focuses on direct mechanisms denoted by $M=(\bq,t)$. Here in the allocation rule $\bq:\cV\to [0,1]^J$, $[\bq(\bv)]_j$ specifies the probability that the seller allocates product $j$ to a buyer who reports valuation $\bv$ and in the payment rule $t: \cV \to \R_+$, $t(\bv)$ is the expected payment that the seller requests from a buyer who reports valuation $\bv$. The buyer is risk-neutral with quasilinear utility. Hence, a buyer whose true valuation is 
$\bv$ and who reports $\bv'$ receives expected utility $\bq(\bv')^\top \bv- t(\bv')$. The selling mechanism must satisfy the incentive compatibility (IC) and individual rationality (IR) constraints:
\begin{equation*}
    \begin{cases}
      \bq(\bv)^\top  \bv  - t(\bv) \ge  \bq(\bv')^\top \bv- t(\bv'), \quad \forall \bv,\bv'\in \cV & \quad \text{(IC)}\\
        \bq(\bv)^\top  \bv  - t(\bv) \ge 0, \quad \forall \bv \in \cV & \quad \text{(IR)}.
    \end{cases}
\end{equation*}
Incentive compatibility (IC) ensures that each buyer maximizes their utility by truthfully reporting their valuation, and individual rationality (IR) guarantees that each buyer receives a non-negative utility from participation.
In the following, we denote the set of all feasible mechanisms by $\cM$, which is the collection of all direct mechanisms satisfying the IC and IR constraints. 

Knowing only the support \(\cV\), the seller faces the ambiguity set \(\Delta(\cV)\) of all probability distributions on \(\cV\). We adopt a benchmark as the optimal revenue achieved by a clairvoyant who knows the exact distribution $\F$ of the buyer's valuation, which is denoted by $\Rev(\OPT,\F)$. 
The seller's expected revenue from a mechanism $M\in \cM$ under a specific valuation distribution $\F$ is denoted by $\Rev(M,\F) = \int_{\bv\in \cV}t^M(\bv) \ d\F(\bv)$. The performance of a mechanism $M$ is evaluated by the ratio of the expected revenue achieved by $M$, i.e., $\Rev(M,\F)$ to the optimal revenue achievable by the clairvoyant, i.e., $\Rev(\OPT,\F)$, given the same valuation distribution $\F$. After the seller determines the selling mechanism $M$, the adversarial nature will choose a distribution $\F\in \Delta(\cV)$ to minimize the performance ratio obtained by the seller's chosen mechanism $M$. 
The seller seeks a \emph{robustly optimal} mechanism, meaning it maximizes the worst-case performance ratio across all possible distributions $\F$ that nature could choose. 
Formally, the seller solves
\begin{align}
    \cR^* =  \sup_{M\in \cM}\inf_{\F\in \cF} \frac{\Rev(M,\F)}{\Rev(\OPT,\F)} = \sup_{M\in \cM}\inf_{\F\in \cF} \frac{\Rev(M,\F)}{\sup_{M'\in \cM}\Rev(M',\F)}, \label{eq:original}
\end{align}
where the ambiguity set includes all possible distributions defined on support $\cV$:
\begin{align*}
    \cF = \Delta(\cV) =  \Delta([\underline{v}_1,\overline{v}_1]\times\cdots \times [\underline{v}_J,\overline{v}_J]).
\end{align*}
We denote $\cR^*$ as the maximum performance ratio that the seller can achieve.
We say a mechanism $M$ is \emph{robustly optimal} if it is an optimal solution to Problem \eqref{eq:original}.
To simplify Problem \eqref{eq:original}, we demonstrate that for any seller's mechanism, nature's best response is a single-point distribution that characterizes both the buyer's valuations and the hindsight optimal selling mechanism. In other words, nature strategically picks a specific valuation $\bv$ within the feasible range $\cV$ and treats it as the buyer’s type. Under this valuation, the clairvoyant benchmark sells the entire bundle with probability one and charges the total value $\allone^\top \bv$. This is formally proved in \Cref{lemma:single}.
\begin{lemma}
    For any seller's mechanism $M$,  nature's optimal strategy $(\F,M')$  is a single-point strategy, i.e., $\F$ is a one-point distribution at a valuation $\bv \in \cV$ and $M'$ sells the full bundle at price $\allone^\top \bv$, so Problem \eqref{eq:original} has the same objective value as 
   \begin{align} 
\sup_{M\in \cM}\min\limits_{\bv\in\cV} \frac{t(\bv)}{\allone^\top \bv}.
\label{eq:simplify}
  \end{align}   
    \label{lemma:single}
\end{lemma}
 Based on \Cref{lemma:single},  we hope to find the optimal solution to Problem \eqref{eq:simplify}. Since $t(\cdot)$ could be a complicated function of $\bv$, it is challenging to derive the optimal $t$ directly. We therefore focus on a tractable and interpretable subclass, \emph{separable mechanisms}, defined as follows.
    \begin{definition}[Separable Mechanism]
    \label{def-sep}
    A mechanism $M=(\bq,t)$ is \textit{separable} if there exist single-dimension mechanisms $\{(q_j,t_j)\}_{j\in \cJ}$, such that $\bq(\bv) = \bigof{q_1(v_1), \dots, q_J(v_J)}$ and $t(\bv) = \sum_{j\in \cJ} t_j(v_j)$,  where $q_j: [\underline{v}_j,\overline{v}_j] \to [0,1],\, t_j:  [\underline{v}_j,\overline{v}_j] \to \R_+$. 
Moreover, $(q_j(v_j),t_j(v_j))$ satisfies incentive compatibility and individual rationality constraints for selling product $j$:
    \begin{equation*}
    \begin{cases}
      q_j(v_j) \cdot v_j  - t_j(v_j) \ge   q_j(v'_j) \cdot v_j  - t_j(v'_j), & \quad \forall\, v_j, v'_j \in [\underline{v}_j, \overline{v}_j],\, \forall j\in \cJ \\
        q_j(v_j) \cdot v_j  - t_j(v_j) \ge  0 , & \quad \forall\,  v_j \in [\underline{v}_j, \overline{v}_j], \, \forall j\in \cJ .
    \end{cases}
    \end{equation*}
\end{definition}
\Cref{semi-feasible} shows that a separable mechanism is feasible for the multi-item screening problem.
\begin{lemma}
\label{semi-feasible}
Any separable mechanism $(\bq,t)$ defined in \Cref{def-sep} satisfies the incentive compatibility and individual rationality constraints.
\end{lemma}
 
  Denoting $\cM_j$ the set of all single-dimensional incentive compatible and individually rational mechanisms $m_j$ for product $j$, where $m_j=(q_j(\cdot), t_j(\cdot))$ with $q_j: [\underline{v}_j,\overline{v}_j] \to [0,1],\, t_j:  [\underline{v}_j,\overline{v}_j] \to \R_+$, then a feasible solution to \eqref{eq:simplify} can be found by solving the following problem: 
     \begin{align} 
      \cR_{\text{Sep}} = 
\sup_{\gamma, \, \{m_j\in \cM_j\}_{j\in\cJ}} & \gamma \label{eq:sep}\\
\mbox{s.t.}\quad & \sum_{j\in \cJ}t_j(v_j)- \gamma \sum_{j\in \cJ} v_j \ge 0,\quad \forall \bv\in \cV \nonumber
  \end{align}  
  Since Problem~\eqref{eq:sep} imposes an additional restriction that the payment rule is separable, i.e., $t(\bv) = \sum_{j\in \cJ} t_j(v_j)$, the optimal objective value of Problem \eqref{eq:sep} is no greater than that of \eqref{eq:simplify}, i.e. $\cR_{\text{Sep}}\le \cR^*$.

To build intuition for constructing solutions to Problem~\eqref{eq:sep}, we present a candidate
family of feasible solutions and defer the formal proof of the solution's optimality to \Cref{sec:separable}.
First, to satisfy the constraint $\sum_{j\in \cJ}t_j(v_j)- \gamma \sum_{j\in \cJ} v_j\ge 0$, it suffices that each single-dimensional payment satisfies $t_j(v_j) \ge \gamma v_j, \forall \, v_j\in[\underline{v}_j,\overline{v}_j], j\in \cJ$.
Motivated by this, consider a simple candidate solution: $t_j(v_j) = \gamma v_j$ for all $v_j\in [\underline{v}_j,\overline{v}_j], \, j\in \cJ$. 
By the single-dimensional incentive compatibility conditions \citep{myerson1981optimal}, this payment rule induces an allocation rule $q_j(v_j) = \gamma(1+ \ln \frac{v_j}{\underline{v}_j})$, for all $v_j\in [\underline{v}_j,\overline{v}_j], \, j\in \cJ$. 
Since feasibility requires $q_j(v_j)\le 1$, we have that $\gamma \le 1/(1+\ln\frac{\overline{v}_j}{\underline{v}_j})$ for all $j\in \cJ$. Thus, this candidate $t_j(v_j) = \gamma v_j$ yields an approximation ratio of at most $\min_{j\in \cJ} \frac{1}{1+\ln \overline{v}_j/\underline{v}_j}$. 
Although not optimal, it
provides useful intuition for a stronger mechanism.
To improve upon it, fix the allocation probability $q_j(\overline{v}_j)=1$ for all $j\in\cJ$.
For each dimension $j$, as $v_j$ decreases,  maintain a constant slope $\gamma$ for the payment function $t_j(v_j)$ so that the constraint $\sum_{j\in \cJ} t_j(v_j) - \gamma \sum_{j\in \cJ} v_j \ge 0$ remains binding, until the allocation probability $q_j(v_j)$ reaches zero. By Myerson's Lemma \citep{myerson1981optimal}, $\gamma = \frac{d t_j(v_j)}{d v_j} = v_j\cdot\frac{d q_j(v_j)}{d v_j}$, which yields $q_j(v_j) = \of{1+\gamma \ln(v_j/\overline{v}_j)}^+$ for all $j\in \cJ$. Finally, applying the incentive compatibility constraints, we explicitly derive the corresponding payment function and formally propose the following mechanism for 
    $\gamma\in [0,1]$:
 \begin{align}
\begin{aligned}
  \bq(\bv)& =\bigof{q_1(v_1), \dots, q_J(v_J)}, \ \text{where } q_j(v_j) = \bigof{\gamma \cdot\ln (v_j/\overline{v}_j)+1}^+  \\
     t(\bv) & = \sum_{j\in \cJ} t_j(v_j), \ \text{where } t_j(v_j) =\begin{cases}
       \gamma\cdot  \bigof{v_j -e^{-1/\gamma} \cdot \overline{v}_j}^+  & \text{if } e^{-1/\gamma}\cdot \overline{v}_j>\underline{v}_j\\
         \gamma\cdot v_j +\underline{v}_j\cdot\bigof{\gamma(\ln(\underline{v}_j/\overline{v}_j)-1)+1} &  \text{if } e^{-1/\gamma}\cdot \overline{v}_j\le \underline{v}_j
     \end{cases} 
     \end{aligned}
        \label{eq:q}
     \tag{$M_\gamma$}
\end{align}
    By construction, the per-product allocation \(q_j(v_j)\) and payment \(t_j(v_j)\) in \ref{eq:q} depend
only on the item’s own valuation \(v_j\) and are independent of \(\bv_{-j}\). The scalar parameter
\(\gamma\), however, may be chosen as a function of the support bounds
\(\bigofff{\underline{v}_j,\overline{v}_j}_{j\in\cJ}\) (but not of the realized valuations).
We next show that \ref{eq:q} is feasible for Problem~\eqref{eq:simplify} in \Cref{lemma:semi} and establish its
approximation guarantee in \Cref{prop:ratio-feasible}; the proof of optimality is deferred to \Cref{sec:separable}. 

 \begin{lemma}
     The mechanism \ref{eq:q} is separable.
     \label{lemma:semi}
 \end{lemma}
\Cref{lemma:semi} shows that \ref{eq:q} satisfies single-dimensional (coordinate-wise) incentive compatibility and individual rationality, and hence defines a separable mechanism.
According to \Cref{semi-feasible} and \Cref{lemma:semi}, the separable mechanism defined in \ref{eq:q} is feasible for the multi-item screening problem. 
A primary benefit of separable mechanisms is that the allocation probability of one product does not depend on the valuation realization for other products, significantly simplifying the design and implementation complexity typically associated with probabilistic mixed-bundling. Specifically, the separable mechanism \ref{eq:q} can be implemented by a randomized posted price mechanism for each product $j\in \cJ$ independently, with a price density function of $$\pi_j(v_j) = \frac{\gamma}{v_j} \text{ for } v_j\in \bigoff{\max\{\underline{v}_j,e^{-1/\gamma}\cdot \overline{v}_j\},\, \overline{v}_j},$$ together with a point mass of  $1+\gamma\ln(\underline{v}_j/\overline{v}_j)$ at $\underline{v}_j$, if $e^{-1/\gamma}\cdot \overline{v}_j\le \underline{v}_j$. Hence, under separable mechanisms, buyers do not need to evaluate and compare utilities from numerous randomized bundles, simplifying their decision-making process considerably.
In addition to its implementation simplicity, in \Cref{prop:ratio-feasible}, we evaluate the performance ratio achieved by mechanism \ref{eq:q}.

\begin{proposition}
For any $\gamma$, denote $\cS(\gamma)=\bigofff{j\in \cJ \mid \underline{v}_j/\overline{v}_j<e^{ -1/\gamma}}$. Let $\gamma^*\in (0,1]$ be the unique solution to $\phi(\gamma)=\gamma\cdot e^{-1/\gamma} \cdot \sum_{j\in \cS(\gamma)} \overline{v}_j - \sum_{j\in \cJ\setminus\cS(\gamma)} \bigof{\underline{v}_j\cdot\bigof{\gamma\ln(\underline{v}_j/\overline{v}_j)-\gamma+1}} = 0$. Then the mechanism $M_{\gamma^*}$ achieves an approximation ratio of $\gamma^*$, i.e., $\min\limits_{\bv\in\cV} \frac{t^{M_{\gamma^*}}(\bv)}{\allone^\top \bv}= \gamma^*$, where $t^{M_{\gamma^*}}$ denotes the payment rule under mechanism $M_{\gamma^*}$.
\label{prop:ratio-feasible}
\end{proposition}
For completeness, in \Cref{increasing-g}, we establish that there is a unique solution $\gamma^*\in (0,1]$ to $\phi(\gamma)=0$, so the performance ratio $\gamma^*$ in \Cref{prop:ratio-feasible} is well-defined.
\begin{lemma}
Function $\phi(\gamma) = \gamma\cdot e^{-1/\gamma} \cdot \sum_{j\in \cS(\gamma)} \overline{v}_j - \sum_{j\in \cJ\setminus\cS(\gamma)} \bigof{\underline{v}_j\cdot\bigof{\gamma\ln(\underline{v}_j/\overline{v}_j)-\gamma+1}}$ is increasing in $\gamma$ over the interval $[0,1]$ and there is a unique solution $\gamma^*\in(0,1]$ to $\phi(\gamma)=0$.
\label{increasing-g}
\end{lemma}

\begin{figure}[htbp]
\caption{\small Illustration of the Separable Mechanism in \Cref{cor:2item} for $\underline{v}_1=0.01$, $\underline{v}_2=0.5$, $\overline{v}_1 = \overline{v}_2=1$}
\centering
\begin{subfigure}{0.32\textwidth}
\includegraphics[width=\linewidth]{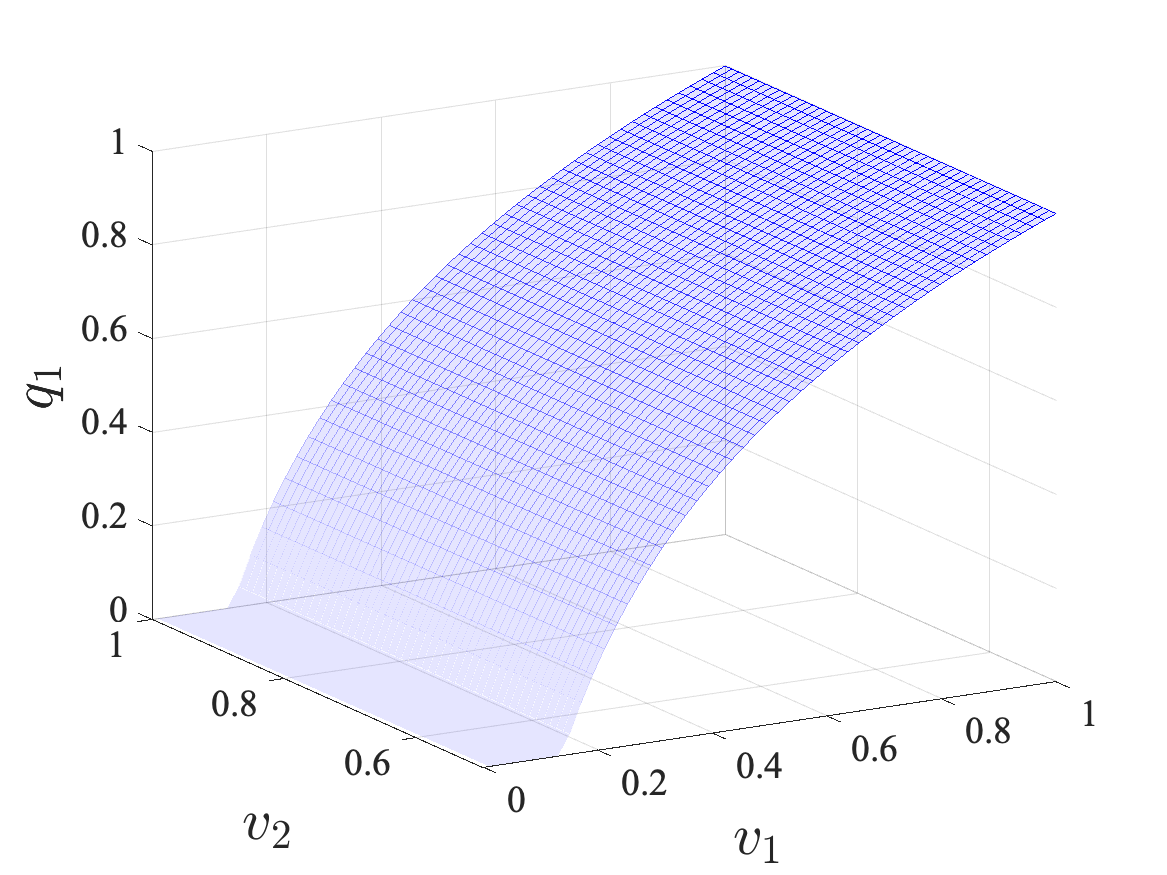}
  \caption{Allocation Probability $q_1$}
  \label{fig:sub1}
\end{subfigure}
\begin{subfigure}{0.32\textwidth}
\includegraphics[width=\linewidth]{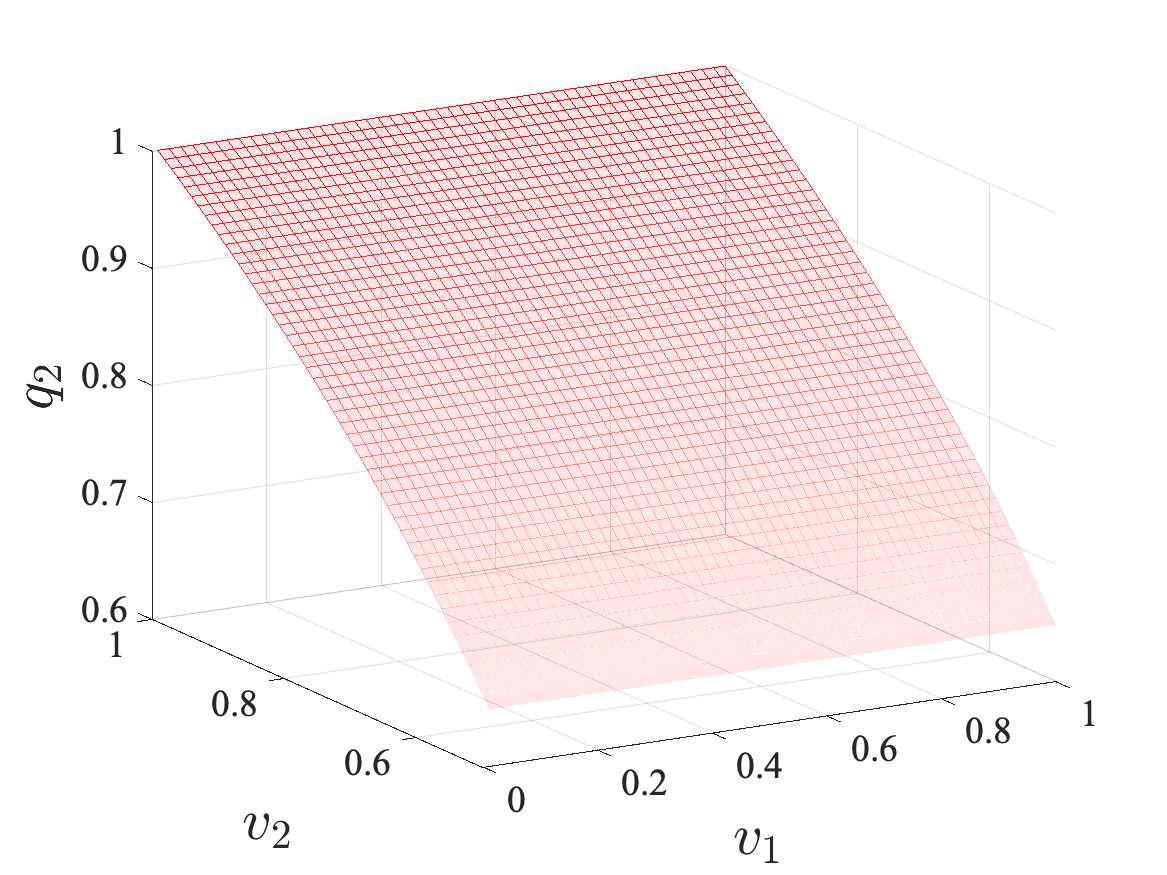}
  \caption{Allocation Probability $q_2$}
  \label{fig:sub2}
\end{subfigure}
\begin{subfigure}{0.33\textwidth}
\includegraphics[width=\linewidth]{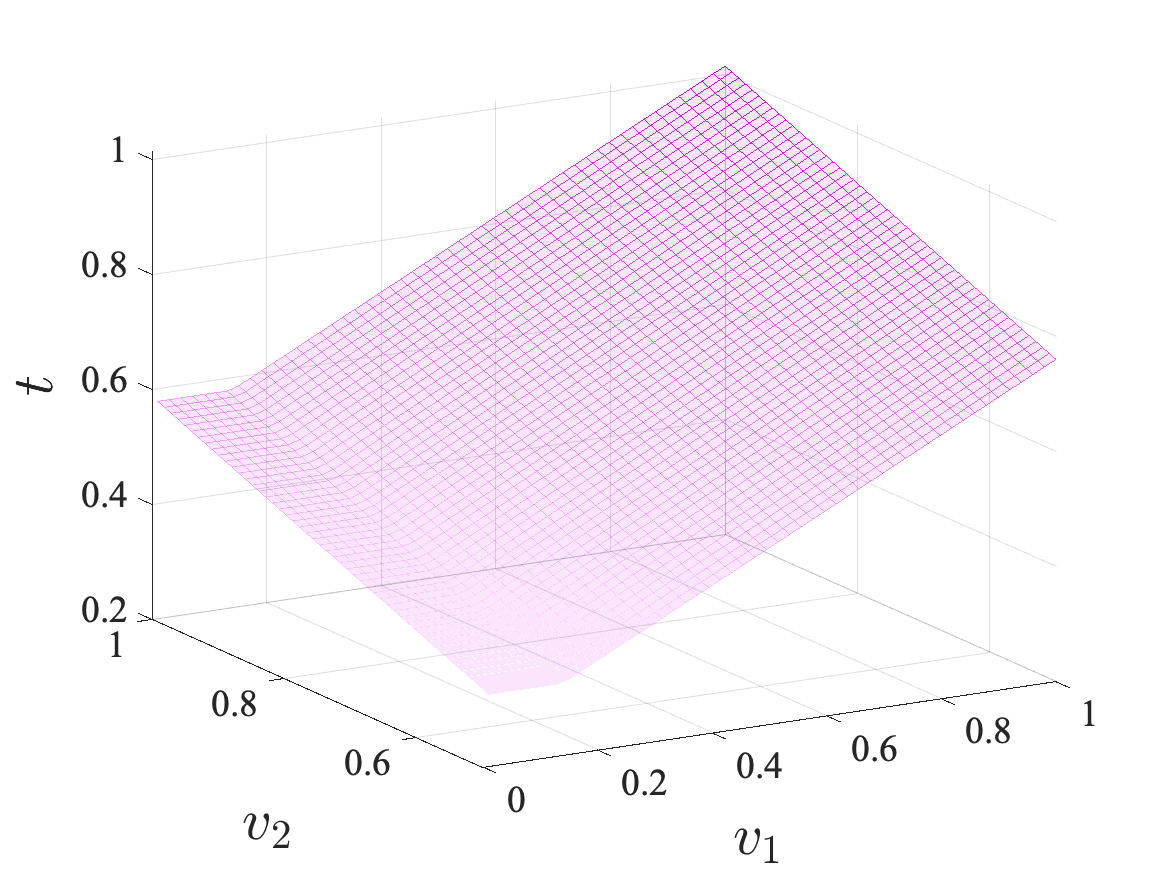}
  \caption{Payment $t$}
  \label{fig:sub3}
\end{subfigure}
\label{fig:2d}
\end{figure}

\Cref{prop:ratio-feasible} provides a performance guarantee for the mechanism $M_{\gamma^*}$, which serves as a lower bound of the optimal performance ratio in Problem \eqref{eq:simplify}, since 
\begin{align*}
\sup_{M\in \cM}\min\limits_{\bv\in\cV} \frac{t^M(\bv)}{\allone^\top \bv} 
\ge \min\limits_{\bv\in\cV} \frac{t^{M_{\gamma^*}}(\bv)}{\allone^\top \bv} =\gamma^*.
\end{align*}
To provide a more intuitive illustration of mechanism $M_{\gamma^*}$, we present an example of the two-item case as follows.
Without loss of generality, suppose $\underline{v}_1/\overline{v}_1\le \underline{v}_2/\overline{v}_2$ and $\underline{v}_2>0$. Then according to \Cref{prop:ratio-feasible}, the mechanism $M_{\gamma^*}$ when $J=2$ is constructed in \Cref{cor:2item}.
\begin{corollary}[Two Products]
When there are two products where $\underline{v}_1/\overline{v}_1\le \underline{v}_2/\overline{v}_2$ and $\underline{v}_2>0$, the following selling mechanism and its performance ratio are feasible for Problem \eqref{eq:simplify}.
\begin{enumerate}
    \item If $\underline{v}_{2}\,\overline{v}_{1}\;>\;\overline{v}_{2}\,\underline{v}_{1} e^{\,1+\underline{v}_{1}/\underline{v}_{2}}$, the approximation ratio $\gamma =\left(W(\frac{\overline{v}_1}{e\overline{v}_2}) + \ln \frac{\overline{v}_2}{\underline{v}_2}+1\right)^{-1}$, and the selling mechanism is defined as
    \begin{align*}
        \bq(\bv) =\of{ \of{\gamma \cdot\ln (v_1/\overline{v}_1)+1}^+, \of{\gamma \cdot\ln (v_2/\overline{v}_2)+1}}, \quad
     t(\bv)  = \gamma\cdot\of{\max\offf{v_1,\, e^{-1/\gamma} \cdot \overline{v}_1}+v_2},
    \end{align*}
    where $W$ is the Lambert-W function defined as the inverse function of $f(W) = We^W$.
    \item If $\underline{v}_{2}\,\overline{v}_{1}\;\le \;\overline{v}_{2}\,\underline{v}_{1} e^{\,1+\underline{v}_{1}/\underline{v}_{2}}$, the approximation ratio is 
$\gamma = \frac{\sum_{j\in \cJ} \underline{v}_j}{\sum_{j\in \cJ} \of{ \underline{v}_j \cdot \of{1+\ln(\overline{v}_j/\underline{v}_j)} }}$, and the  selling mechanism is defined as
 \begin{align*}
     \bq(\bv) =\bigof{\of{\gamma \cdot\ln (v_1/\overline{v}_1)+1},\of{\gamma \cdot\ln (v_2/\overline{v}_2)+1}},  \quad
     t(\bv)  =  \gamma\cdot \bigof{v_1+v_2}.
 \end{align*}
\end{enumerate}
    \label{cor:2item}
\end{corollary}
\Cref{fig:2d} provides a graphical illustration of the separable mechanism defined in \Cref{cor:2item} for $\underline{v}_1=0.01$, $\underline{v}_2=0.5$, $\overline{v}_1 = \overline{v}_2=1$. \Cref{fig:sub1} and \Cref{fig:sub2} present the allocation probability for product 1 and product 2, respectively. 
The allocation probability depends only on the valuation in the corresponding dimension but not on the valuation in the other dimension. \Cref{fig:sub3} visualizes that the payment function is a piecewise affine function of the buyer's valuations. 

\subsection{Discussion of the Other Separable Mechanisms}
There are different ways to construct separable mechanisms satisfying \Cref{def-sep}. For instance, a commonly adopted separable mechanism is the optimal solution in the decomposed one-dimensional mechanism for each product \citep{carroll2017robustness,koccyiugit2022robust}. 
According to \cite{eren2010monopoly}, when the support of valuation for product $j$ is between $[\underline{v}_j,\overline{v}_j]$, where $\underline{v}_j>0,\, \overline{v}_j<\infty$, the optimal screening mechanism for the one-dimensional problem is 
\begin{align}
    q_j^\dag (v_j) = \frac{1+\ln (v_j/\underline{v}_j) }{1+\ln(\overline{v}_j/\underline{v}_j)}, \quad t_j^\dag(v_j) = \frac{v_j }{1+\ln(\overline{v}_j/\underline{v}_j)}, \quad \forall v_j\in[\underline{v}_j,\overline{v}_j]
    \label{eq:separate}
\end{align}
which achieves an approximation ratio of $r_j^\dag =  \frac{1 }{1+\ln(\overline{v}_j/\underline{v}_j)}$ in dimension $j$. This mechanism can be implemented as a randomized pricing scheme with a price density function of $\frac{1 }{\of{1+\ln(\overline{v}_j/\underline{v}_j)}v_j}$ for $v_j\in (\underline{v}_j,\overline{v}_j]$ and a probability mass of $\frac{1}{1+\ln(\overline{v}_j/\underline{v}_j)}$ at $\underline{v}_j$. 
\Cref{prop:separable} provides the approximation ratio achieved by mechanism defined in \eqref{eq:separate}.

\begin{figure}[htbp]
    \centering
\caption{Performance Ratios of $M_{\gamma^*}$ and Decomposed Separable Mechanism in \eqref{eq:separate} for $\underline{v}_2=0.5$, $\overline{v}_1 = \overline{v}_2=1$ with Different $\underline{v}_1$}
\includegraphics[width=0.4\linewidth]{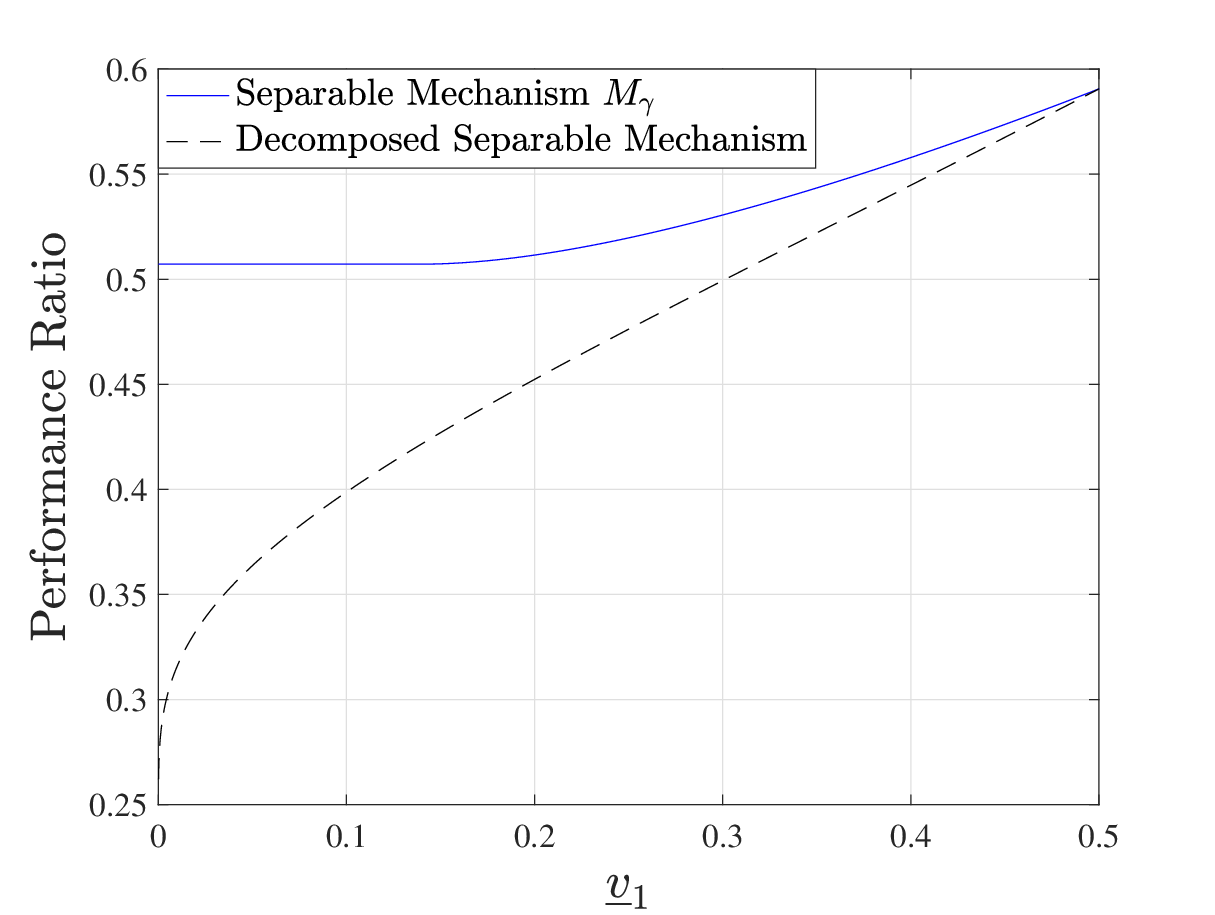}
    \label{fig:sub4}
\end{figure}

\begin{proposition}
\label{prop:separable}
    The approximation ratio achieved by the decomposed separable mechanism defined in \eqref{eq:separate} is calculated as $
    \min\limits_{j=2,\dots, J} 
 \frac{\sum_{j'=1}^{ j-1}r_{j'}^\dag \, \overline{v}_{j'} +\sum_{j'=j}^{J}r_{j'}^\dag \, \underline{v}_{j'}  }{\sum_{j'=1}^{j-1} \, \overline{v}_{j'} +\sum_{j'=j}^{J} \, \underline{v}_{j'} }$,
where $j\in \cJ$ is sorted in increasing order of $\{\underline{v}_j/\overline{v}_j\}$, i.e. $\underline{v}_1/\overline{v}_1\le \dots, \underline{v}_J/\overline{v}_J$, and $r_j^\dag =  \frac{1 }{1+\ln(\overline{v}_j/\underline{v}_j)}$ for all $j\in \cJ$.
\end{proposition}
\Cref{fig:sub4} compares the approximation ratios of the separable mechanism defined in \Cref{cor:2item} (depicted in blue solid line) and that of the optimal decomposed separable mechanism defined in \eqref{eq:separate} (depicted in black dashed line), for different $\underline{v}_1$. It shows that when $\underline{v}_1$ is small, the performance ratio of the decomposed separable mechanism \eqref{eq:separate} deteriorates significantly while that of the separable mechanism in \Cref{cor:2item} is stable. 
This illustrates the robustness of the separable mechanism established in \Cref{prop:ratio-feasible}. 
\Cref{cor:semi-sep} provides an instance where the approximation ratio achieved by the separable mechanism defined in \Cref{prop:ratio-feasible} can be significantly larger than that achieved by the decomposed separable mechanism in \eqref{eq:separate}.
\begin{corollary}
\label{cor:semi-sep}
    Suppose only one product has a positive lower bound for the valuation, i.e., $\cV=[0,\overline{v}_1]\times[0,\overline{v}_2]\dots \times[0,\overline{v}_{J-1}]\times[\underline{v}_J,\overline{v}_J]$. The approximation ratio achieved by the decomposed separable mechanism in \eqref{eq:separate} is $\cR_{dec} = \Bigof{\bigof{(\sum_{j=1}^{J-1}\overline{v}_j)/{\underline{v}_J}+1}\cdot \of{1+\ln(\overline{v}_J/\underline{v}_J)} }^{-1}$ and the approximation ratio achieved by the separable mechanism in \Cref{prop:ratio-feasible} is $\cR_{M_{\gamma^*}} = \bigof{1+\ln(\overline{v}_J/\underline{v}_J)+W(\frac{\sum_{j=1}^{J-1}\overline{v}_j}{e\overline{v}_J}) }^{-1}$, where $W$ is the Lambert-W function defined as the inverse function of $f(W) = We^W$.

\end{corollary}
Consider a simple example where $\cV = [0,1]\times [0,1]\times \dots\times [\epsilon,1]$. Then \Cref{cor:semi-sep} implies that $\cR_{dec} = \frac{\epsilon}{(J-1+\epsilon)(1+\ln(1/\epsilon))}\xrightarrow{\substack{J \to \infty \\ \epsilon \to 0}}\frac{\epsilon}{J(1+\ln(1/\epsilon))}$ and $\cR_{M_{\gamma^*}} = \bigof{1+\ln(1/\epsilon)+W(\frac{J-1}{e}) }^{-1}\xrightarrow{\substack{J \to \infty \\ \epsilon \to 0}} \frac{1}{1+\ln(1/\epsilon)+\ln J}$.  As $\epsilon$ decreases, and the number of products $J$ increases, the performance of $\cR_{M_{\gamma^*}}$ degrades much more slowly than that of the decomposed separable mechanism $\cR_{dec}$.
 \Cref{cor:semi-sep} highlights that the performance ratio of the decomposed separable mechanism as shown in \Cref{prop:separable} relative to that of the separable mechanism provided in \Cref{prop:ratio-feasible} can become arbitrarily small. This occurs when there are many products with negligible lower bounds and large upper bounds.
While \Cref{prop:ratio-feasible} demonstrates that mechanism $M_{\gamma^*}$ achieves a performance ratio of $\gamma^*$, it does not prove its optimality among all IC and IR mechanisms. In the next section, we will prove that the separable mechanism $M_{\gamma^*}$ proposed in \Cref{prop:ratio-feasible} is optimal across all IC and IR mechanisms.

\section{Optimality of the Separable Mechanism $M_{\gamma^*}$}
\label{sec:separable}
In this section, we establish the optimality of the separable mechanism defined in \Cref{prop:ratio-feasible} via a saddle point approach.
Notice that Problem \eqref{eq:simplify} is equivalent to the following problem, where nature takes a mixed strategy in the inner problem:
   \begin{align}
\cR^* = \sup_{M\in \cM}\min\limits_{\bv\in \cV}  \Bigoff{\frac{t(\bv)}{\allone^\top \bv}} = \sup_{M\in \cM}\min\limits_{\F\in \cF} \E_{\bv\sim \F} \Bigoff{\frac{t(\bv)}{\allone^\top \bv}}
\label{eq:saddle}
    \end{align}
Therefore, finding a robustly optimal mechanism amounts to solving a zero-sum game between the seller and nature where the seller chooses a mechanism $M\in \cM$ against the adversarial nature, who chooses a distribution $\F$ over $\bv$ to minimize the seller's expected performance ratio. Here $\bv$ simultaneously represents the buyer's valuation and the corresponding hindsight optimal posted price for all products. With a little abuse of notation, denote the performance ratio achieved by the seller's mechanism $M$ and nature's strategy $\F$ as
$$\cR(M,\F) = \E_{\bv\sim \F} \Bigoff{\frac{t^M(\bv)}{\allone^\top \bv}},$$
where $t^M(\cdot)$ is the payment rule in mechanism $M$.
A saddle point of Problem \eqref{eq:saddle} is defined below.
\begin{definition}
    The solution $(M^*,\F^*)$ is a saddle point to problem \eqref{eq:saddle} if and only if $\cR(M, \F^*)\le \cR(M^*, \F^*)\le \cR(M^*, \F)$, for all $M\in \cM$ and $\F\in \cF$.
    \label{def-saddle}
\end{definition}
If there exists a saddle point $(M^*,\F^*)$, then $\sup_{M\in \cM}\inf_{\F\in \cF}\cR(M, \F) \le \sup_{M\in \cM}\cR(M, \F^*) \le \cR(M^*, \F^*) \le \inf_{\F\in \cF}\cR(M^*, \F) \le \sup_{M\in \cM}\inf_{\F\in \cF}\cR(M, \F) $. 
Therefore, the optimal approximation ratio $\cR^*$ is equal to $\cR(M^*, \F^*)$, and $M^*$ and $ \F^*$ are the robustly optimal strategies for the seller and nature, respectively. Hence, identifying a saddle point as described in \Cref{def-saddle} suffices for solving our problem.
The conditions given in \Cref{def-saddle} can equivalently be restated as:
    \begin{align*}
 \sup_{M\in \cM}\E_{\bv\sim \F^*} \Bigoff{\frac{t^M(\bv)}{\allone^\top \bv}}  \le \E_{\bv\sim \F^*} \Bigoff{\frac{t^{M^*}(\bv)}{\allone^\top \bv}} \le 
   \min\limits_{\F\in \cF} \E_{\bv\sim \F} \Bigoff{\frac{t^{M^*}(\bv)}{\allone^\top \bv}}. 
    \end{align*}
In \Cref{prop:ratio-feasible}, we verified that the separable mechanism $M_{\gamma^*}$ defined in \Cref{prop:ratio-feasible} achieves $\min\limits_{\F\in \cF} \E_{\bv\sim \F} \Bigoff{\frac{t^{M_{\gamma^*}}(\bv)}{\allone^\top \bv}} = \gamma^*$. Given the optimal approximation ratio $\cR^* = \sup\limits_{M\in \cM}\min\limits_{\F\in \cF} \E_{\bv\sim \F} \Bigoff{\frac{t^M(\bv)}{\allone^\top \bv}} $, we thus have $\gamma^*\le \cR^*$. 
In order to prove the optimality of the solution provided in \Cref{prop:ratio-feasible}, our next step is to identify a worst-case distribution $\F^*$ such that $\sup\limits_{M\in \cM}\E_{\bv\sim \F^*} \Bigoff{\frac{t^M(\bv)}{\allone^\top \bv}} =\gamma^*$, which will imply that $\gamma^*\ge \cR^*$.
Constructing nature's strategy $\F^*$ is often challenging in the multi-item setting since it entails optimizing over joint distributions and, for each candidate distribution, solving the seller's best-response mechanism. 
Our goal is to choose $\F^*$ so that the seller’s optimal response coincides with our proposed mechanism $M_{\gamma^*}$. We proceed in two steps. First, because any negative correlation (or even independence or weak positive correlation) in $\F^*$ could make bundling preferable for the seller while our proposed mechanism in $M_{\gamma^*}$ is separable, natural candidates of $\F^*$ would be comonotonic distributions. Under comonotonicity, valuations admit a single-index representation, which reduces the design of $(\bq,t)$ to a vector-valued functional optimization problem defined on a scalar variable. 
Second, since our proposed mechanism $M_{\gamma^*}$ is implemented as randomized pricing, we then carefully design the comonotonic distribution so that the seller is indifferent among different prices for each dimension, and the functional optimization's kernel function is unimodal. Under unimodality, the functional optimization problem further reduces to a scalar optimization with a closed-form solution. We illustrate the idea below, and the formal construction and proofs are verified in \Cref{sec:multi}. 

Considering comonotonic distributions, suppose the valuation profile $\bv$ can be represented as a vector of monotonically increasing functions of a common scalar variable $\xi$, i.e. $\bv=\of{v_1(\xi), v_2(\xi),\dots, v_J(\xi)}$, where each component $v_j(\xi)$ is an increasing function of $\xi$ with $v_j(0)=0,\, \forall j\in \cJ$ and $\xi$ follows a distribution of $\G$. 
Then the allocation rule $\bq(\bv)$ and payment rule $t(\bv)$ can also be represented by functions of the scalar variable $\xi$. 
Denote $\boldsymbol{\alpha}(\xi)$ and $\tau(\xi)$ the allocation and payment at $\bv(\xi)$, respectively, i.e., $(\boldsymbol{\alpha}(\xi),\tau(\xi))= (\bq(\bv(\xi)), t(\bv(\xi)))$. 
Let $U(\bv)$ be the utility of a buyer with valuation $\bv$ under this mechanism.
By incentive compatibility, the envelope theorem \citep{milgrom2002envelope} implies $\frac{\partial U}{\partial v_j} =[\bq(\bv)]_j$ for each $j$, so $dU\of{\bv(\xi)} = \bq(\bv(\xi))^\top \,  d\bv(\xi)$. Integrating from $0$ to $\xi$ yields $U\of{\bv(\xi)} =U(\bv\of{0}) + \int_{0}^{\xi}  \bq(\bv(x))^\top \,  d\bv(x) = \int_{0}^{\xi}  \boldsymbol{\alpha}(x)^\top \,  d\bv(x)$ with standard normalization $U(\bv(0))=U(\mathbf{0})=0$ (by IR). Hence, the payment function can be represented as
\begin{align*}
    \tau(\xi) = \boldsymbol{\alpha}(\xi)^\top \cdot \bv(\xi) -\int_{0}^{\xi}  \boldsymbol{\alpha}(x)^\top \,  d\bv(x)
 = \sum_{j\in \cJ} \bigof{\int_{0}^\xi v_j(x)\, d\alpha_j(x)},
\end{align*}
where the second equality follows from integration by parts. Thus, the approximation ratio becomes
{\small
\begin{align*}
    \mathop{\E}\limits_{\bv\sim \F} \Bigoff{\frac{t(\bv)}{\allone^\top \bv}} & = \mathop{\E}\limits_{\xi \sim \G} \Bigoff{\frac{\tau(\xi)}{ \sum\limits_{i\in\cJ}v_i(\xi)}} = \int_{0}^{\infty} \Bigoff{\frac{ \sum\limits_{j\in \cJ} \bigof{\int_{0}^\xi v_j(x) d\alpha_j(x)}}{ \sum_{i\in\cJ}v_i(\xi)}} d\G(\xi) =\sum_{j\in \cJ} \Bigoff{\int_{0}^{\infty} \Bigof{v_j(\xi) \int_{\xi}^\infty \frac{d\G(x)}{ \sum\limits_{i\in\cJ}v_i(x)}}\, d\alpha_j(\xi) }. 
\end{align*}
}
Hence, the mechanism design problem reduces to the following scalar functional optimization: 
\begin{equation}
    \max_{\boldsymbol{\alpha}(\cdot)}\sum_{j\in \cJ} \Bigoff{\int_{0}^{\infty} \Bigof{v_j(\xi) \int_{\xi}^\infty \frac{d\G(x)}{ \sum_{i\in\cJ}v_i(x)}}\, d\alpha_j(\xi) } 
    \label{eq:scalar}
\end{equation}
subject to incentive compatibility, individual rationality, and feasibility constraints ($\alpha_j(\xi)\in [0,1],\, \forall \xi\in[0,\infty),\, j\in \cJ$). 
Though this function $\boldsymbol{\alpha}(\xi)$ is defined on a scalar variable, which is already simplified from the original multi-dimensional mechanism design problem, the optimal structure of $\boldsymbol{\alpha}(\xi)$ is not straightforward to characterize. 
The optimal form of $\alpha_j(\xi)$ can vary substantially and need not be monotone, depending on the specific choices of $v_j(\xi)$ and $\G$.

  We now sketch the intuition behind the construction of the worst-case distribution $\F^*$, and the formal construction and proofs are in \Cref{sec:multi}. In \eqref{eq:scalar}, the kernel function $v_j(\xi) \int_{\xi}^\infty \frac{d\G(x)}{ \sum_{i\in\cJ}v_i(x)}$ can be interpreted as the expected contribution to the approximation ratio from posting price $v_j(\xi)$ for item $j$. To induce randomized pricing in \ref{eq:q} as the seller’s best response under nature’s strategy, we select $v_j(\xi)$ and $\G$ so that $v_j(\xi) \int_{\xi}^\infty \frac{d\G(x)}{ \sum_{i\in\cJ}v_i(x)}$ is constant on the region where the price distribution has positive density. Denote this constant by $c_j = v_j(\xi) \int_{\xi}^\infty \frac{d\G(x)}{ \sum_{i\in\cJ}v_i(x)}$. Since $\int_{\xi}^\infty \frac{d\G(x)}{ \sum_{i\in\cJ}v_i(x)} $ depends only on $\xi$, not on $j$, we have that $v_j(\xi)/v_{j'}(\xi) = c_j/c_{j'}$, which is independent in $\xi$. Thus, all coordinates are proportional along the path as $\xi$ changes.
  Hence, a simple representation of $v_j(\xi)$ is to take valuations linear in $\xi$ within the region of positive price density. After specifying the format of $\bv$, we can determine $\G$ by making  $v_j(\xi) \int_{\xi}^\infty \frac{d\G(x)}{ \sum_{i\in\cJ}v_i(x)}$ constant in $\xi$, although the constant may differ for different $j$.
Note that the preceding discussion provides an intuitive illustration of the construction of $\F^*$, but it is not a rigorous proof. Specifically, the exact form of $v_j(\xi)$ and the distribution of the scalar variable $\xi$ still need to be explicitly characterized. Moreover, it remains to rigorously verify that the optimal mechanism under these comonotonic distributions indeed achieves the performance ratio $\gamma^*$ provided as the lower bound in the previous section.
In \Cref{sec:2item}, we first discuss a two-dimensional problem to offer an intuitive illustration of the nature's strategy, and then we provide a rigorous proof for general $J\ge 2$ in \Cref{sec:multi}.

 \subsection{Warm Up: Two-Item Problem}
 \label{sec:2item}
\begin{figure}[htbp]
    \centering  
    \caption{Support of the Joint Distribution of $(v_1,v_2)$ when $\underline{v}_1=2,\, \underline{v}_2=4,\, \overline{v}_1=\overline{v}_2=12,\, \omega_1 = 3$}\includegraphics[width=0.4\textwidth]{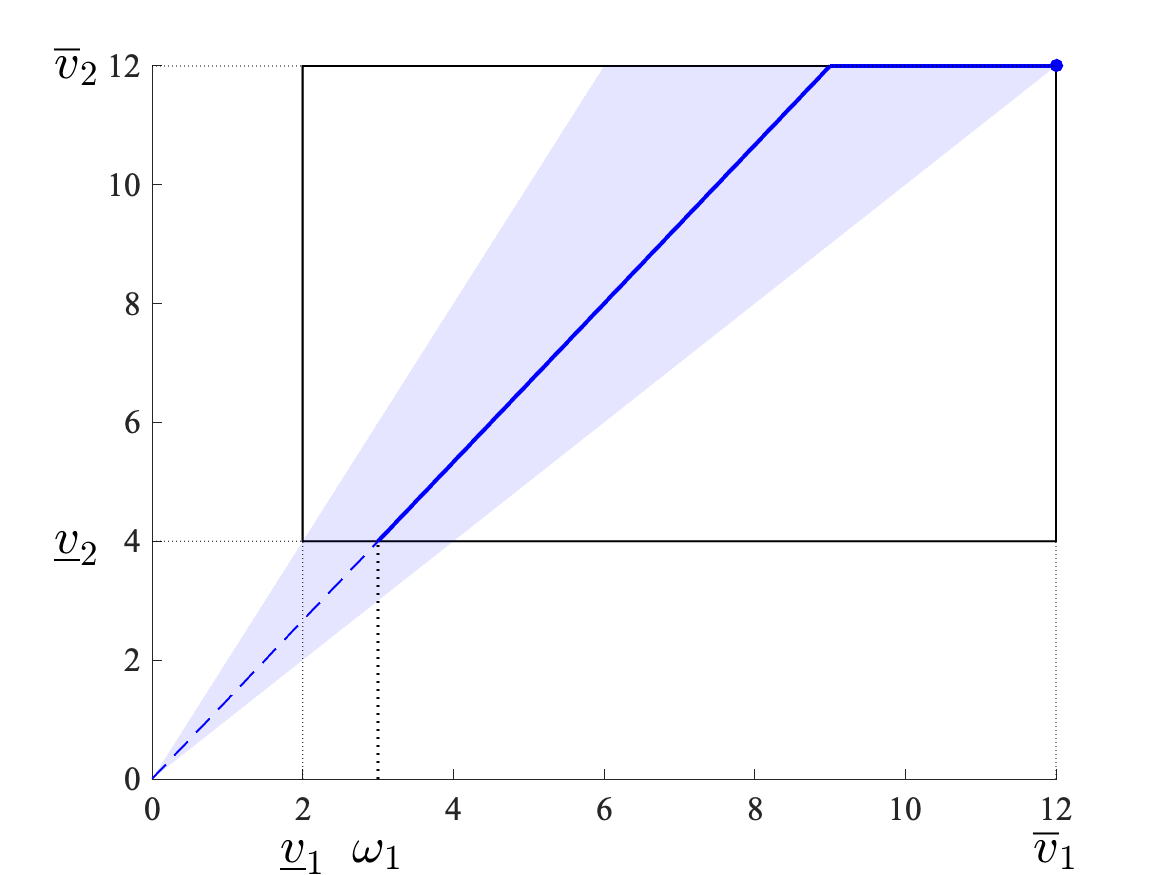}
    \label{fig_support_2D}
\end{figure}

Suppose there are two products $\cJ=\{1,2\}$ with $\underline{v}_1/\overline{v}_1\le \underline{v}_2/ \overline{v}_2$ and $\underline{v}_2>0$. 
Based on the intuition discussed above, we construct a simple valuation distribution such that $\xi \int_{\xi}^\infty \frac{d\G(x)}{ \sum_{i\in\cJ}v_i(x)}$ is constant in $\xi$ and each valuation function $v_j(\xi)$ is piecewise linear in $\xi$.
Formally, the valuation distribution $\F_{\boldsymbol{\omega}}$ of 
$\bv= (v_1,v_2)$, parametrized by $\boldsymbol{\omega}$, is supported on a one-dimensional ray defined by $\bv =\xi\cdot \boldsymbol{\omega}$ within the feasible support $\cV=[\underline{v}_1,\overline{v}_1]\times [\underline{v}_2,\overline{v}_2]$. 
Here $\boldsymbol{\omega} = (\omega_1, \omega_2)\in \R^2_+$ represents the direction of the ray and $\xi\ge 1$ is the scaling factor indicating magnitude.
In particular, the valuations of product $j=1,2$ are given by:
\begin{align}
    v_1(\xi) = \min\{\omega_1\cdot \xi,\,  \overline{v}_1\},\quad   v_2(\xi) = \min\{ \omega_2\cdot \xi, \, \overline{v}_2\}
    \label{eq:v2}
\end{align}
where $\omega_1\in [\underline{v}_1, \frac{\overline{v}_1\underline{v}_2}{\overline{v}_2}]$, $\omega_2 =\underline{v}_2$, and $\xi$ is a random variable distributed from $[1,\infty)$.
As $\xi$ grows, the path follows the ray until one coordinate first hits its cap, and then proceeds vertically or horizontally along the boundary until both coordinates reach $(\overline{v}_1,\overline{v}_2)$; mass beyond that point induces a point mass at $(\overline{v}_1,\overline{v}_2)$.
For a more intuitive illustration, \Cref{fig_support_2D} shows the support of $\F_{\boldsymbol{\omega}}$ when $\underline{v}_1=2,\, \underline{v}_2=4,\, \overline{v}_1=\overline{v}_2=12$. 
The distribution $\F_{\boldsymbol{\omega}}$ for $\omega_1=3$ is on the solid blue piecewise linear curve with a point mass on $(\overline{v}_1,\overline{v}_2)$. 
Allowing $\omega_1$ to vary from $\underline{v}_1=2$ to $\frac{\overline{v}_1\underline{v}_2}{\overline{v}_2} = 4$ sweeps out the light-blue region in \Cref{fig_support_2D}. Each choice of $\omega_1$ yields a distinct comonotone distribution $\F_{\boldsymbol{\omega}}$. 

According to the definition of $\bv$ in \eqref{eq:v2}, the distribution of $\bv$ is uniquely determined by the distribution of the scaling factor $\xi$. Define a cumulative distribution function $\G(\xi)$ as
\begin{align}
 \G(\xi) = \int_1^\xi \frac{\zeta\cdot (v_1(x)+v_2(x))}{x^2} dx,\, \, \forall \, \xi\in [1, \infty)
\label{eq:g2}
 \end{align}
 where $\zeta$ is a normalization constant ensuring that $\int_1^\infty d\G(\xi)=1$.
The definition of $\G$ ensures $$v_j(\xi)\cdot \int_{\xi}^\infty \frac{d\G(x)}{ v_1(x)+v_2(x)} \, =\begin{cases}
\omega_j\xi \cdot \int_{\xi}^\infty  \frac{\zeta\cdot d x}{x^2}\,  = \omega_j \zeta, & \xi\le \overline{v}_j/\omega_j\\
\overline{v}_j\cdot  \zeta\cdot \frac{1}{\xi}& \xi> \overline{v}_j/\omega_j.
\end{cases}$$
Since the second segment $\overline{v}_j\cdot \zeta\cdot \frac{1}{\xi} < \omega_j \zeta$ for $\xi> \overline{v}_j/\omega_j$ and is decreasing in $\xi$, we have that $v_j(\xi)\cdot \int_{\xi}^\infty \frac{d\G(x)}{ v_1(x)+v_2(x)}$ is constant up to $\overline{v}_j/\omega_j$, and strictly decreasing thereafter. 
Thus, the kernel function is unimodal in $\xi$. 
Given this unimodality, in \Cref{lemma: unimodal}, we show that an optimal solution to the functional optimization problem \eqref{eq:scalar} (ignoring IC/IR constraints for the moment) is a step function. 
\begin{lemma}
Let $\kappa:[0,\infty)\to\mathbb{R}_+$ be unimodal and continuous at its mode $x_0$, where $x_0\in\arg\max_x \kappa(x)$. Then an optimizer of
$\max\limits_{\alpha:[0,\infty)\to[0,1]} \int_{0}^{\infty} \kappa(x)\, d\alpha(x)$
is a threshold rule $\alpha(x)=\mathbbm{1}_{x\ge x_0}$. 
    \label{lemma: unimodal}
\end{lemma}
Based on \Cref{lemma: unimodal}, the optimal allocation rule solved by \eqref{eq:scalar} is an indicator function $\alpha_j(\xi) =\mathbbm{1}_{\xi\ge \tilde{\xi}_j}$, where $\tilde{\xi}_j$ can be any value between 1 and $ \overline{v}_j/\omega_j$.  This is because this optimal $\boldsymbol{\alpha}(\cdot)$ obtained under the feasibility constraints $\alpha_j(\xi)\in [0,1]$ automatically satisfies the IC and IR constraints. 
Moreover, because the kernel is flat on $[1, \overline{v}_j/\omega_j]$, any threshold within this interval attains the same objective value; hence any convex combination of such indicator rules is also optimal.
The essential insight in the construction is that, under the $\G$ and $\bv$ defined above, the seller's best response is a simple posted price $v_j(\tilde{\xi}_j)$ in each dimension, yielding a closed-form solution. 
\begin{lemma}
    If $\bv$ follows the distribution defined by \eqref{eq:v2} and \eqref{eq:g2}, the seller's optimal mechanism posts price $\omega_1$ for product 1 and $\underline{v}_2$ for product 2. 
    \label{lemma:2item}
\end{lemma}
\Cref{lemma:2item} is a special case of \Cref{prop:nitem} that will be proved in \Cref{sec:multi}.
With the optimal mechanism in hand, we can compute its performance ratio under nature’s distribution $\F_{\boldsymbol{\omega}}$. This serves as an upper bound on the optimal performance ratio $\cR^*$ in Problem \eqref{eq:saddle}.
\begin{proposition}
Under nature's strategy $\F_{\boldsymbol{\omega}}$ defined in \eqref{eq:v2} and \eqref{eq:g2}, the separate posted-price mechanism in \Cref{lemma:2item} obtains a performance ratio of $\bigof{\omega_1 \ln \frac{\overline{v}_1}{\omega_1} +  {\omega_1} +\underline{v}_2 \ln \frac{\overline{v}_2}{\underline{v}_2} +  {\underline{v}_2}}^{-1}   \cdot (\omega_1+\underline{v}_2)$ .
    \label{prop:2item}
\end{proposition}
Following \Cref{prop:2item}, nature will choose $\omega_1$ to minimize the seller's best achievable ratio. By optimizing $\omega_1$, we find the minimum ratio coincides with the performance ratio achieved by the separable mechanism in \Cref{cor:2item}, which implies the optimality of the mechanism in \Cref{cor:2item}. 
\begin{proposition}
For $J=2$, the mechanism in \Cref{cor:2item} is robustly optimal among all incentive-compatible and individually rational (IC/IR) mechanisms.
    \label{prop:2item-opt}
\end{proposition}
\Cref{prop:2item-opt} indicates that the robustly optimal selling mechanism is separable in the two-item screening problem. 
When the relative range $\frac{\underline{v}_1}{\overline{v}_1}$ and $\frac{\underline{v}_2}{\overline{v}_2}$ of the two dimensions are close to each other, or when $\underline{v}_1/\underline{v}_2$ is large, i.e., $\underline{v}_{2}\,\overline{v}_{1}\;\le \;\overline{v}_{2}\,\underline{v}_{1} e^{\,1+\underline{v}_{1}/\underline{v}_{2}}$,  the robustly optimal mechanism can be implemented as randomized posted prices over the entire range $[\underline{v}_j,\overline{v}_j]$ for each product $j$. 
On the other hand, if the relative range $\frac{\underline{v}_2}{\overline{v}_2}$ is much higher than $\frac{\underline{v}_1}{\overline{v}_1}$, i.e., $\underline{v}_{2}\,\overline{v}_{1}\;> \;\overline{v}_{2}\,\underline{v}_{1} e^{\,1+\underline{v}_{1}/\underline{v}_{2}}$, then the optimal price distribution for product 1 concentrates only on the upper segment $[\omega_1,\overline{v}_1]$, skipping the region of $v_1\in[\underline{v}_1,\omega_1)$. 
Intuitively, when one product has a substantially wider support, the robust policy may assign zero allocation probability to its low valuations—especially if its minimum value is not much larger than that of the other product.
According to \cite{eren2010monopoly,wang2024minimax}, the inverse of the optimal performance ratio achieved in the one-item problem is $1/r_j^\dag = 1+\ln(\overline{v}_j/\underline{v}_j)$. 
Interestingly, the inverse of the performance ratio for the two-item problem in \Cref{prop:2item}, i.e., $ \bigof{\omega_1 \ln \frac{\overline{v}_1}{\omega_1} +  {\omega_1} +\underline{v}_2 \ln \frac{\overline{v}_2}{\underline{v}_2} +  {\underline{v}_2}}   \cdot (\omega_1+\underline{v}_2)^{-1}$ can be interpreted as a ``weighted average'' of the terms $1+\ln(\overline{v}_j/\omega_j)$ with weights $\omega_j$ for each product.
When the relative range $\frac{\underline{v}_1}{\overline{v}_1}$ and $\frac{\underline{v}_2}{\overline{v}_2}$ of the two dimensions are close, or when $\underline{v}_1$ is much higher than $\underline{v}_2$, i.e., $\underline{v}_{2}\,\overline{v}_{1}\;\le \;\overline{v}_{2}\,\underline{v}_{1} e^{\,1+\underline{v}_{1}/\underline{v}_{2}}$, the inverse of the performance ratio is the ``weighted average'' of the inverses of the performance ratios with weights $\underline{v}_j$ in the decomposed one-dimensional mechanisms. That is, $
    \frac{1}{\cR^*} = \frac{\sum_{j=1}^2\of{\underline{v}_j /r_j^\dag}}{\sum_{j=1}^2 \underline{v}_j}$,
where $r_j^\dag$ is the optimal performance ratio in the one-dimensional problem for product $j$, i.e., $r_j^\dag = \bigof{1+\ln(\overline{v}_j/\underline{v}_j)}^{-1}$.
The insights from this two-item analysis guide the construction of nature’s strategy for the general multi-item setting with $J\ge 2$.

\subsection{General Multi-Item Problem}
\label{sec:multi}
In this subsection, we prove the optimality of the selling mechanism defined in \Cref{prop:ratio-feasible} for general $J\ge 2$. 
In particular, we construct a nature's strategy $\F$ for which the best achievable approximation ratio equals $\gamma^*$. 
Two distinct challenges arise.
First, as noted in previous literature \citep{hart2013menu, daskalakis2014complexity,briest2015pricing, babaioff2020simple}, characterizing the optimal multi-item mechanism under a given joint distribution $\F$ can be difficult, since the optimal mechanism can be non-monotonic, randomized, infinite-dimensional, or computationally intractable. 
Our \Cref{lemma:single} helps mitigate this complexity in a robust setting with support information.
Second, because the maximin ratio objective couples items through the sum of values in the denominator, it induces a different adversarial response. Hence, the elegant approaches from the maximin-revenue settings \citep{carroll2017robustness,che2021robustly} do not directly transfer.
Fortunately, the previous discussions at the beginning of \Cref{sec:separable} and our finding in \Cref{prop:2item-opt} inspire us to start with a comonotonic distribution $\F$ parameterized by a single index $\xi$.
Consider a distribution $\F$ supported on a one-dimensional ray $\bv = \xi \cdot \boldsymbol{\omega}$, projected on the feasible support $\cV=\prod_{j\in \cJ}[\underline{v}_j,\overline{v}_j]$, 
where $\boldsymbol{\omega} = (\omega_1,\omega_2,\dots,\omega_J)\in \R^J_+$ is a fixed direction, and 
$\xi \ge 1$ is a randomized scalar representing the magnitude. In particular,
\begin{align}
    v_j(\xi) = \min\{\omega_j \xi,\, \overline{v}_j\}, \quad \forall j\in \cJ
    \label{eq:supportdef}
\end{align}
Hence, each coordinate $v_j(\xi)$ is increasing in $\xi$, so the valuations are comonotonic. 
Furthermore, the support of valuations defined in \eqref{eq:supportdef} is uniquely determined by the constant coefficient vector $\boldsymbol{\omega} = (\omega_1,\omega_2,\dots,\omega_J)$. We parameterize $\boldsymbol{\omega}$ by a single scalar $\eta\in (0, 1]$ and set
\begin{align}
    \omega_j =\begin{cases}
    \overline{v}_j\cdot e^{-\frac{1}{\eta}} & \text{if }  \underline{v}_j/\overline{v}_j< e^{-1/\eta}\\
    \underline{v}_j & \text{if }  \underline{v}_j/\overline{v}_j\ge e^{-1/\eta}.
    \end{cases} 
    \label{eq:def-w}
\end{align}
Since $\boldsymbol{\omega}$ is determined by parameter $\eta$, the induced distribution $\F_\eta$ of $\bv$ also depends on $\eta$. For notational simplicity, sometimes we skip the subscript of $\eta$ in $\boldsymbol{\omega}$ and $\F$ when there is no confusion.

Based on \eqref{eq:supportdef}, each coordinate $v_j(\xi)$ grows linearly in $\xi$ when $\xi<\frac{\overline{v}_j}{\omega_j}$, and then remains constant at $\overline{v}_j$ for $\xi\ge \frac{\overline{v}_j}{\omega_j}$. Without loss of generality, we sort $j\in\cJ$ in increasing order of $\underline{v}_j/\overline{v}_j$, i.e. $\underline{v}_1/\overline{v}_1\le \dots \le \underline{v}_J/\overline{v}_J$.  Formally, $\bv(\xi)$ is determined as follows:
\begin{align}
\bv(\xi) = \begin{cases}
    \boldsymbol{\omega}\cdot \xi & \xi \in [1,\frac{\overline{v}_J}{\underline{v}_J}]\\
     \bigof{\boldsymbol{\omega}_{1:J-1}\cdot \xi, \, \overline{v}_J}& \xi \in (\frac{\overline{v}_J}{\underline{v}_J}, \frac{\overline{v}_{J-1}}{\underline{v}_{J-1}}]\\  
     \dots\\
       \bigof{\boldsymbol{\omega}_{1:k}\cdot \xi,\,  \overline{\bv}_{k+1:J}} & \xi \in (\frac{\overline{v}_{k+1}}{\underline{v}_{k+1}}, \frac{\overline{v}_{k}}{\underline{v}_{k}}]\\  
       \dots \\
       \bigof{\boldsymbol{\omega}_{1:\tilde{j}(\eta)}\cdot \xi,\,  \overline{\bv}_{\tilde{j}(\eta)+1:J}} & \xi \in (\frac{\overline{v}_{\tilde{j}(\eta)+1}}{\underline{v}_{\tilde{j}(\eta)+1}}, e^{\frac{1}{\eta}}]\\  
       \overline{\bv} & \xi \in (e^{\frac{1}{\eta}}, \infty)
\end{cases}
\label{eq:v}
\end{align}
where $\tilde{j}(\eta)=\max\{j\in \cJ \mid \underline{v}_j/\overline{v}_j< e^{-1/\eta}\}$ and $\boldsymbol{\omega}$ is explicitly defined by parameter $\eta$ as in \eqref{eq:def-w}. 
Based on the definition of $\bv$ in \eqref{eq:v}, nature's strategy $\F_\eta$ of $\bv$ is determined by the distribution $\G$ of $\xi$. 
Following our intuition outlined at the start of \Cref{sec:separable}, \Cref{prop:nitem} specifies a distribution $\G(\xi)$ such that the kernel $v_j(\xi) \int_{\xi}^\infty \frac{d\G(x)}{ \sum_{i\in\cJ}v_i(x)}$ in \eqref{eq:scalar} is constant in $\xi$ within a certain range $\xi\in[1,\overline{v}_j/\omega_j]$.
Then we prove that the seller's optimal mechanism is a posted-price mechanism under the corresponding nature's strategy, which significantly simplifies the analysis in obtaining an upper bound of the approximation ratio.
\begin{proposition}
    \label{prop:nitem}
    Suppose the distribution $\F_\eta$ of $\bv$ is defined in \eqref{eq:v}, with the cumulative distribution function $\G$ of $\xi$ satisfying
     \begin{align*}
\G(\xi) = \zeta \int_1^\xi  \frac{\sum_{j\in\cJ}v_j(x)}{x^2} \, dx,\, \, \forall \, \xi\in [1, \infty)
 \end{align*}
where $\zeta$ is a normalization constant such that $\int_1^\infty d\G(\xi)=1$. Then the seller's optimal mechanism is to separately charge a price of $\omega_j$ for each product $j\in\cJ$, which achieves an approximation ratio of $\Bigof{\sum_{j\in\cJ} \bigof{\omega_j \ln \frac{\overline{v}_j}{\omega_j} +  {\omega_j} }}^{-1}\cdot \bigof{\sum_{j\in\cJ}\omega_j}$.
\end{proposition}
\begin{proof}{Proof of \Cref{prop:nitem}.}
For any seller's mechanism $\of{\bq(\bv(\xi)),t(\bv(\xi))}$, let $\boldsymbol{\alpha}(\xi)=\bq(\bv(\xi))$ and $\tau(\xi)=t(\bv(\xi))$ denote the allocation probability and payment at $\bv(\xi)$, respectively. 
By incentive compatibility and the envelope theorem \citep{milgrom2002envelope}, the payment satisfies
\begin{align*}
    \tau(\xi) = \boldsymbol{\alpha}(\xi)^\top \cdot \bv(\xi) -\int_{1}^{\xi}  \boldsymbol{\alpha}(x)^\top \,  d\bv(x)  = \sum_{j\in \cJ} \bigof{\int_{1}^\xi v_j(x) d\alpha_j(x)} 
\end{align*}

Thus, the performance ratio is evaluated as 
{\small
\begin{align*}\displaystyle
    \mathop{\E}\limits_{\bv\sim \F_\eta} \Bigoff{\frac{t(\bv)}{\allone^\top \bv}}  =   \mathop{\E}\limits_{\xi \sim \G} \Bigoff{\frac{\tau(\xi)}{ \sum\limits_{i\in\cJ}v_i(\xi)}} = \int_{1}^{\infty} \frac{ \sum\limits_{j\in \cJ} \bigof{\int_{1}^\xi v_j(x) d\alpha_j(x)}  }{ \sum_{i\in\cJ}v_i(\xi)} d\G(\xi) =\sum_{j\in \cJ} \Bigoff{\int_{1}^{\infty} \Bigof{v_j(\xi) \int_{\xi}^\infty \frac{d\G(x)}{ \sum\limits_{i\in\cJ}v_i(x)}}\, d\alpha_j(\xi) }. 
\end{align*}
}
Incorporating the definition of $\G$ in \Cref{prop:nitem}, we have
{\small
 $$
 v_j(\xi) \int_{\xi}^\infty \frac{d\G(x)}{ \sum_{i\in\cJ}v_i(x)} = v_j(\xi) \cdot \frac{\zeta}{\xi} =\begin{cases}
     \zeta \cdot\omega_j  & \text{when } \xi \le \overline{v}_j/\omega_j \\
     \overline{v}_j\cdot \frac{\zeta}{\xi} & \text{when } \xi > \overline{v}_j/\omega_j.
 \end{cases}
 $$}
It implies that $ v_j(\xi) \int_{\xi}^\infty \frac{d\G(x)}{ \sum_{i\in\cJ}v_i(x)}$ is a constant when $ \xi \in [1, \overline{v}_j/\omega_j]$ and then decreasing in $\xi$ when $\xi \ge \overline{v}_j/\omega_j$.
For any nature's strategy $\G$, the seller will select $\boldsymbol{\alpha}$ in order to maximize $ \E_{\bv\sim \F_\eta} \Bigoff{\frac{t(\bv)}{\allone^\top \bv}} $. 
By feasibility constraint, the allocation probability $\alpha_j(\xi) \in [0,1]$, for all $\xi$. 
Since nature's strategy satisfies that for all $j\in \cJ$, $v_j(\xi) \int_{\xi}^\infty \frac{d\G(x)}{ \sum_{i\in\cJ}v_i(x)}$ is nonnegative, continuous and unimodal, by \Cref{lemma: unimodal}, the maximizer of $\max_{\alpha_j} \int_1^\infty v_j(\xi) \int_{\xi}^\infty \frac{d\G(x)}{ \sum_{i\in\cJ}v_i(x)} d\alpha_j(\xi)$ is a step function $\alpha_j(\xi) = \mathbbm{1}_{\xi\ge 1}$. This allocation rule also respects the incentive compatibility and individual rationality constraints and can be simply implemented by a posted price mechanism. 
The maximum approximation ratio the seller could achieve under nature's strategy $\G$ is calculated as 
{\small
\begin{align*}
    \underset{\bv\sim \F_\eta}{\E} \Bigoff{\frac{t(\bv)}{\allone^\top \bv}}  =&\sum_{j\in \cJ} \Bigoff{\int_{1}^{\infty} \Bigof{v_j(\xi) \int_{\xi}^\infty \frac{d\G(x)}{ \sum_{i\in\cJ}v_i(x)}}\, d\alpha_j(\xi) }  =\sum_{j\in \cJ} \Bigoff{\int_{1}^{\infty} \min\{\zeta \omega_j, \frac{\zeta \overline{v}_j }{\xi} \} \, d\alpha_j(\xi) }\\
    \le & \sum_{j\in \cJ} \zeta\cdot \omega_j \int_{1}^{\infty}  \, d\alpha_j(\xi) \le \zeta\sum_{j\in \cJ}  \omega_j.
\end{align*}
}
The last inequality is due to $\alpha_j(\xi)\in [0,1]$ for all $j\in \cJ$.
The inequalities becomes equalities if each allocation rule is a unit step function $\alpha_j(\xi)=\mathbbm{1}_{\xi\ge 1}$ for all $j\in \cJ$, which implements separate posted prices $\omega_j$ for product $j\in \cJ$. The factor $\zeta$ in the approximation ratio above $\zeta\sum_{j\in \cJ}  \omega_j$ is obtained from the normalization condition $\int_{1}^{\infty} d\G(\xi) = 1$ in the definition of $\G$ as specified in \Cref{prop:nitem}:
{\footnotesize
\begin{align*}
    1= &\int_{1}^{\infty} d\G(\xi)
    = \zeta\int_{1}^{\infty} \, \frac{\sum_{j\in\cJ}v_j(\xi)}{\xi^2}\, d\xi
=   \zeta\cdot\Bigof{ \int_{1}^{\frac{\overline{v}_J}{\underline{v}_J}} \frac{\sum_{j\in\cJ}\omega_j }{\xi}\, d\xi + \sum_{k=2}^J \int_{\frac{\overline{v}_{k}}{\omega_{k}}}^ {\frac{\overline{v}_{k-1}}{\omega_{k-1}}} 
   \bigof{ \frac{\sum_{j=1}^{k-1}\omega_j }{\xi}\, + 
    \frac{ \sum_{j=k}^{J}\overline{v}_j}{\xi^2}}\, d\xi  + \int_{\frac{\overline{v}_{1}}{\omega_{1}}}^{\infty} 
  \frac{ \sum_{j=1}^{J} \overline{v}_j }{\xi^2}\, d\xi } \\
   =& \zeta\cdot \sum_{j=1}^J \bigof{\int_{1}^{\frac{\overline{v}_j}{\omega_j}}  \frac{\omega_j}{\xi} \,d\xi + \int_{\frac{\overline{v}_j}{\omega_j}}^{\infty}  \frac{\overline{v}_j}{\xi^2}\, d\xi } 
     = \zeta\cdot \sum_{j=1}^J \bigof{\omega_j \ln \frac{\overline{v}_j}{\omega_j} +  {\omega_j} }.
\end{align*}}
Hence, the optimal performance ratio is $\of{\sum_{j=1}^J \bigof{\omega_j \ln \frac{\overline{v}_j}{\omega_j} +  {\omega_j} }}^{-1}\cdot \bigof{\sum_{j\in\cJ}\omega_j}$.
 \QED
 \end{proof}
Given the expression of the seller's optimal performance ratio in \Cref{prop:nitem}, $\bigof{\sum_{j\in \cJ} \bigof{\omega_j \ln \frac{\overline{v}_j}{\omega_j} +  {\omega_j} }}^{-1}\cdot \bigof{\sum_{j\in\cJ}\omega_j}$,
We now show that when the adversary (nature) selects the parameter $\eta$ to minimize the performance ratio, the resulting ratio coincides with the lower bound in \Cref{prop:ratio-feasible}.
\begin{proposition}
\label{prop:ratio-upper bound}
Let $\eta^*$ be the unique solution to $\phi(\eta) = \eta\cdot e^{-1/\eta} \cdot \sum_{j\in \cS(\eta)} \overline{v}_j - \sum_{j\in \cJ\setminus \cS(\eta)} \bigof{\underline{v}_j\cdot\bigof{\eta\ln(\underline{v}_j/\overline{v}_j)-\eta+1}} = 0$, where $\cS(\eta)=\bigofff{j\in \cJ \mid \underline{v}_j/\overline{v}_j< e^{-1/\eta}}$. Under nature's strategy $\F_{\eta^*}$ (constructed in \Cref{prop:nitem}), the optimal performance ratio the seller can achieve is $\eta^*$. Formally,
\begin{align*}
\sup_{M\in\cM} \cR(M,\F_{\eta^*}) = \eta^*. 
\end{align*}
\end{proposition}
\Cref{prop:ratio-upper bound} is proved by substituting the representation of $\boldsymbol{\omega}$ given in \eqref{eq:def-w}, and then optimizing over $\eta$ for the adversary. We postpone the proof to the appendix.
\Cref{prop:ratio-upper bound} indicates that under nature's policy $\F_{\eta^*}$, the optimal performance ratio obtained by the seller is exactly $\eta^*$. We summarize the construction of nature's strategy 
$\F_{\eta^*}$ as follows.
\begin{enumerate}
    \item For any  support information $\cV=\prod_{j\in \cJ}[\underline{v}_j,\overline{v}_j]$, let $\eta^*$ be the unique solution to $\phi(\eta) = \eta\cdot e^{-1/\eta} \cdot \sum_{j\in \cS(\eta)} \overline{v}_j - \sum_{j\in \cJ\setminus \cS(\eta)} \bigof{\underline{v}_j\cdot\bigof{\eta\ln(\underline{v}_j/\overline{v}_j)-\eta+1}} = 0$, with $\cS(\eta)=\bigofff{j\in \cJ \mid \underline{v}_j/\overline{v}_j< e^{-1/\eta}}$.
    \item  Define $\boldsymbol{\omega}$ as follows: \begin{align*}
    \omega_j =\begin{cases}
    \overline{v}_j\cdot e^{-\frac{1}{\eta^*}} & \text{if }  \underline{v}_j/\overline{v}_j< e^{-1/\eta^*}\\
    \underline{v}_j & \text{if }  \underline{v}_j/\overline{v}_j\ge e^{-1/\eta^*}.
    \end{cases} 
\end{align*}
\item
The nature's strategy $\F_{\eta^*}$ of $\bv$ is determined by 
\begin{align*}
    v_j(\xi) = \min\{\omega_j \xi,\, \overline{v}_j\} \quad \forall j\in \cJ
\end{align*}
where $\xi\ge 1$ follows a distribution $\G$ such that 
 \begin{align*}
\G(\xi)=\int_1^\xi \biggof{  \frac{1}{\eta^*}\sum_{j\in \cJ\setminus \cS(\eta^*)} \Bigof{\underline{v}_j\cdot\bigof{\ln(\underline{v}_j/\overline{v}_j)+\frac{1}{\eta^*}}} }^{-1}\, \frac{\sum_{j\in\cJ}v_j(x)}{x^2}dx,\, \, \forall \, \xi\in [1, \infty).
 \end{align*}
\end{enumerate}
Since $\phi(\eta)=0$ has a unique solution (by \Cref{increasing-g}), the $\eta^*$ defined in \Cref{prop:ratio-upper bound} is equal to the $\gamma^*$ defined in \Cref{prop:ratio-feasible}.
Thus, \Cref{prop:ratio-upper bound} implies that under nature's policy $\F_{\eta^*}$, the seller's optimal performance ratio exactly matches the feasible performance ratio in \Cref{prop:ratio-feasible}.
Hence, by the saddle point argument, this proves that the performance ratio obtained in \Cref{prop:ratio-feasible} is tight and therefore optimal for Problem \eqref{eq:original}. 

\begin{theorem}
Let $\gamma^*$ be the unique solution to $\phi(\gamma)=\gamma\cdot e^{-1/\gamma} \cdot \sum_{j\in \cS(\gamma)} \overline{v}_j - \sum_{j\in \cJ\setminus\cS(\gamma)} \of{\underline{v}_j\cdot\of{\gamma\ln(\underline{v}_j/\overline{v}_j)-\gamma+1}} = 0$, where $\cS(\gamma)=\bigofff{j\in \cJ \mid \underline{v}_j/\overline{v}_j<e^{ -1/\gamma}}$.
   Then $(M_{\gamma^*}, \F_{\gamma^*})$ forms a saddle point of Problem \eqref{eq:saddle}, and thus, $M_{\gamma^*}$ is optimal for Problem \eqref{eq:original}
which achieves an optimal performance ratio of $\gamma^*$.
    \label{thm:optimal}
\end{theorem}
\begin{proof}{Proof of \Cref{thm:optimal}.}
    \Cref{prop:ratio-feasible} proves that 
$\inf_{\F\in \cF} \cR(M_{\gamma^*},\F) = \gamma^*$, and
    \Cref{prop:ratio-upper bound} demonstrates that $\sup_{M\in \cM}  \cR(M,\F_{\gamma^*}) = \gamma^*$.
    It follows that 
    \begin{align*}
   \gamma^*  =\inf_{\F\in \cF} \cR(M_{\gamma^*},\F)  \le \sup_{M\in \cM} \inf_{\F\in \cF} \cR(M,\F) \le \sup_{M\in \cM} \cR(M,\F_{\gamma^*})= \gamma^*.
    \end{align*} 
    Therefore, the optimal approximation ratio achieved by Problem \eqref{eq:original} is $\gamma^*$. Since $\inf_{\F\in \cF} \cR(M_{\gamma^*},\F) = \gamma^*$, it implies the optimality of $M_{\gamma^*}$. Moreover, because $\gamma^*  =\inf_{\F\in \cF} \cR(M_{\gamma^*},\F)  \le \cR(M_{\gamma^*},\F_{\gamma^*}) \le \sup_{M\in \cM} \cR(M,\F_{\gamma^*})= \gamma^*$, we have that  $(M_{\gamma^*}, \F_{\gamma^*})$ forms a saddle point to Problem \eqref{eq:saddle}.
    \QED
\end{proof}

\Cref{thm:optimal} indicates that the robustly optimal selling mechanism is separable. 
The optimal mechanism allocates only to valuations above certain thresholds $\omega_j$, effectively truncating the lower end of each product's valuation support. Intuitively, the mechanism avoids selling at very low prices when there is large uncertainty about the buyer's valuation for a product.
Moreover, similar to the two-item case, the inverse of the performance ratio for the multi-item problem 
$ \frac{\sum_{j\in \cJ}\bigof{\omega_j\cdot (\ln\frac{\overline{v}_j}{\omega_j}+1)}}{\sum_{j\in \cJ}\omega_j}$
 can be interpreted as a weighted average of $1+\ln(\overline{v}_j/\omega_j)$ with weight $\omega_j$ for each product.
When the relative ranges $\frac{\underline{v}_j}{\overline{v}_j}$ are balanced across dimensions so that $\omega_j=\underline{v}_j, \, \forall\, j\in \cJ$, the inverse of the performance ratio is the weighted average of the inverse of the performance ratios in the decomposed separable problem for each dimension, i.e.,
$
    \frac{1}{\cR^*} = \frac{\sum_{j\in \cJ}\of{\underline{v}_j /r_j^\dag}}{\sum_{j\in \cJ}\underline{v}_j}
$
where $r_j^\dag$ is the performance ratio in the one-dimensional problem for product $j$, i.e., $r_j^\dag = \bigof{1+\ln(\overline{v}_j/\underline{v}_j)}^{-1}$.
By \Cref{thm:optimal}, the separable mechanism attains the optimal performance ratio while remaining easy to interpret and implement.
Furthermore, we can also extend the application of \Cref{thm:optimal} to more general ambiguity sets beyond the rectangular supports.
\begin{corollary}
Consider an ambiguity set $\widetilde{\cF}\subseteq \cF$ with $\F_{\gamma^*}\in\widetilde{\cF}$. Then $M_{\gamma^*}$ from \Cref{prop:ratio-feasible} remains robustly optimal on ambiguity set $\widetilde{\cF}$ and $(M_{\gamma^*}, \F_{\gamma^*})$ is a saddle point of $\sup_{M\in \cM} \inf_{\F\in \widetilde{\cF}} \cR(M,\F)$.
\label{cor-subsetF}
\end{corollary}
\begin{proof}{Proof.}
    \Cref{cor-subsetF} follows directly from 
 $\sup_{M\in \cM} \inf_{\F\in \widetilde{\cF}} \cR(M,\F) \le \sup_{M\in \cM} \cR(M,\F_{\gamma^*}) = \gamma^* = \sup_{M\in \cM} \inf_{\F\in {\cF}} \cR(M,\F) \le \sup_{M\in \cM} \inf_{\F\in \widetilde{\cF}} \cR(M,\F) .$
    \QED
\end{proof}
\begin{figure}[htbp]
\caption{Support for Some Examples of $\widetilde{\cF}\subseteq \cF$ with $\F_{\gamma^*}\in\widetilde{\cF}$}
\label{fig:fig_cor1}
\begin{subfigure}[t]{0.329\textwidth}
\includegraphics[width=\textwidth,keepaspectratio]{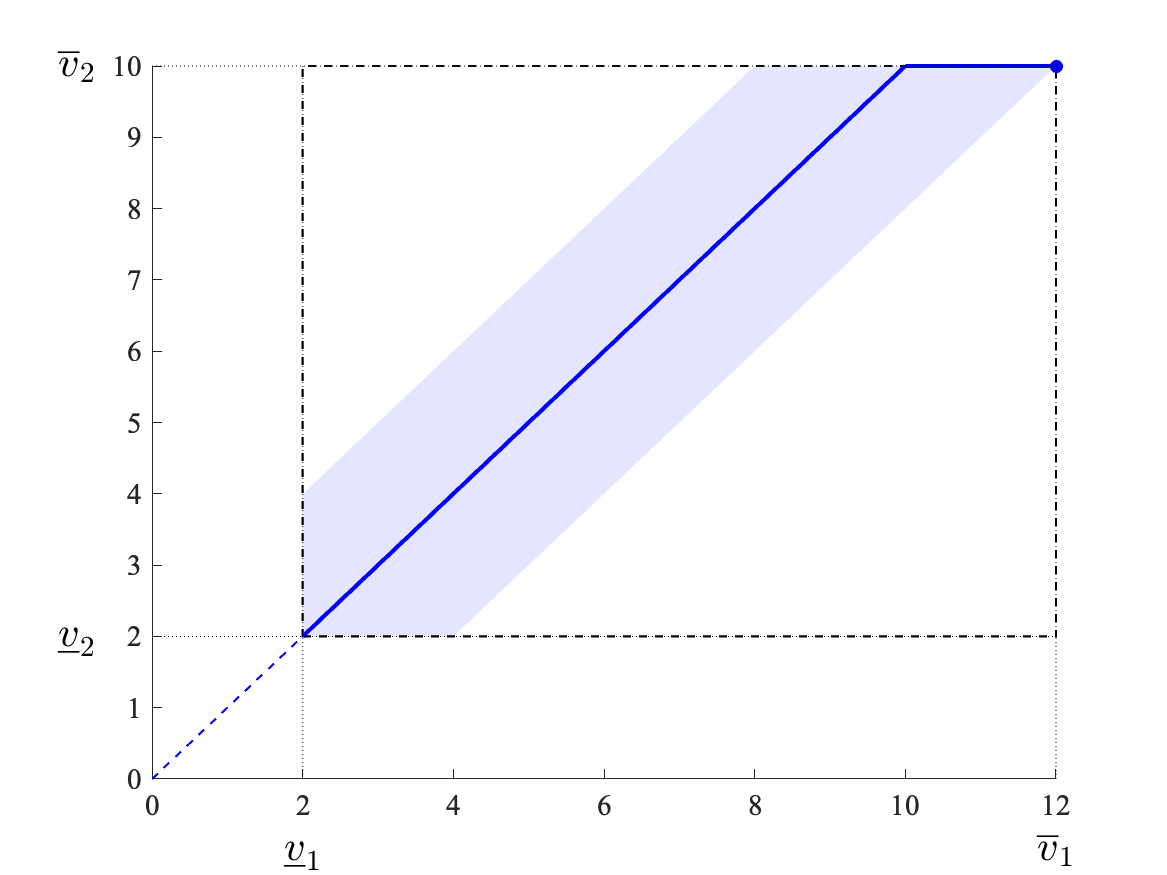}
\caption{\footnotesize Similar Items}
\label{fig:fig_support_cor1_2D}
\end{subfigure}
\begin{subfigure}[t]{0.329\textwidth}
\includegraphics[width=\textwidth,keepaspectratio]{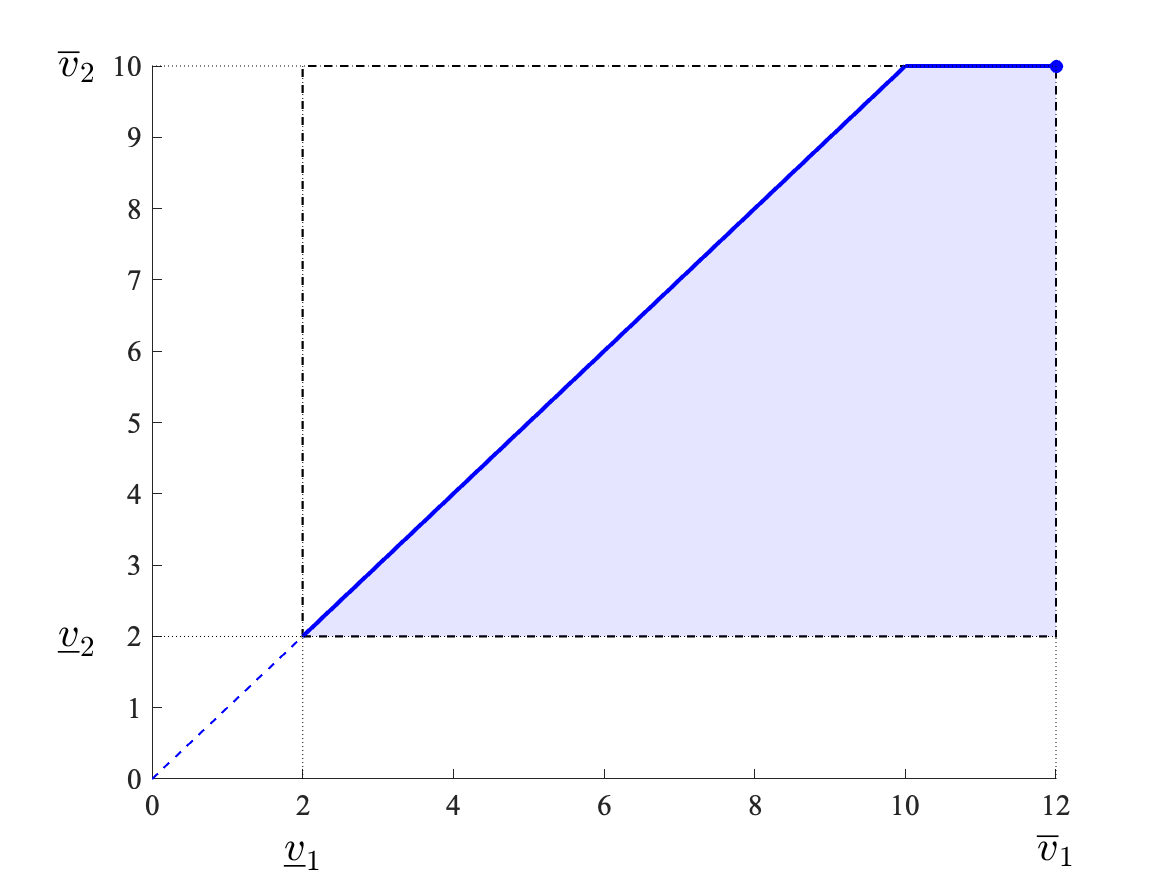}
\caption{\footnotesize  Premium and Standard Items}
\label{fig:fig_support_cor1_2D_triangle}
\end{subfigure}
\begin{subfigure}[t]{0.33\textwidth}
\includegraphics[width=\textwidth,keepaspectratio]{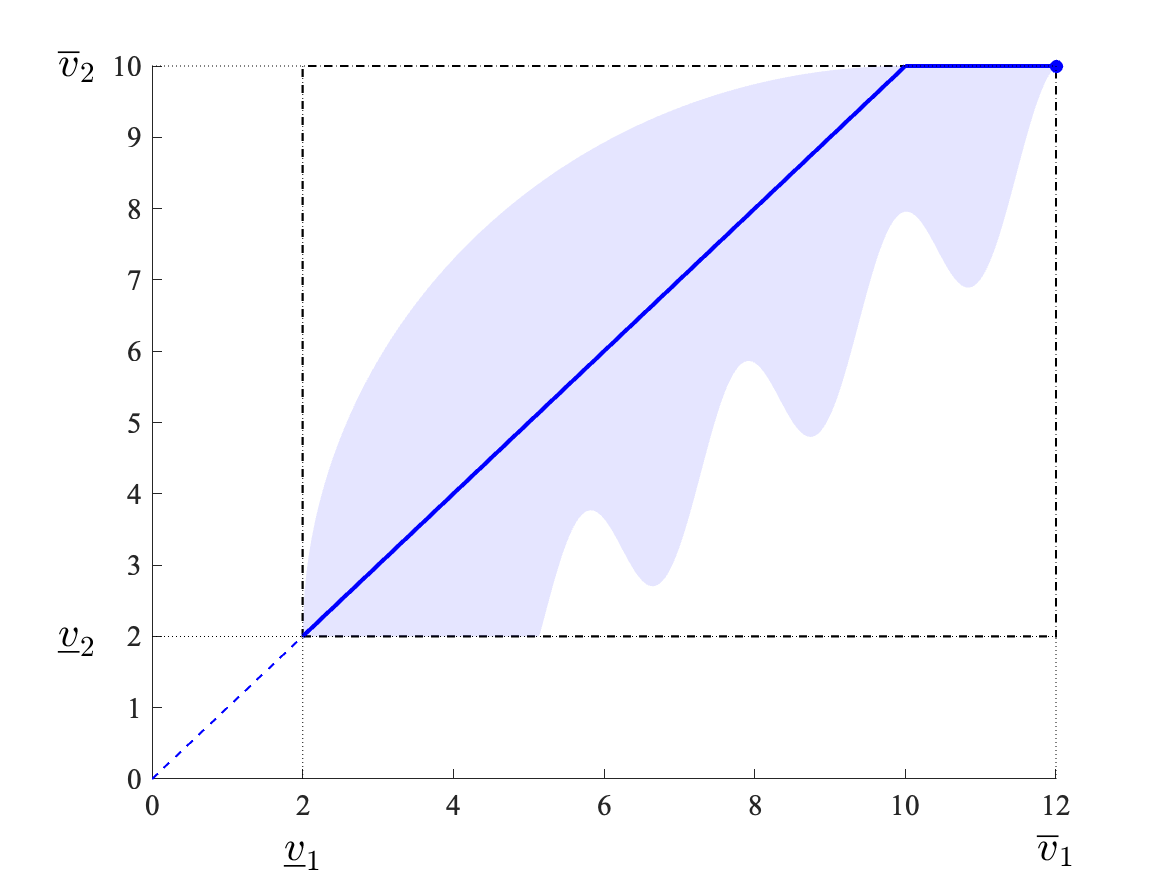}
\caption{\footnotesize  Mildly Comparable Items}
\label{fig:fig_support_cor1_2D_irregular}
\end{subfigure}
\end{figure}
\Cref{cor-subsetF} shows that the separable mechanism $M_{\gamma^*}$ is still robustly optimal under any ambiguity set $\widetilde{F}\subseteq \cF$ that includes $\F_{\gamma^*}$. 
In \Cref{fig:fig_cor1}, we plot the support of $\F_{\gamma^*}$ in the solid blue line for the two-item case. Now we discuss the following three practical scenarios.
First, though the seller does not know the buyer's valuation distribution of each product, they may know that for each buyer, the valuations of the two products are similar. 
The light blue area in \Cref{fig:fig_support_cor1_2D} represents all possible pairs of valuations in which one product’s valuation remains within a bounded range of the other’s. 
Second, the seller may offer a standard product and a premium product. Though the exact valuation of each product is unknown, it is reasonable to assume the standard product is always valued no more than the premium product. This ambiguity set's support is captured in the light blue area in \Cref{fig:fig_support_cor1_2D_triangle}. 
Finally, sometimes the seller knows only that the two valuations are mildly comparable. Then as long as one product's valuation fluctuates near the other’s, as shown in the light blue area in \Cref{fig:fig_support_cor1_2D_irregular}, the separable mechanism remains robustly optimal.

\begin{figure}[htbp]
\centering
\caption{Feasible Valuation Set $\cV_{\cB} = \bigofff{\bv:  v_1+v_2 \in [2,4],\, v_3\in[1,2 ]}$  }
\includegraphics[width=0.36 \textwidth,keepaspectratio]{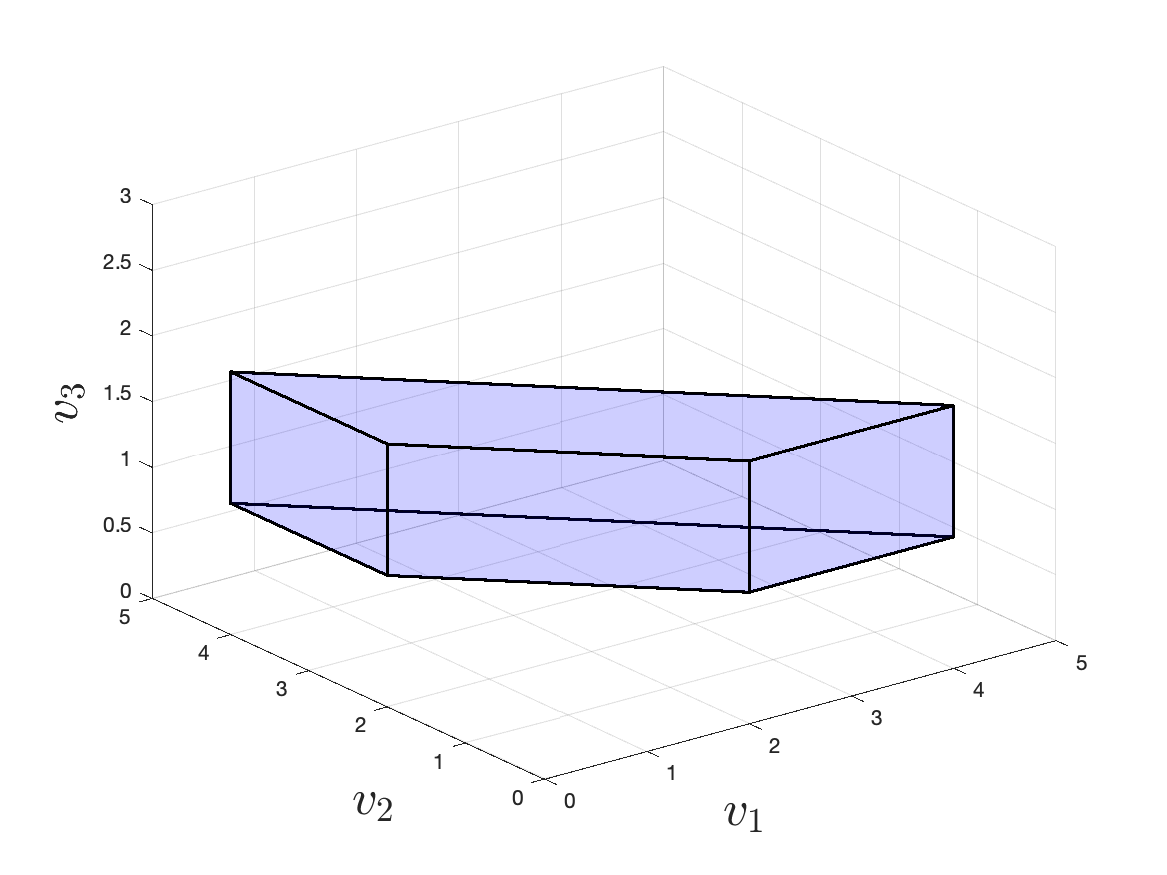}
\label{fig:fig_cor3_bundle}
\end{figure}

Building on \Cref{cor-subsetF} (and \Cref{fig:fig_cor1}), the separable mechanism from \Cref{thm:optimal} remains optimal for any smaller ambiguity set that still includes the worst-case distribution. 
In the following, we explore more general ambiguity sets.
Suppose the seller has prior information about the sum of valuations over certain product groups. 
Let $\cB$ be an arbitrary partition of all products $\cJ$, where each element $b\in \cB$ represents a subset of products in $\cJ$ and all elements in $\cB$ are mutually exclusive and collectively exhaustive subsets. The region of feasible valuations is defined as:
\begin{align*}
    \cV_{\cB} = \bigofff{\bv: \sum_{j\in b} v_j \in [\underline{v}_b,\overline{v}_b],\quad \forall b\in \cB}.
\end{align*}
This allows us to incorporate prior knowledge about the total value within each bundle $b\in \cB$. 
When $\cB$ is the finest partition of $\cJ$, i.e. $\cB = \bigofff{\{1\},\{2\},\dots, \{J\} }$, then $\cV_{\cB}$ becomes the box uncertainty set. \Cref{fig:fig_cor3_bundle} depicts the feasible valuation 
 $\cV_{\cB} = \bigofff{\bv:  v_1+v_2 \in [2,4],\, v_3\in[1,2 ]}$, given $\cB = \bigofff{\{1, 2\},\{3\}}$.
 
Assuming nature selects any distribution within $\Delta(\cV_{\cB})$, the seller can design a \emph{bundle-wise separable} mechanism based on partition $\cB$.
Effectively, the seller can treat each bundle $b \in \cB$ as a ``pseudo-item'' and interpret the ambiguity set $\cF = \Delta(\cV_{\cB})$ as an ambiguity set with box support defined on all bundles $b\in \cB$. 
Ordering the bundles $b \in \cB$ according to increasing values of $\underline{v}_b/\overline{v}_b$ - i.e., $\underline{v}_{b_1}/\overline{v}_{b_1} \le \dots \le \underline{v}_{b_B}/\overline{v}_{b_B}$, where $B = |\cB|$, we introduce a bundle-wise separable mechanism:
 \begin{align}
 \begin{aligned}
     \bq(\bv)& =\Bigof{q_j\bigof{\sum_{i\in b(j)}v_i}}_{j\in \cJ}, \ \text{where } q_j(\sum_{i\in b(j)}v_i)  = \bigof{\gamma \cdot\ln \frac{\sum_{i\in b(j)}v_i}{\overline{v}_{b(j)}}+1}^+  \label{eq:q-bundle}\\
     t(\bv) & = \sum_{b\in \cB} t_b(\sum_{j\in b}v_j), \ \text{where } t_b(\sum_{j\in b}v_j)=\begin{cases}
       \gamma\cdot  \bigof{\sum_{j\in b}v_j -e^{-1/\gamma} \cdot \overline{v}_b}^+  & \text{if } e^{-1/\gamma}\cdot \overline{v}_b>\underline{v}_b\\
         \gamma\cdot \sum_{j\in b}v_j +\underline{v}_b\cdot\bigof{\gamma\ln\frac{\underline{v}_b}{e\overline{v}_b}+1} &  \text{if } e^{-1/\gamma}\cdot \overline{v}_b\le \underline{v}_b
     \end{cases} 
     \end{aligned}
 \end{align}
 where $b(j)$ denotes the bundle $b\in \cB$ that includes item $j\in \cJ$, and $\gamma\in (0,1]$ is a constant that only depends on $\{\underline{v}_b\}_{b\in \cB}$ and $\{\overline{v}_b\}_{b\in \cB}$. 
 Denoting $M^{\cB}_{\gamma}$  the mechanism defined in \eqref{eq:q-bundle}, we can directly leverage \Cref{thm:optimal} to obtain the following results for ambiguity sets defined by partition $\cB$. 
\begin{corollary}
    Consider ambiguity set $\cF = \Delta(\cV_{\cB})$. Let $\gamma^*_{\cB}$ be the unique solution to $\phi(\gamma)=\gamma\cdot e^{-1/\gamma} \cdot \sum_{b\in \cS(\gamma)} \overline{v}_b - \sum_{b\in \cB\setminus\cS(\gamma)} \bigof{\underline{v}_b\cdot\bigof{\gamma\ln(\underline{v}_b/\overline{v}_b)-\gamma+1}} = 0$, where $\cS(\gamma)=\bigofff{b\in \cB \mid \ln(\underline{v}_b/\overline{v}_b)< -1/\gamma}$.
Then $M^{\cB}_{\gamma_{\cB}^*}$ defined in \eqref{eq:q-bundle} is optimal for problem $\sup_{M\in \cM} \inf_{\F\in \Delta(\cV_{\cB})} \frac{\Rev(M,\F)}{\sup_{M'\in \cM}\Rev(M',\F)}$, and it achieves an approximation ratio of $\gamma^*_{\cB}$.
\label{cor-bundle}
\end{corollary}
By exploiting the estimated lower and upper bounds on the aggregate value in every bundle, the seller implements a bundle‑wise separable mechanism that uses only marginal support information at the bundle level. Similar to the itemwise separable mechanism, this bundle-wise separable mechanism can be implemented as a randomized pricing mechanism for each bundle $b\in\cB$, which can be viewed as a randomized version of the ``single bundle with the rest'' (SBR) proposed in  \cite{sun2025partition}. \Cref{cor-bundle} demonstrates that the bundle-wise separable selling mechanism is robustly optimal when the seller knows the valuation support for a partition of the products.
We extend these findings in \Cref{cor-approx} to scenarios where the seller's information goes beyond a single partition. Suppose the seller knows the lower and upper bounds for valuations of bundles within a collection $\cC \subseteq 2^{\cJ}$. 
Each element $c\in \cC$ is a subset of products in $\cJ$, but $\cC$ itself can be a superset of a partition of $\cJ$. Formally, the seller considers a region of valuations as follows:
\begin{align*}
    \cV_{\cC} = \bigofff{\bv: \sum_{j\in c}v_j \in [\underline{v}_c,\overline{v}_c], \quad \forall c\in \cC }
\end{align*}
Denote $\cP(\cC)$ the family of all collections of subsets in $\cC$ that form a partition of $\cJ$. For instance, suppose $\cJ=\bigofff{1,2,3}$ and $\cC = \bigofff{\{1\}, \{2\},\{3\}, \{1,2\} , \{2,3\}, \{1,2,3\} }$. This means the seller may have the support information on each product, $v_1,v_2,v_3$, together with the support information on bundles $v_1 + v_2$, $v_2 + v_3$ and $v_1 + v_2 + v_3$. 
Then $\cP(\cC) =\bigofff{\offf{ \{1\},   \{2,3\} }, \offf{ \{1,2\}, \{3\} },  \allowbreak \offf{\{1\}, \{2\},\{3\}},  \allowbreak \offf{\{1,2,3\}} }$, where each element in $\cP(\cC) $ is a subset of $\cC$ that forms a partition of $\cJ$. Then \Cref{cor-approx} provides an approximation guarantee (not necessarily tight) for this ambiguity set.
\begin{corollary}
    \label{cor-approx}
    Let $\cC\subseteq 2^\cJ$ be a collection of product bundles, and let $\Delta(\cV_{\cC})$ denote the support-only ambiguity set defined by the bounds on $\sum_{j\in c} v_j$ for $c\in \cC$.
    For each partition $\cB\in \cP(\cC)$, let $\gamma_{\cB}^*$ be the optimal ratio obtained by the bundle-wise separable mechanism defined in \eqref{eq:q-bundle} specialized to $\cB$.
  Then mechanism $M^{\cB^*}_{\gamma^*_{\cB^*}}$, where $\cB^* \in \argmax_{\cB\in \cP(\cC)} \gamma_{\cB}^*$, obtains an approximation ratio $\max_{\cB\in \cP(\cC)}\, \gamma_{\cB}^*$.
\end{corollary}

\Cref{cor-approx} demonstrates that if the seller knows more information than the support of a partition of $\cJ$, they can achieve an approximation ratio generated by the optimal partition $\cB$, which is a subset of $\cC$. 
Specifically, the seller can leverage the knowledge of lower and upper bounds for various bundle collections to find an optimal partition of $\cJ$ that yields the highest approximation ratio.
Moreover, if the seller's ambiguity set has a non-standard or non-convex support, one can delineate the tightest boundaries for the total valuations within each subset of products and then choose the optimal partition that obtains the highest approximation ratio. This provides a feasible mechanism that achieves a clear performance guarantee for a general and potentially irregular ambiguity set.

\section{Optimality Condition for Bundling}
We have studied the ambiguity set where the valuation range of each product is independent of the valuation of other products. In practice, a product's valuation range may depend on the valuations of other products. We capture a special form of such dependence with the following class.

\begin{figure}[htbp]
    \centering
    \caption{Geometric Illustration of $\boldsymbol{\rho}$-scaled Invariant Set}
\includegraphics[width=0.55\linewidth]{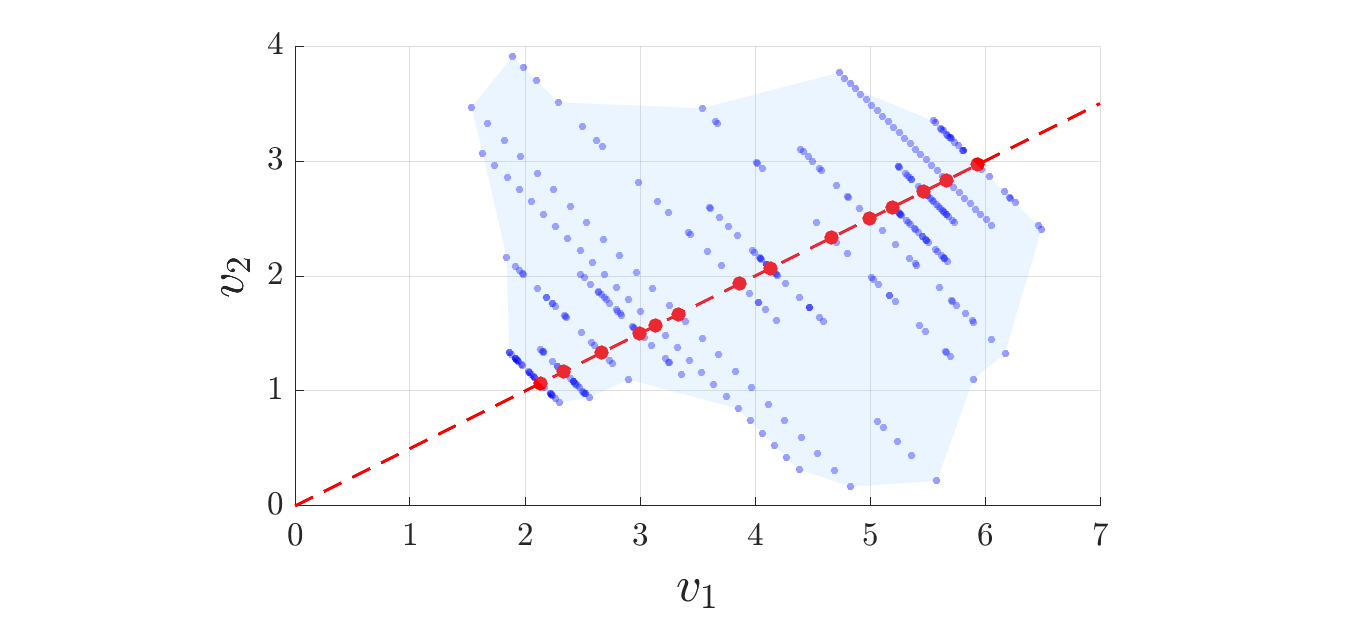}
    \label{fig:rho}
\end{figure}

\begin{definition}[$\boldsymbol{\rho}$-Scaled Invariant Uncertainty Set and Ambiguity Set]
Let $\boldsymbol{\rho}\in \R_+^J$ satisfy $\sum_{j=1}^J \rho_j = 1$.
     An uncertainty set $\cV\subseteq \R_+^J$ is \textit{$\boldsymbol{\rho}$-scaled invariant} if for any $\bv=(v_1,\dots,v_J)\in \cV$, the proportional vector 
     $\tilde{\bv} = \bigof{\sum_{j\in\cJ}v_j}\cdot \boldsymbol{\rho} \in \cV$ as well. The associated support-only ambiguity set $\Delta\of{\cV}$ is called a $\boldsymbol{\rho}$-scaled invariant ambiguity set.
\label{def-rho}
\end{definition}
\Cref{def-rho} requires that whenever the total value of all items equals $\xi$, the proportional valuation $\xi\cdot \boldsymbol{\rho}$ is also feasible. Geometrically, the support of ambiguity set needs to include a radial segment of the ray starting from the origin, in direction $\boldsymbol{\rho}$, as illustrated in \Cref{fig:rho}. In \Cref{fig:rho}, the uncertainty set (depicted by the light blue shaded area) is a perturbation around a radial segment parametrized by $\boldsymbol{\rho}=(1/3,2/3)$. For each feasible valuation with total value $\xi$ (depicted by small blue dots), a red dot $\xi\cdot \boldsymbol{\rho}$ also lies in $\cV$. For notational simplicity, we make the following assumption.
\begin{assumption}
    Denote $\underline{v} = \min_{\bv\in\cV} \sum_{j\in\cJ} v_j$ and $\overline{v} = \max_{\bv\in\cV} \sum_{j\in\cJ} v_j$. For any $\xi\in [\underline{v},\overline{v}]$, there exists $\bv\in\cV$ such that $\sum_{j\in\cJ}v_j=\xi$.
    \label{assum:ambiguityset}
\end{assumption}
\begin{remark}
    A broad class of ambiguity sets satisfy the $\boldsymbol{\rho}$-scaled invariance in \Cref{def-rho} and  \Cref{assum:ambiguityset}. 
    For instance, we say an uncertainty set $\cV$ is \textit{symmetric} if  for every valuation vector $\bv=(v_1,\dots,v_J)\in\cV$ and every permutation $\sigma:\cJ\to\cJ$, the permuted vector $\bv^{\sigma} = (v_{\sigma(1)}, v_{\sigma(2)}, \dots, v_{\sigma(J)})$ also belongs to $\cV$. Then any convex and symmetric uncertainty set is $\boldsymbol{\rho}$-scaled invariant with $\boldsymbol{\rho}=\of{1/J, \dots, 1/J}$ and satisfies \Cref{assum:ambiguityset}. The proof is straightforward: for any $\bv\in \cV$, since $\cV$ is convex, the center of $\offf{\bv^\sigma}$ for all $\sigma\in \Sigma(\cJ)$, which is $\tilde{\bv} = \of{\frac{\sum_{j\in\cJ}v_j}{J}, \dots,\frac{\sum_{j\in\cJ}v_j}{J}}$ is also in $\cV$. Hence, this set $\cV$ is $\of{1/J,\dots, 1/J}$-scaled invariant. In the following, we provide some examples of convex and symmetric uncertainty sets:
\end{remark}

\begin{figure}[htbp]
    \centering
    \caption{Examples of $\boldsymbol{\rho}$-Scaled Invariant Uncertainty Sets}
    \begin{subfigure}[b]{0.2\textwidth}
        \centering      \includegraphics[width=0.8\textwidth,keepaspectratio]{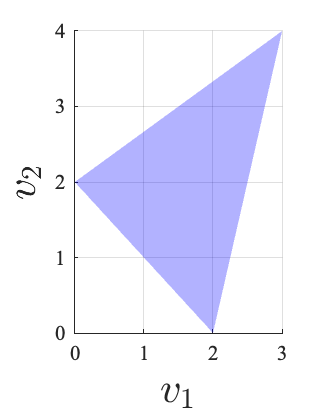}
         \caption{$\boldsymbol{\rho}=(\frac{3}{7},\frac{4}{7})$}
\end{subfigure}
        \begin{subfigure}[b]{0.23\textwidth}
        \centering      \includegraphics[width=0.9\textwidth,keepaspectratio]{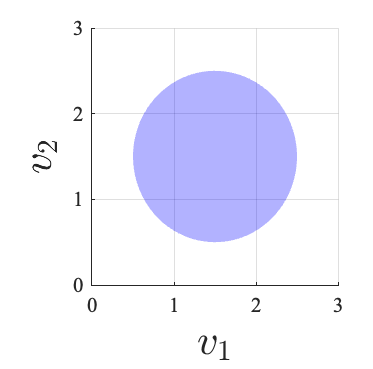}
         \caption{$\boldsymbol{\rho}=(\frac{1}{2},\frac{1}{2})$}
        \end{subfigure}
          \begin{subfigure}[b]{0.25\textwidth}
        \centering      \includegraphics[width=\textwidth,keepaspectratio]{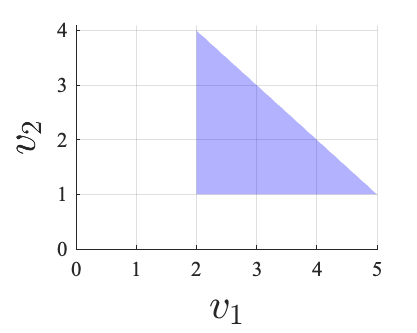}
\caption{$\boldsymbol{\rho}=(\frac{2}{3},\frac{1}{3})$}
          \label{fig:ambiguity set-c}
        \end{subfigure}
        \begin{subfigure}[b]{0.25\textwidth}
        \centering      \includegraphics[width=\textwidth,keepaspectratio]{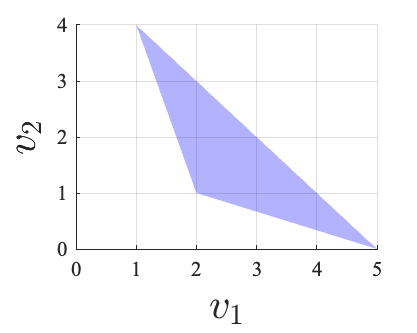}
       \caption{$\boldsymbol{\rho}=(\frac{2}{3},\frac{1}{3})$}
\end{subfigure}
          \begin{subfigure}[b]{0.28\textwidth}
        \centering      \includegraphics[width=\textwidth,keepaspectratio]{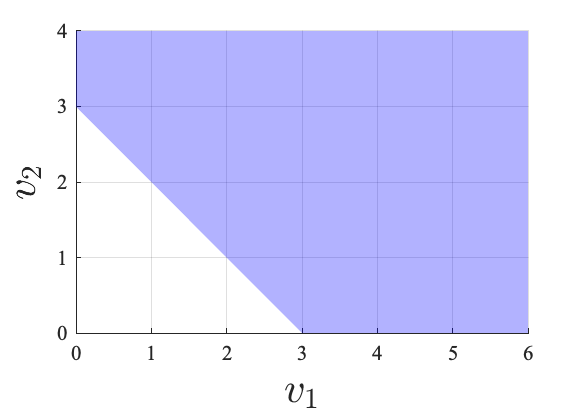}
\caption{$\boldsymbol{\rho}=(\frac{3}{5},\frac{2}{5})$}
        \end{subfigure}
        \begin{subfigure}[b]{0.33\textwidth}
        \centering      \includegraphics[width=\textwidth,keepaspectratio]{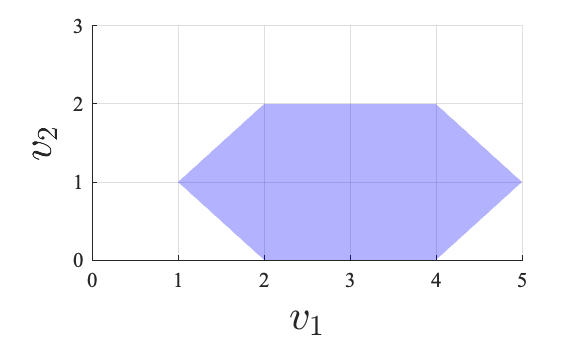}
\caption{$\boldsymbol{\rho}=(\rho_1,1-\rho_1), \rho_1\in[\frac{2}{3},\frac{5}{6}]$}
        \end{subfigure}
         \begin{subfigure}[b]{0.3\textwidth}
        \centering      \includegraphics[width=0.85\textwidth,keepaspectratio]{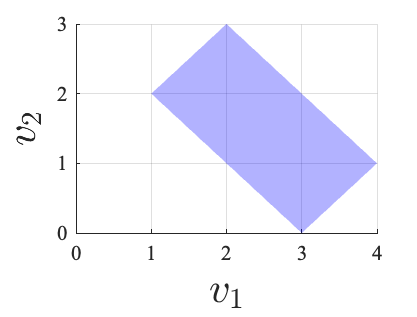}
        \caption{$\boldsymbol{\rho}=(\rho_1,1-\rho_1), \rho_1\in[\frac{2}{5},\frac{4}{5}]$}
        \end{subfigure}
        
        \label{fig:ambiguity set}
\end{figure}

\begin{itemize}
    \item Triangle uncertainty set: $\cV=\bigofff{\bv\mid\sum\limits_{j\in\cJ}v_j\le v_{\text{sum}}, \, v_j\ge v_{\min}, \, \forall j\in \cJ}$ with $0<v_{\min}<v_{\text{sum}}<\infty$.
    \item (Truncated) ellipsoidal uncertainty set: $\cV = \bigofff{\bv\mid\sum_{j\in\cJ}(v_j-v_0)^2 \le r, \, v_{\min} \le v_j\le v_{\max},\, \forall j\in \cJ}$ for some $r, v_{\min}, v_{\max} \in \R_+$.
    \item (Truncated) symmetric $\ell_1$-ball uncertainty set: $\cV = \bigofff{\bv\mid\sum_{j\in\cJ}|v_j-v_0| \le r, \,  v_{\min} \le v_j\le v_{\max}, \, \forall j\in \cJ}$ for some $r, v_{\min}, v_{\max} \in \R_+$.
\end{itemize}
\Cref{fig:ambiguity set} provides  visual illustrations of different examples of $\boldsymbol{\rho}$-scaled invariant uncertainty sets.
In \Cref{thm:bundle-symmetric}, we prove that a randomized bundling mechanism is optimal for all $\boldsymbol{\rho}$-scaled invariant ambiguity sets satisfying \Cref{assum:ambiguityset}, including the examples provided above.
\begin{theorem}
    For a $\boldsymbol{\rho}$-scaled invariant ambiguity set $\Delta\of{\cV}$ whose support $\cV$ satisfies \Cref{assum:ambiguityset}, there is a randomized grand bundling mechanism that is robustly optimal. In particular, denoting $\underline{v} = \min_{\bv\in\cV} \sum_{j\in\cJ} v_j$ and $\overline{v} = \max_{\bv\in\cV} \sum_{j\in\cJ} v_j$, 
the following randomized bundling mechanism is optimal and achieves a performance ratio $\frac{1}{1+\ln(\overline{v}/\underline{v})}$.
    \begin{align}
        [\bq(\bv)]_j = \frac{1+\ln\bigof{{\sum_{i\in\cJ}v_i}/{\underline{v}}}}{1+\ln(\overline{v}/\underline{v})} , \, \forall j\in \cJ, \quad t(\bv) = \frac{\sum_{i\in\cJ}v_i}{1+\ln(\overline{v}/\underline{v})} 
        \label{eq:bundle-symmetric}
    \end{align}
    In addition, nature's optimal strategy is supported on the ray $\bv(\xi) = \of{\xi\rho_1,\dots, \xi\rho_J}, \forall \, \xi\in [\underline{v},\overline{v}]$, where $\xi$ follows a distribution $\G(\xi) =
\frac{\ln\xi-\ln \underline{v}}{1+\ln(\overline{v}/\underline{v})}, \, \forall \xi\in[\underline{v},\overline{v})$ and  $\G(\overline{v})= 1$.
    \label{thm:bundle-symmetric}
\end{theorem}
The mechanism in \Cref{thm:bundle-symmetric} can be implemented by selling all products in a single bundle at a randomized price $p$, with a price density function of $\pi(p) = \frac{1}{\of{1+\ln(\overline{v}/\underline{v})}\cdot p}$ for $p\in (\underline{v},\overline{v}],$ and a probability mass of $\frac{1}{1+\ln(\overline{v}/\underline{v})}$ at $\underline{v}$. Notice that the worst-case distribution is comonotonic as well, similar to the worst-case distribution in \Cref{sec:separable} where the seller's optimal mechanism is separable.
It is well established in standard Bayesian mechanism design that positive correlation typically favors separable mechanisms. In view of this, it may appear counterintuitive that nature's optimal strategy is positively correlated while the seller's optimal response is bundling. Actually, under the comonotonic distribution in \Cref{thm:bundle-symmetric}, a separable mechanism attains the same performance ratio. Nevertheless, the separable mechanism is not robustly optimal, as a slight deviation from nature’s optimal strategy diminishes its performance ratio. The example below illustrates this point.
\begin{example}
\label{eg1}
Consider a triangular uncertainty set in \Cref{fig:ambiguity set-c}, where the valuation of product $j$ is bounded below by $\underline{v}_j$ and the total valuation is at most $\overline{v}$, i.e.,
$\cV = \bigofff{\bv: v_j\ge \underline{v}_j, \forall j\in \cJ, \, \text{and} \, \sum_{j\in \cJ}v_j\le \overline{v}}$. By \Cref{thm:bundle-symmetric}, the optimal performance ratio is $\bigof{1+\ln\of{{\overline{v}}/{\underline{v}}}}^{-1}$, where $\underline{v} = \sum_{j\in \cJ} \underline{v}_j$.
Nature's optimal strategy is comonotonic on the ray $\bv(\xi) = \xi\bigof{\rho_1, \dots, \rho_J} $, with $\rho_j=\frac{\underline{v}_j}{\underline{v}}$ and $\xi\sim\G$ where $\G(\xi) =
\frac{\ln\xi-\ln \underline{v}}{1+\ln(\overline{v}/\underline{v})} $ for $ \xi\in [\underline{v},\overline{v})$ and $\G(\overline{v})=1$. 
Suppose the seller adopts a separable randomized pricing policy, with price distribution $\alpha_j$ for item $j$, i.e., $\P[p_j\le v_j(\xi)]=\alpha_j(\xi)$, where $p_j$ is the price for item $j$. Then $\alpha_j$ is nondecreasing and $\alpha_j(\overline{v})= 1$ for all $j\in\cJ$. 
A buyer with valuation $v_j(\xi)$ purchases and pays $v_j(x)$ when a random draw $x\le \xi$, so the expected payment collected from this buyer for item $j$ is $\int_{\underline{v}}^{\xi}v_j(x) d\alpha_j(x)$. The performance ratio under nature's optimal strategy is:
{$\mathop{\E}\limits_{\bv\sim \F} \Bigoff{\frac{t(\bv)}{\allone^\top \bv}}   =   \mathop{\E}\limits_{\xi \sim \G} \Bigoff{\frac{ \sum\limits_{j\in \cJ} \of{\int_{\underline{v}}^\xi v_j(x) d\alpha_j(x)}  }{ \sum_{i\in\cJ}v_i(\xi)}} =\sum\limits_{j\in \cJ} \Bigoff{\int_{\underline{v}}^{\overline{v}} \Bigof{v_j(\xi) \int_{\xi}^{\overline{v}} \frac{d\G(x)}{ \sum\limits_{i\in\cJ}v_i(x)}}\, d\alpha_j(\xi) }=
\sum\limits_{j\in\cJ}\int_{\underline{v}}^{\overline{v}}\bigof{\rho_j \xi\int_{\xi}^{\overline{v}}\frac{d\G(x)}{x}
    }d\alpha_j(\xi)  = \sum\limits_{j\in\cJ}\int_{\underline{v}}^{\overline{v}}  \frac{\rho_j\cdot \xi \cdot d\alpha_j(\xi) }{(1+\ln(\overline{v}/\underline{v}))\cdot \xi}
    = \frac{1}{1+\ln(\overline{v}/\underline{v})}$},
matching the optimal performance ratio. 
\begin{figure}[htbp]
\caption{Support of Nature's Optimal and Perturbed Strategies}
\label{fig:triangle}
\begin{subfigure}[t]{0.48\textwidth}
\centering
\includegraphics[width=0.6\textwidth,keepaspectratio]{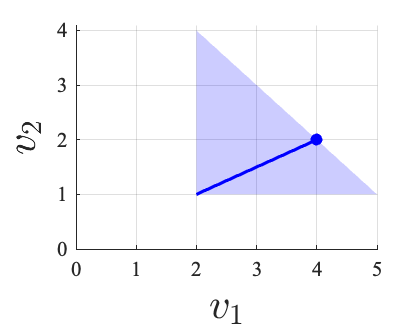}
\caption{\footnotesize Nature's Optimal Strategy (Dark Blue Segment)}
\label{fig:triangle0}
\end{subfigure}
\hfill
\begin{subfigure}[t]{0.48\textwidth}
\centering
\includegraphics[width=0.6\textwidth,keepaspectratio]{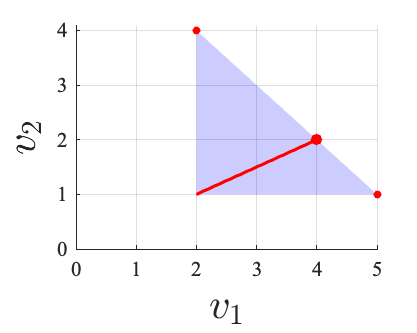}
\caption{\footnotesize  Nature's Perturbed Strategy (Red)}
\label{fig:triangle1}
\end{subfigure}
\end{figure}

However, we will demonstrate that despite achieving this optimal performance ratio under nature’s optimal strategy, the separable mechanism’s performance deteriorates under slight perturbations. Consider a small perturbation of nature’s strategy: shift a small probability mass $\epsilon$ away from the valuation $(\rho_1\overline{v},\dots, \rho_J\overline{v})$  to vertices of the uncertainty set $(\overline{v}-\underline{v}+\underline{v}_1,\underline{v}_2,\dots,\underline{v}_J)$, $(\underline{v}_1,\overline{v}-\underline{v}+\underline{v}_2,\dots,\underline{v}_J), \dots$, $(\underline{v}_1,\dots,\underline{v}_{J-1}, \overline{v}-\underline{v}+\underline{v}_J)$, assigning probability $\rho_j\epsilon$ to the $j$-th vertex. 
In \Cref{fig:triangle1}, the support of the perturbed distribution is marked by the red segment and the red point masses; for comparison, the blue segment with its endpoint mass in \Cref{fig:triangle0} illustrates the original worst‑case distribution.
This perturbation maintains the distribution of total value $\sum_{j\in \cJ}v_j$ unchanged (each vertex still sums to $\overline{v}$), so it does not affect the performance ratio of randomized bundling in \Cref{thm:bundle-symmetric}. However, it increases dispersion in each individual product's valuation, which affects the performance ratio of the separable mechanism. For any separable mechanism with randomized pricing policy $\alpha_j(\xi)$ on the original support for coordinate $j$, along with an additional pricing mass $1-\alpha_j(\overline{v})$ at the perturbed value $v_j=\overline{v}-\underline{v}+\underline{v}_j$, the performance ratio is 
{\small
\begin{align*}
\sum_{j\in\cJ}\Bigof{ \underbrace{ \rho_j  \frac{\alpha_j(\overline{v})}{(1+\ln(\overline{v}/\underline{v}))} }_{\text{baseline under the original support}} - \underbrace{ \int_{\xi\in(\underline{v},\overline{v}]}(1-\rho_j) \ \epsilon  \ \frac{\rho_j \ \xi }{\overline{v} } \  d\alpha_j(\xi)}_{\text{penalty from probability mass shifted to }\underline{v}_j} +\underbrace{ 
\of{1-\alpha_j(\overline{v})}\cdot \rho_j\epsilon \cdot \frac{\underline{v}_j+\overline{v}-\underline{v}}{\overline{v}}}_{\text{gain from pricing at perturbed point } \overline{v}-\underline{v}+\underline{v}_j} }. 
\end{align*}
}For sufficiently small $\epsilon$, the coefficient of $\alpha_j$ in the first term dominates those in the second and third terms. Hence, assigning any positive pricing probability at the perturbed point $\overline{v}-\underline{v}+\underline{v}_j$ strictly lowers the performance ratio. Optimality therefore requires placing no mass there, i.e., $\alpha_j(\overline{v})=1$, and then the third term is zero. 
Meanwhile, as a probability mass $(1-\rho_j)\epsilon$ is moved from $\rho_j\overline{v}$ to $\underline{v}_j$, the second term $- \int_{\xi\in(\underline{v},\overline{v}]}(1-\rho_j)  \epsilon   \frac{\rho_j  \xi }{\overline{v} }   d\alpha_j(\xi)$ is strictly negative unless $d\alpha_j(\xi)=0$ for all $\xi\in(\underline{v}, \overline{v}]$, which can  occur only if all pricing mass for item $j$ is placed at the lower bound $\underline{v}_j$. 
But concentrating price at $\underline{v}_j$ is strictly suboptimal if the total value is $\overline{v}$ with probability one. Consequently, every separable mechanism suffers a strictly lower worst-case ratio under this perturbation, whereas randomized bundling remains at the optimum.
\end{example}

As illustrated by \Cref{eg1}, no (randomized) separable mechanism is robustly optimal. \Cref{eg1} also clarifies the seemingly counterintuitive optimality of a positively correlated strategy for nature.  While positively correlated valuations emerge as optimal for nature and make separate selling and bundling tie, the possibility of slight perturbations toward partial negative correlation degrade separable mechanisms and ultimately motivates the robust optimality of bundling.

\section{Summary}
In this paper, we study multi‑item mechanism design when the seller knows only the support of the buyer’s valuation. Incorporating the maximin ratio as our performance criterion, we fully characterize the robustly optimal mechanism.
For a rectangular-supported ambiguity set, we prove the optimality of a separable mechanism, which is easy to interpret and straightforward to implement.
The proof proceeds via a saddle-point argument that identifies a comonotonic worst-case distribution and reduces the multi-dimensional design problem to a tractable scalar functional program. Extending the analysis to the broader class of $\boldsymbol{\rho}$-scaled invariant ambiguity sets, we show that a randomized bundling mechanism becomes optimal.
These results provide practical guidance for mechanism design under non‑standard ambiguity structures. A promising direction for future research is to incorporate additional information, such as moment or quantile constraints and partial correlation structures, into the ambiguity set and examine how this affects the achievable approximation ratio and the optimal mechanism.

\bibliographystyle{informs2014}
\bibliography{myref.bib}

\clearpage
\begin{APPENDICES}
\section{Omitted Proofs}
\begin{proof}{Proof of \Cref{lemma:single}.}
Problem \eqref{eq:original} can be equivalently formulated as follows.
    \begin{eqnarray*} \sup_{M\in \cM}\inf_{\F\in \cF} \frac{\Rev(M,\F)}{\sup_{M'\in \cM}\Rev(M',\F)} = \sup_{M\in \cM}\inf_{M'\in \cM}\inf_{\F\in \cF} \frac{\Rev(M,\F)}{\Rev(M',\F)} 
 = \sup_{M\in \cM}\inf_{M'\in \cM}\inf_{\F\in \cF} \frac{\int_{\bv\in \cV} t(\bv)d \F(\bv)}{\int_{\bv\in \cV} t'(\bv)d \F(\bv)}.
    \end{eqnarray*}
    For any given $M$ and $M'$, the numerator and denominator are both linear in $\F$. The coefficient of $d\F(\bv)$ in the numerator is $t(\bv)$ and coefficient of $d\F(\bv)$ in the denominator is $t'(\bv)$. Hence, for any optimal solution for the adversary $(\F,M')$, we have that $\frac{\int_{\bv\in \cV} t(\bv)d \F(\bv)}{\int_{\bv\in \cV} t'(\bv)d \F(\bv)}\ge \min\limits_{\bv\in\cV} \frac{t(\bv)}{t'(\bv)}$. Thus, there is an optimal solution for the adversary such that $\F$ reduces to a one-point distribution, due to
    $$
\sup_{M\in \cM}\inf_{M'\in \cM}\inf_{\F\in \cF} \frac{\int_{\bv\in \cV} t(\bv)d \F(\bv)}{\int_{\bv\in \cV} t'(\bv)d \F(\bv)} = \sup_{M\in \cM}\inf_{M'\in \cM}\min\limits_{\bv\in\cV} \frac{t(\bv)}{t'(\bv)} = \sup_{M\in \cM}\min\limits_{\bv\in\cV} \inf_{M'\in \cM}\frac{t(\bv)}{t'(\bv)}.
    $$
Since $M'=(\bq', t')$ satisfies the IR constraints, we have $t'(\bv)\le \allone^\top \bv$. Then for any given $\bv$, the adversary's optimal strategy is to allocate all products with probability one and collect a payment of $\allone^\top \bv$. Therefore, the adversary's optimal strategy is a point mass with a performance ratio of
 \begin{eqnarray*}\inf_{\F\in \cF} \frac{\Rev(M,\F)}{\sup_{M'\in \cM}\Rev(M',\F)} =\inf_{M'\in \cM}\inf_{\F\in \cF} \frac{\int_{\bv\in \cV} t(\bv)d \F(\bv)}{\int_{\bv\in \cV} t'(\bv)d \F(\bv)} = \inf_{M'\in \cM}\min\limits_{\bv\in\cV} \frac{t(\bv)}{t'(\bv)} = \min\limits_{\bv\in\cV} \inf_{M'\in \cM}\frac{t(\bv)}{t'(\bv)} = \min\limits_{\bv\in\cV} \frac{t(\bv)}{\allone^\top \bv}.
    \end{eqnarray*}  \QED  \end{proof}
\begin{proof}{Proof of \Cref{semi-feasible}.}

For any $\bv\in \cV$, since the IR constraints hold for the one-dimensional mechanisms $\offf{\of{q_j,t_j}}_{j\in\cJ}$, we have that
$\bq(\bv)^\top \bv - t(\bv)= \sum_{j\in \cJ} q_j(v_j)\cdot v_j - \sum_{j\in\cJ} t_j(v_j) \ge 0$.
Hence, the IR constraint for the separable mechanism is satisfied.
On the other hand, for any $\bv\in \cV$ and $\bv'\in \cV$, we have that 
{\small
\begin{align*}
&\bigof{\bq(\bv)^\top \bv - t(\bv)} -\bigof{\bq(\bv')^\top \bv - t(\bv')} 
= \sum_{j\in \cJ} q_j (v_j) \cdot v_j - \sum_{j\in\cJ} t_j(v_j)  - \bigof{\sum_{j\in \cJ} q_j (v_j') \cdot v_j - \sum_{j\in\cJ} t_j(v_j') }\\
=&\sum_{j\in \cJ}\bigof{
q_j (v_j) \cdot v_j -  t_j(v_j) - q_j (v_j') \cdot v_j + t_j(v_j')
} \ge  0
\end{align*}
}
where the inequality is due to the IC constraints for the one-dimensional mechanisms $\offf{\of{q_j,t_j}}_{j\in\cJ}$.
Hence, the IC constraint for the separable mechanism is satisfied. 
\QED
\end{proof}

 \begin{proof}{Proof of \Cref{lemma:semi}.}
     Since $(\bq(\bv), t(\bv))$ can be represented as 
$\bq(\bv) = \bigof{q_1(v_1), \dots, q_J(v_J)}$, $t(\bv) = \sum_{j\in \cJ} t_j(v_j)$, 
where $q_j(v_j)= \bigof{\gamma\cdot\ln (v_j/\overline{v}_j)+1}^+$ and $t_j(v_j) = \gamma\cdot \bigof{v_j -e^{-1/\gamma} \cdot \overline{v}_j }^+$ or $t_j(v_j) = \gamma\cdot v_j +\underline{v}_j\cdot\bigof{\gamma(\ln(\underline{v}_j/\overline{v}_j)-1)+1}$ are both independent of $\bv_{-j}$, we only need to verify that the one-dimensional rule $(q_j, t_j)$ satisfies IC and IR constraints in dimension $j$. 
\begin{enumerate}
    \item If $e^{-1/\gamma}\cdot \overline{v}_j\le \underline{v}_j$, then for all $v_j\in [\underline{v}_j,\overline{v}_j], \gamma\cdot\ln (v_j/\overline{v}_j)+1 \ge 0$, so 
    $q_j(v_j) =\gamma\cdot \ln (v_j/\overline{v}_j)+1$ and 
    $t_j(v_j) = \gamma\cdot v_j +\underline{v}_j\cdot\bigof{\gamma(\ln(\underline{v}_j/\overline{v}_j)-1)+1} $. We have that the buyer's utility $u_j(v_j) =  q_j(v_j)v_j - t_j(v_j) = \gamma \cdot \bigof{v_j\cdot \ln(v_j/\overline{v}_j)-\underline{v}_j\cdot \ln(\underline{v}_j/\overline{v}_j)}+(1-\gamma)(v_j-\underline{v}_j)$. Taking the derivative of $u_j(v_j)$ with respect to $v_j$, we have that $\frac{\partial u_j}{\partial v_j} = \gamma\cdot\bigof{\ln(v_j/\overline{v}_j)+1}+1-\gamma = \gamma\cdot\bigof{\ln(v_j/\overline{v}_j)}+1 \ge 0$ and $\frac{\partial^2 u_j}{\partial v_j^2} = \gamma/v_j\ge 0$, which implies that $u_j$ is increasing and convex, so $(q_j(v_j), t_j(v_j))$ satisfies the IC constraint. Besides, since $u_j(\underline{v}_j) = 0$,  $(q_j(v_j), t_j(v_j))$ satisfies the IR constraint.
    \item If $e^{-1/\gamma}\cdot \overline{v}_j>\underline{v}_j$, then 
    \begin{align*}
        (q_j(v_j), t_j(v_j)) = \begin{cases}
            (0,\, 0) & \text{if } v_j \le e^{-1/\gamma} \cdot \overline{v}_j \\
            \bigof{\gamma\cdot\ln (v_j/\overline{v}_j)+1, \, \gamma\cdot \bigof{v_j -e^{-1/\gamma} \cdot \overline{v}_j}} & \text{if } v_j > e^{-1/\gamma} \cdot \overline{v}_j
        \end{cases}
    \end{align*}
    Hence, $u_j(v_j) =0$ when  $v_j \le e^{-1/\gamma} \cdot \overline{v}_j$ and $u_j(v_j) =  q_j(v_j)v_j - t_j(v_j) =\bigof{\gamma\cdot\ln (v_j/\overline{v}_j)+1}\cdot v_j - \bigof{\gamma\cdot \bigof{v_j -e^{-1/\gamma} \cdot \overline{v}_j}} $  when  $v_j > e^{-1/\gamma} \cdot \overline{v}_j$. Taking the derivative of the second part (when $v_j > e^{-1/\gamma} \cdot \overline{v}_j$) of $u_j(v_j)$ with respect to $v_j$, we have that  $\frac{\partial u_j}{\partial v_j} = \gamma\cdot \bigof{1+\ln (v_j/\overline{v}_j)} + 1-\gamma = 1+\gamma \ln (v_j/\overline{v}_j)>0$ and $\frac{\partial^2 u_j}{\partial v_j^2} = \gamma/v_j\ge 0$. Considering $u_j(v_j) =0$ when  $v_j \le e^{-1/\gamma} \cdot \overline{v}_j$, we can see that $u_j(v_j)$ is nonnegative, increasing and convex within $[\underline{v}_j,\overline{v}_j]$, so $(q_j(v_j), t_j(v_j))$ satisfied the IC and IR constraints. \QED
\end{enumerate}
 \end{proof}

 \begin{proof}{Proof of \Cref{increasing-g}.}

  When $\gamma \neq -1/\ln(\underline{v}_j/\overline{v}_j)$ for any $j\in\cJ$, the derivative of $\phi$ with respect to $\gamma$ is
$\phi'(\gamma)=e^{-1/\gamma}(1+  \frac{1}{\gamma}) \sum_{j\in \cS(\gamma)} \overline{v}_j - \sum_{j\in \cJ\setminus\cS(\gamma)} \bigof{\underline{v}_j\cdot\bigof{\ln(\underline{v}_j/\overline{v}_j)-1}}> e^{-1/\gamma}(1+  \frac{1}{\gamma}) \sum_{j\in \cS(\gamma)} \overline{v}_j - \sum_{j\in \cJ\setminus\cS(\gamma)} \bigof{\underline{v}_j\cdot\bigof{-\frac{1}{\gamma}-1}}
> 0$, so $\phi(\gamma)$ is strictly increasing in $\gamma$ when $\gamma$ is within the intervals where $\cS(\gamma)$ does not change. 
At the $\gamma$ values where $\gamma = -1/\ln(\underline{v}_j/\overline{v}_j)$ for some $j\in\cJ$, as $\gamma$ increases, the set $\cS(\gamma)$ could be larger. The only possible points of discontinuity occur at $\gamma=-1/\ln(\underline{v}_j/\overline{v}_j) $ for some $j\in \cJ$.    
    For any $j\in \cJ$, now we examine whether $\phi(\gamma)$ is continuous in $\gamma$ at $\gamma_0=-1/\ln(\underline{v}_j/\overline{v}_j) $. Since $\cS(\gamma_0-) = \cS(\gamma_0)$, we have that $\lim_{\gamma\to\gamma_0-}\phi(\gamma) = \phi(\gamma_0)$, indicating left continuity. Then by $\cS(\gamma_0+) = \cS(\gamma_0)\cup \{j\}$, we have that $\lim_{\gamma\to\gamma_0+}\phi(\gamma) - \phi(\gamma_0) = \gamma_0\cdot e^{-1/\gamma_0} \cdot  \overline{v}_j + \bigof{\underline{v}_j\cdot\bigof{\gamma_0\ln(\underline{v}_j/\overline{v}_j)-\gamma_0+1}} = \overline{v}_j \cdot \bigof{\gamma_0\cdot e^{-1/\gamma_0}  + \underline{v}_j/
\overline{v}_j\cdot\bigof{\gamma_0\ln(\underline{v}_j/\overline{v}_j)-\gamma_0+1}} = \overline{v}_j \cdot \bigof{\gamma_0\cdot e^{-1/\gamma_0}  - e^{-1/\gamma_0}\cdot\gamma_0} = 0$, indicating right continuity.
Therefore, $\phi$ continuously increases in $\gamma$. Since $\phi(0) = - \sum_{j\in \cJ} \bigof{\underline{v}_j} <0$ and $\phi(1) = e^{-1} \cdot \sum_{j\in \cS(1)} \overline{v}_j - \sum_{j\in \cJ\setminus\cS(1)} \bigof{\underline{v}_j\cdot\ln(\underline{v}_j/\overline{v}_j)}\ge 0 $, there is a unique solution to $\phi(\gamma)=0$ at $\gamma\in(0,1]$. 
\QED
\end{proof}

\begin{proof}{Proof of \Cref{prop:ratio-feasible}.}

Since $\cS(\gamma)=\bigofff{j\in \cJ \mid \underline{v}_j/\overline{v}_j< e^{-1/\gamma}}$,
by \Cref{lemma:single}, the performance ratio of \ref{eq:q} is
    \begin{align*}
        \min_{\bv\in \cV} \frac{t(\bv)}{\allone^\top\bv} = \min_{\bv\in \cV} 
 \frac{\sum_{j\in \cS(\gamma)}  \gamma\cdot \of{v_j -e^{-1/\gamma} \cdot \overline{v}_j}^+ + \sum_{j\in \cJ\setminus\cS(\gamma)}
      \bigof{ \gamma\cdot v_j +\underline{v}_j\cdot\bigof{\gamma(\ln(\underline{v}_j/\overline{v}_j)-1)+1}}  }{\sum_{j\in \cJ}v_j}.
    \end{align*}
    Denote $\Rad_{\gamma}(\bv)=\frac{t(\bv)}{\allone^\top\bv} 
$. We can see that $\Rad_{\gamma}(\bv)$ is continuous in $\bv$. Now we aim to find the worst-case $\bv$ to minimize $\Rad_{\gamma}(\bv)$ within $\cV$. 

We proceed in two steps. We first reduce the candidate range of the worst-case $\bv$ by showing $\min_{\bv\in \cV}\Rad_{\gamma}(\bv) =\min_{\bv\in \underline{\cV}}\Rad_{\gamma}(\bv) $ where $ \underline{\cV} = \offf{\bv\in \cV\mid v_j\ge  e^{-1/\gamma} \cdot \overline{v}_j, \forall j\in \cS(\gamma)}$. Second, we find the worst-case $\bv$ within $\underline{\cV}$.

Step 1. Taking the derivative of $\Rad_{\gamma}(\bv)$ with respect to $v_i$ for a product $i\in \cS(\gamma)$, 
    {\footnotesize
            \begin{align*}
          \frac{\partial \Rad_{\gamma}(\bv)}{\partial v_i} = \frac{ \gamma\cdot \mathbbm{1}\off{v_i \ge e^{-1/\gamma} \cdot \overline{v}_i}\cdot  \of{\sum_{j\in \cJ}v_j} - 
          \sum_{j\in \cS(\gamma)}  \gamma\cdot \of{v_j -e^{-1/\gamma} \cdot \overline{v}_j}^+ - \sum_{j\in \cJ\setminus\cS(\gamma)} \of{ \gamma\cdot v_j +\underline{v}_j\cdot\of{\gamma(\ln(\underline{v}_j/\overline{v}_j)-1)+1}}  }{\of{\sum_{j\in \cJ}v_j}^2}.
          \end{align*}
     }
      
The denominator is always positive, so we consider the numerator.
When $v_i \le e^{-1/\gamma} \cdot \overline{v}_i$, the numerator is 
{\small $
  - 
  \sum_{j\in \cS(\gamma)}  \gamma\cdot \bigof{v_j -e^{-1/\gamma} \cdot \overline{v}_j}^+ - \sum_{j\in \cJ\setminus\cS(\gamma)} \bigof{ \gamma\cdot v_j +\underline{v}_j\cdot\bigof{\gamma(\ln(\underline{v}_j/\overline{v}_j)-1)+1}}  \le 0,
$}
which implies that the ratio $\Rad_{\gamma}(\bv)$ is decreasing in $v_i$.
Since the support $\cV$ is rectangular, for any given $\bv_{-i}$, increasing $v_i$ up to $e^{-1/\gamma} \cdot \overline{v}_i$ will not increase the performance ratio. 
      Thus, in order to minimize $\Rad_{\gamma}(\bv)$, it is without loss of optimality to consider $\bv$ with $v_i\ge e^{-1/\gamma} \cdot \overline{v}_i$, for all $i\in  \cS(\gamma)$. Therefore, $\min_{\bv\in \cV}\Rad_{\gamma}(\bv) =\min_{\bv\in \underline{\cV}}\Rad_{\gamma}(\bv) $. 

      Step 2. Next, we find the optimal $\bv\in \underline{\cV}$ via first-order condition.
      
      \begin{enumerate}
          \item For $i\in \cS(\gamma)$, we have that    $v_i \ge e^{-1/\gamma} \cdot \overline{v}_i$, for all $i\in  \cS(\gamma)$, so the numerator of the derivative becomes
      {\small
      \begin{align*}
        & \, \gamma\cdot  \bigof{\sum_{j\in \cJ}v_j} - 
          \sum_{j\in \cS(\gamma)}  \gamma\cdot \bigof{v_j -e^{-1/\gamma} \cdot \overline{v}_j} - \sum_{j\in \cJ\setminus\cS(\gamma)} \bigof{ \gamma\cdot v_j +\underline{v}_j\cdot\bigof{\gamma(\ln(\underline{v}_j/\overline{v}_j)-1)+1}}  \\
          =& \,  
           \gamma\cdot e^{-1/\gamma} \cdot \sum_{j\in \cS(\gamma)} \overline{v}_j - \sum_{j\in \cJ\setminus\cS(\gamma)} \bigof{\underline{v}_j\cdot\bigof{\gamma\ln(\underline{v}_j/\overline{v}_j)-\gamma+1}} = \phi(\gamma)
      \end{align*}
      }
      which is a constant independent of $i$ for all $i\in \cS(\gamma)$.

      \item For $i\in \cJ\setminus\cS(\gamma)$, we have that
      {\small
          $
          \frac{\partial \Rad_{\gamma}(\bv)}{\partial v_i} = \frac{ \gamma\cdot  \of{\sum_{j\in \cJ}v_j} - 
          \sum_{j\in \cS(\gamma)}  \gamma\cdot \of{v_j -e^{-1/\gamma} \cdot \overline{v}_j}^+ - \sum_{j\in \cJ\setminus\cS(\gamma)} \of{ \gamma\cdot v_j +\underline{v}_j\cdot\of{\gamma(\ln(\underline{v}_j/\overline{v}_j)-1)+1}}  }{\of{\sum_{j\in \cJ}v_j}^2}.
      $
      }
      Sicne $v_j \ge e^{-1/\gamma} \cdot \overline{v}_j$ for all $j\in  \cS(\gamma)$, the numerator is equivalent to
     {\small \begin{align*}
          & \gamma\cdot  \bigof{\sum_{j\in \cJ}v_j} - 
          \sum_{j\in \cS(\gamma)}  \gamma\cdot \bigof{v_j -e^{-1/\gamma} \cdot \overline{v}_j} - \sum_{j\in \cJ\setminus\cS(\gamma)} \bigof{ \gamma\cdot v_j +\underline{v}_j\cdot\bigof{\gamma(\ln(\underline{v}_j/\overline{v}_j)-1)+1}}  \\
          =&  
           \gamma\cdot e^{-1/\gamma} \cdot \sum_{j\in \cS(\gamma)} \overline{v}_j - \sum_{j\in \cJ\setminus\cS(\gamma)} \bigof{\underline{v}_j\cdot\bigof{\gamma\ln(\underline{v}_j/\overline{v}_j)-\gamma+1}} = \phi(\gamma)
      \end{align*}}
      which is the same constant as for $i\in \cS(\gamma)$.
      \end{enumerate}
Hence, we observe that the sign of the partial derivative  $\frac{\partial \Rad_{\gamma}(\bv)}{\partial v_i}$ is the same as the sign of $\phi(\gamma)$, which is the same across all dimensions $i\in \cJ$. 
When $\phi(\gamma) =0$, for any $\bv \in \underline{\cV}$, the derivative of $\Rad_{\gamma}(\bv)$ with respect to $v_i$ is $0$  for all $i\in \cJ$. Hence, 
the performance ratio becomes:
{
\footnotesize
\begin{align*}
    \Rad_{\gamma^*}(\bv)=&\frac{\sum_{j\in \cS(\gamma^*)}  \gamma^*\cdot \bigof{v_j -e^{-1/\gamma^*} \cdot \overline{v}_j} + \sum_{j\in \cJ\setminus \cS(\gamma^*)}
      \bigof{ \gamma^*\cdot v_j +\underline{v}_j\cdot\bigof{\gamma^*(\ln(\underline{v}_j/\overline{v}_j)-1)+1}}  }{\sum_{j\in \cJ}v_j} 
      = \frac{\sum_{j\in \cJ}  \gamma^*\cdot v_j -\phi(\gamma^*)}{\sum_{j\in \cJ}v_j} =\gamma^*
\end{align*}
}
which completes our proof that the performance ratio obtained by mechanism $M_{\gamma^*}$ is exactly $\gamma^*$. 
\QED
\end{proof}
\begin{proof}{Proof of \Cref{cor:2item}.}

According to \Cref{prop:ratio-feasible}, the performance ratio is solved by $\phi(\gamma)=\gamma\cdot e^{-1/\gamma} \cdot \sum_{j\in \cS(\gamma)} \overline{v}_j - \sum_{j\in \cJ\setminus\cS(\gamma)} \bigof{\underline{v}_j\cdot\bigof{\gamma\ln(\underline{v}_j/\overline{v}_j)-\gamma+1}} = 0$ where $\cS(\gamma)=\bigofff{j\in \cJ \mid \underline{v}_j/\overline{v}_j<e^{ -1/\gamma}}$. For different values of $\gamma$, $\cS(\gamma)$ can be $\offf{}, \offf{1}$ or $\offf{1,2}$. Considering these three scenarios, we list the following three possible formulations of $\phi$. By \Cref{increasing-g}, since $\phi(\gamma)$ is increasing, only one of  the following three equations has a positive solution:
\begin{align*}
   &\phi(\gamma)= -\sum_{j\in \cJ} \bigof{\underline{v}_j\cdot\bigof{\gamma\ln(\underline{v}_j/\overline{v}_j)-\gamma+1}} = 0,\quad \gamma\in (0, \frac{1}{\ln(\overline{v}_1/\underline{v}_1)}] \,& (  \cS(\gamma) \text{ is empty})\\
  &  \phi(\gamma)= \gamma\cdot e^{-1/\gamma} \cdot \overline{v}_1 - \bigof{\underline{v}_2\cdot\bigof{\gamma\ln(\underline{v}_2/\overline{v}_2)-\gamma+1}} = 0, \quad \gamma\in (\frac{1}{\ln(\overline{v}_1/\underline{v}_1)}, \frac{1}{\ln(\overline{v}_2/\underline{v}_2)}] \,& ( \cS(\gamma) = \{1\}) \\
 & \phi(\gamma)=   \gamma\cdot e^{-1/\gamma} \cdot (\overline{v}_1 +\overline{v}_2)= 0, \quad\gamma\in (\frac{1}{\ln(\overline{v}_2/\underline{v}_2)} ,1] \, & (\cS(\gamma) =  \{1,2\})
\end{align*}
Notice that the third equation can not hold, so we only need to analyze the first two.

To identify whether $\cS(\gamma)$ is empty or $\cS(\gamma)=\{1\}$, we need to check the value of $\phi(\gamma)$ at $\gamma = \frac{1}{\ln(\overline{v}_1/\underline{v}_1)}$, i.e. $\phi(\frac{1}{\ln(\overline{v}_1/\underline{v}_1)})=-\sum_{j\in \cJ} \bigof{\underline{v}_j\cdot\bigof{\frac{1}{\ln(\overline{v}_1/\underline{v}_1)}\ln(\underline{v}_j/\overline{v}_j)-\frac{1}{\ln(\overline{v}_1/\underline{v}_1)}+1}}$.

\begin{enumerate}
    \item  If $\underline{v}_{2}\,\overline{v}_{1}\;>\;\overline{v}_{2}\,\underline{v}_{1} e^{\,1+\underline{v}_{1}/\underline{v}_{2}}$, then $\phi(\frac{1}{\ln(\overline{v}_1/\underline{v}_1)})= - \sum_{j\in \cJ} \bigof{\underline{v}_j\cdot\bigof{\frac{1}{\ln\overline{v}_1/\underline{v}_1}(\ln(\underline{v}_j/\overline{v}_j)-1)+1}} = \frac{\underline{v}_1}{\ln\overline{v}_1/\underline{v}_1}-\underline{v}_2\cdot \of{\frac{1}{\ln\overline{v}_1/\underline{v}_1}\cdot \of{\ln(\underline{v}_2/\overline{v}_2)-1}+1}<\frac{\underline{v}_1}{\ln\overline{v}_1/\underline{v}_1}-\underline{v}_2\cdot \of{\frac{1}{\ln\overline{v}_1/\underline{v}_1}\cdot \of{1+\underline{v}_1/\underline{v}_2+\ln(\underline{v}_1/\overline{v}_1)-1}+1} = 0$, where the inequality is due to $\underline{v}_{2}\,\overline{v}_{1}\;>\;\overline{v}_{2}\,\underline{v}_{1} e^{\,1+\underline{v}_{1}/\underline{v}_{2}}$.
    Since $\phi(\gamma) $ is increasing in $\gamma$, this implies that when $\gamma\le \frac{1}{\ln\overline{v}_1/\underline{v}_1}$ the function $\phi(\gamma)< 0$. On the other hand, when $\gamma = \frac{1}{\ln(\overline{v}_2/\underline{v}_2)}$, we have $\phi(\frac{1}{\ln(\overline{v}_2/\underline{v}_2)}) = \gamma\cdot e^{-1/\gamma} \cdot \overline{v}_1 - \bigof{\underline{v}_2\cdot\bigof{\gamma\ln(\underline{v}_2/\overline{v}_2)-\gamma+1}} = \gamma\cdot e^{-1/\gamma} \cdot (\overline{v}_1 +\overline{v}_2) > 0 $, so there exists $\gamma\in (\frac{1}{\ln(\overline{v}_1/\underline{v}_1)}, \frac{1}{\ln(\overline{v}_2/\underline{v}_2)})$, such that $ \phi(\gamma)=\gamma\cdot e^{-1/\gamma} \cdot \overline{v}_1 - \bigof{\underline{v}_2\cdot\bigof{\gamma\ln(\underline{v}_2/\overline{v}_2)-\gamma+1}} = 0$. By solving this equation, the performance ratio is \begin{eqnarray*} 
\gamma =\left(W(\frac{\overline{v}_1}{e\overline{v}_2}) + \ln \frac{\overline{v}_2}{\underline{v}_2}+1\right)^{-1}
\end{eqnarray*} 
where $W$ is the Lambert-W function defined as the inverse function of $f(W) = We^W$. Thus, by the definition of \ref{eq:q}, $q_1(v_1)=\of{1+\gamma\ln (v_1/\overline{v}_1)}^+$ and $q_2(v_2)=\of{1+\gamma\ln (v_2/\overline{v}_2)}$, since $\gamma\in (\frac{1}{\ln(\overline{v}_1/\underline{v}_1)}, \frac{1}{\ln(\overline{v}_2/\underline{v}_2)})$; $t(\bv) =  \gamma\cdot  \bigof{v_1 -e^{-1/\gamma} \cdot \overline{v}_1}^+ + 
         \gamma\cdot v_2 +\underline{v}_2\cdot\bigof{\gamma(\ln(\underline{v}_2/\overline{v}_2)-1)+1}=\gamma\cdot  \bigof{v_1 -e^{-1/\gamma} \cdot \overline{v}_1}^+ + 
         \gamma\cdot v_2 +\gamma e^{-1/\gamma} \cdot \overline{v}_1=\gamma\cdot\of{\max\offf{v_1,\, e^{-1/\gamma} \cdot \overline{v}_1}+v_2}$.

\item If $\underline{v}_{2}\,\overline{v}_{1}\;\le \;\overline{v}_{2}\,\underline{v}_{1} e^{\,1+\underline{v}_{1}/\underline{v}_{2}}$, then $\phi(\frac{1}{\ln(\overline{v}_1/\underline{v}_1)})=  - \sum_{j\in \cJ} \bigof{\underline{v}_j\cdot\bigof{\frac{1}{\ln\overline{v}_1/\underline{v}_1}\ln(\underline{v}_j/\overline{v}_j)-\frac{1}{\ln\overline{v}_1/\underline{v}_1}+1}} = \frac{\underline{v}_1}{\ln\overline{v}_1/\underline{v}_1}-\underline{v}_2\cdot \of{\frac{1}{\ln\overline{v}_1/\underline{v}_1}\cdot \of{\ln(\underline{v}_2/\overline{v}_2)-1}+1}\ge  \frac{\underline{v}_1}{\ln\overline{v}_1/\underline{v}_1}-\underline{v}_2\cdot \of{\frac{1}{\ln\overline{v}_1/\underline{v}_1}\cdot \of{1+\underline{v}_1/\underline{v}_2+\ln(\underline{v}_1/\overline{v}_1)-1}+1} = 0$, where the inequality is due to $\underline{v}_{2}\,\overline{v}_{1}\;\le \;\overline{v}_{2}\,\underline{v}_{1} e^{\,1+\underline{v}_{1}/\underline{v}_{2}}$. Since $\phi(\gamma)$ is increasing, there exists a unique solution to 
$\phi(\gamma) =- \sum_{j\in \cJ} \bigof{\underline{v}_j\cdot\bigof{\gamma\ln(\underline{v}_j/\overline{v}_j)-\gamma+1}} = 0$ for $\gamma\in(0, \frac{1}{\ln(\overline{v}_1/\underline{v}_1)}]$. Solving this equation, the performance ratio is
 \begin{eqnarray*} 
\gamma = \frac{\sum_{j\in \cJ} \underline{v}_j}{\sum_{j\in \cJ} \bigof{ \underline{v}_j \cdot \bigof{1+\ln(\overline{v}_j/\underline{v}_j)} }}
\end{eqnarray*} Thus, by the definition of \ref{eq:q}, $q_j(v_j)=1+\gamma\ln (v_j/\overline{v}_j)$, since $\gamma\le \frac{1}{\ln(\overline{v}_j/\underline{v}_j)}$ for $j=1,2$; $t(\bv) = \sum_{j\in\cJ}\of{ 
         \gamma\cdot v_j +\underline{v}_j\cdot\bigof{\gamma(\ln(\underline{v}_j/\overline{v}_j)-1)+1}}=\gamma\cdot\of{v_1+v_2}$.
\end{enumerate}
These two cases complete our proof of \Cref{cor:2item}. \QED
\end{proof}

\begin{proof}{Proof of \Cref{prop:separable}.}
    It is straightforward that the decomposed separable mechanism $M=(\bq,t)$ with $\bq(\bv) = \bigof{q_j^\dag (v_j)}_{j\in \cJ}$ and $t(\bv) = \sum_{j\in \cJ} t_j^\dag(v_j)$ defined in \eqref{eq:separate} satisfies the incentive‐compatibility (IC) and individual‐rationality (IR) constraints, so it is a feasible mechanism for the multi-item problem. The approximation ratio achieved by this mechanism is $\min_{\bv\in \cV} \Rad_{dec}(\bv)$ where
$
   \Rad_{dec}(\bv)=\frac{t(\bv)}{\allone^\top\bv} =  \frac{\sum_{j\in \cJ} t_j^\dag(v_j)}{\allone^\top\bv} =
 \frac{\sum_{j\in \cJ}r_j^\dag v_j }{\sum_{j\in \cJ}v_j}
$, with $r_j^\dag =\frac{1 }{1+\ln(\overline{v}_j/\underline{v}_j)}$ denoting the approximation ratio obtained by the optimal single-dimensional mechanism for product $j$. 
Notice that $\Rad_{dec}(\bv)= \frac{\sum_{j\in \cJ}r_j^\dag v_j }{\sum_{j\in \cJ}v_j}$ can be interpreted as the weighted average of $r_j^\dag $ with weight $ \frac{ v_j }{\sum_{i\in \cJ}v_j}$. Hence, we can find the worst-case $\bv$ in the following process. If $r_j^\dag< \Rad_{dec}(\bv)$, then nature will raise $v_j$ (up to $\overline{v}_j$) to drive $\Rad_{dec}(\bv)$ down; if $r_j^\dag> \Rad_{dec}(\bv)$, then nature will lower $v_j$ (down to $\underline{v}_j$). Hence, the worst-case $\bv$ must lie on the boundary of the hyper-rectangle. 
Let us sort the products in increasing order of $r_j^\dag$, i.e., $r_1^\dag\le r_2^\dag\le \dots\le r_J^\dag$. 
The optimal $\bv$ is in the form of $\bigof{\overline{v}_1,\dots,\overline{v}_{j-1},\underline{v}_{j},\dots,\underline{v}_J}$ for some $j=2,3,\dots, J$. 

Now let us take an alternative and more formal way to prove this result.
Taking the derivative of $\Rad_{dec}(\bv)$ with respect to $v_j$, 
$
    \frac{\partial \Rad_{dec}}{\partial v_j} = \frac{\sum_{j'\in \cJ}(r_j^\dag-r_{j'}^\dag)\cdot v_{j'} }{\of{\sum_{i\in \cJ}v_i}^2}.
$
Since the denominator is positive, and the numerator $\sum_{j'\in \cJ}(r_j^\dag-r_{j'}^\dag)\cdot v_{j'}$ is increasing in $r_j^\dag$, the derivative is first negative and then becomes positive as $j$ increases. Moreover, for any $\bv_{-j}$, since the coefficient $(r_j^\dag-r_{j}^\dag)=0$ for $v_j$, the derivative of $v_j$ has a constant sign for all $v_j\in [\underline{v}_j,\overline{v}_j]$, which implies that $\Rad_{dec}$ is monotonic in $v_j$ for all $\bv_{-j}$. Therefore, for any $j\in \cJ$, the optimal $v_j$ is achieved at $\underline{v}_j$ or $\overline{v}_j$.
Let us sort the products in increasing order of $r_j^\dag$, i.e., $r_1^\dag\le r_2^\dag\le \dots\le r_J^\dag$.  When $j=1$, the derivative with respect to $v_1$ is $    \frac{\partial \Rad_{dec}}{\partial v_1} = \frac{\sum_{j'\in \cJ}(r_1^\dag-r_{j'}^\dag)\cdot v_{j'} }{\of{\sum_{j\in \cJ}v_j}^2}\le 0$, so the optimal $v_1=\overline{v}_1$. 
On the other hand, when $j=J$, the derivative with respect to $v_J$ is $    \frac{\partial \Rad_{dec}}{\partial v_J} = \frac{\sum_{j'\in \cJ}(r_J^\dag-r_{j'}^\dag)\cdot v_{j'} }{\of{\sum_{j\in \cJ}v_j}^2}\ge 0$, so the optimal $v_J$ for nature is $v_J=\underline{v}_J$. 
Analogously, for any $j\in \cJ$, if $\sum_{j'\neq j}(r_j^\dag-r_{j'}^\dag)\cdot v_{j'} <0$, then increasing $v_j$ always decreases the approximation ratio; on the other hand, if $\sum_{j'\neq j}(r_j^\dag-r_{j'}^\dag)\cdot v_{j'} > 0$, then decreasing $v_j$ always decreases the approximation ratio; if $\sum_{j'\neq j}(r_j^\dag-r_{j'}^\dag)\cdot v_{j'} = 0$, then the approximation ratio does not change with value $v_j$. This is because the support $\cV=[\underline{v}_1,\overline{v}_1]\times \dots \times [\underline{v}_n,\overline{v}_n]$ is rectangular and the feasible range of $v_j$ does not depend on the valuations $\bv_{-j}$ of other dimensions. Therefore, as the index $j$ increases and $r_j^\dag$ increases, $\sum_{j\in \cJ}(r_j^\dag-r_{j'}^\dag)\cdot v_{j'}$ is first nonpositive for small $j$ and then nonnegative for large $j$. Hence, the optimal $\bv$ for nature to minimize $\Rad_{dec}(\bv)$ has the form of $\bigof{\overline{v}_1,\dots,\overline{v}_{j-1},\underline{v}_{j},\dots,\underline{v}_J}$ for some $j=2,3,\dots, J$. 
By \Cref{lemma:single}, the performance ratio of the decomposed separable mechanism has the form of $
\min_{\bv} \Rad_{dec}(\bv)= \min_{\bv}\frac{\sum_{j\in \cJ}r_j^\dag v_j }{\sum_{j\in \cJ}v_j} =   \min\limits_{j=2,\dots, J} 
 \frac{\sum_{j'=1}^{ j-1}r_{j'}^\dag \, \overline{v}_{j'} +\sum_{j'=j}^{J}r_{j'}^\dag \, \underline{v}_{j'}  }{\sum_{j'=1}^{j-1} \, \overline{v}_{j'} +\sum_{j'=j}^{J} \, \underline{v}_{j'} }$,
where $j\in \cJ$ is sorted in increasing order of $\{\underline{v}_j/\overline{v}_j\}$.
\QED
\end{proof}
\begin{proof}{Proof of \Cref{cor:semi-sep}.}
By \Cref{prop:separable}, the performance ratio obtained by the decomposed separable mechanism is 
    $\cR_{\emph{dec}}^* =\min_{\bv\in \cV}\Rad_{dec}(\bv) = 
    \min\limits_{j=2,\dots, J} 
 \frac{\sum_{j'=1}^{ j-1}r_{j'}^\dag \, \overline{v}_{j'} +\sum_{j'=j}^{J}r_{j'}^\dag \, \underline{v}_{j'}  }{\sum_{j'=1}^{j-1} \, \overline{v}_{j'} +\sum_{j'=j}^{J} \, \underline{v}_{j'} } = \min\limits_{j=2,\dots, J} 
 \frac{ r_{J}^\dag\underline{v}_{J}  }{\sum_{j'=1}^{j-1} \, \overline{v}_{j'} +\sum_{j'=j}^{J} \, \underline{v}_{j'} }= 
 \frac{ r_{J}^\dag\underline{v}_{J}  }{\sum_{j'=1}^{J-1} \, \overline{v}_{j'} + \, \underline{v}_{J} }= \frac{1/(1+\ln(\overline{v}_J/\underline{v}_J))}{\of{\sum_{j=1}^{J-1}\overline{v}_j}/{\underline{v}_J}+1}$. Now we consider the performance ratio achieved by mechanism $M_{\gamma^*}$. According to \Cref{prop:ratio-feasible}, for any $\gamma>0$, since $e^{-1/\gamma}>0$, we have that $\{1,2,\dots,J-1\} \subseteq \cS(\gamma)$. If $J\in \cS(\gamma)$ as well, then $\phi(\gamma) = \gamma e^{-1/\gamma} \sum_{j\in\cJ}\overline{v}_j>0$, so $J\notin \cS(\gamma)$. Hence, $\gamma$ is solved by $\phi(\gamma)=\gamma\cdot e^{-1/\gamma} \cdot \sum_{j=1}^{J-1} \overline{v}_j - \bigof{\underline{v}_J\cdot\bigof{\gamma\ln(\underline{v}_J/\overline{v}_J)-\gamma+1}} = 0$. Solving this equation and simplifying the solution, we have that $\cR_{M_{\gamma^*}} = \bigof{\frac{1}{r^\dag_J}+W(\frac{\sum_{j=1}^{J-1}\overline{v}_j}{e\overline{v}_J}) }^{-1} =\bigof{1+\ln(\overline{v}_J/\underline{v}_J)+W(\frac{\sum_{j=1}^{J-1}\overline{v}_j}{e\overline{v}_J}) }^{-1} $. 
 \QED
\end{proof}

\begin{proof}{Proof of \Cref{lemma: unimodal}.}
For $T>x_0$, Stieltjes integration by parts gives
\[
\int_{0}^{T} \kappa(x)\, d\alpha(x)
= \kappa(T)\alpha(T) - \kappa(0)\alpha(0) - \int_{0}^{T} \alpha(x)\, d\kappa(x).
\]
Since $\kappa\ge 0$ and $0\le \alpha \le 1$, the boundary terms are maximized by $\alpha(0)=0$ and $\alpha(T)=1$, so it suffices to minimize $\int_{0}^{T} \alpha(x)\, d\kappa(x)$.
For $x\in[0,x_0)$, since $d\kappa(x)\ge 0$, the minimum value of $\int_{0}^{T} \alpha(x)\, d\kappa(x)$ is achieved by setting $\alpha(x)=0$ at $x\in[0,x_0)$. For $x\in[x_0,T]$,  since $d\kappa(x)\le 0$, the minimum value of $\int_{0}^{T} \alpha(x)\, d\kappa(x)$ is achieved by setting $\alpha(x)=1$ for $x\in[x_0,T]$. Since $\kappa$ is decreasing with a finite lower bound $0$ as $T\to\infty$, the integral is well-behaved as $T\to\infty$. Thus, a threshold rule at a mode $x_0\in\arg\max_x \kappa(x)$ maximizes $\int_{0}^{\infty} \kappa(x)\, d\alpha(x)$. \QED
\end{proof}

\begin{proof}{Proof of \Cref{prop:2item}.}
The performance ratio achieved by the posted price mechanism in \Cref{lemma:2item} is
\begin{align*}
    \int_{\bv} \frac{\omega_1+\underline{v}_2}{v_1+v_2}\cdot d\F_{\boldsymbol{\omega}}(\bv) =  \int_{\xi} \frac{\omega_1+\underline{v}_2}{v_1(\xi)+v_2(\xi)}\cdot d\G(\xi)=\int_{1}^{\infty} \frac{\omega_1+\underline{v}_2}{v_1(\xi)+v_2(\xi)}\cdot \frac{\zeta\cdot (v_1(\xi)+v_2(\xi))}{\xi^2} d\xi = \zeta\cdot (\omega_1+\underline{v}_2).
\end{align*}
The remaining step is to solve the normalization constant $\zeta$ by $\int_1^\infty d\G(\xi)=1$, where $\G$ is defined in \eqref{eq:g2}.
Incorporating the support of value $\bv$ defined in \eqref{eq:v2}, i.e.,
\begin{align*}
\bv(\xi) = \begin{cases}
    \bigof{\omega_1\cdot \xi, \, \underline{v}_2\cdot \xi} & \xi \in [1,\frac{\overline{v}_2}{\underline{v}_2}]\\
     \bigof{\omega_1\cdot \xi, \, \overline{v}_2}& \xi \in (\frac{\overline{v}_2}{\underline{v}_2}, \frac{\overline{v}_{1}}{\omega_{1}}]\\
      \bigof{\overline{v}_1, \, \overline{v}_2} & \xi \in (\frac{\overline{v}_{1}}{\omega_{1}}, \infty),
\end{cases}
\end{align*}
we are ready to solve $\zeta$ by
{\small
\begin{align*}
    1= &\int_{1}^{\infty} d\G(\xi)
    = \int_{1}^{\infty}  \frac{{\zeta\cdot (v_1(\xi)+v_2(\xi))}}{\xi^2}\, d\xi 
    = \int_{1}^{\frac{\overline{v}_2}{\underline{v}_2}}(\omega_1 + \underline{v}_2)\, \frac{\zeta}{\xi}\, d\xi +\int_{\frac{\overline{v}_{2}}{\underline{v}_{2}}}^ {\frac{\overline{v}_{1}}{\omega_{1}}} 
   \bigof{\omega_1\, \frac{\zeta}{\xi}\, + 
  \overline{v}_2\frac{\zeta}{\xi^2}}\, d\xi
   + \int_{\frac{\overline{v}_{1}}{\omega_{1}}}^{\infty} 
   \sum_{j=1}^{2}\, \overline{v}_j\frac{\zeta}{\xi^2}\, d\xi \\
   =& \zeta\cdot \bigof{\int_{1}^{\frac{\overline{v}_1}{\omega_1}}  \frac{\omega_1}{\xi}\,  d\xi + \int_{\frac{\overline{v}_1}{\omega_1}}^{\infty} \frac{\overline{v}_1 }{\xi^2}\, d\xi +\int_{1}^{\frac{\overline{v}_2}{\underline{v}_2}} \frac{\underline{v}_2 }{\xi} d\xi + \int_{\frac{\overline{v}_2}{\underline{v}_2}}^{\infty}  \frac{\overline{v}_2}{\xi^2} d\xi } 
     = \zeta\cdot \bigof{\omega_1 \ln \frac{\overline{v}_1}{\omega_1} +  {\omega_1} +\underline{v}_2 \ln \frac{\overline{v}_2}{\underline{v}_2} +  {\underline{v}_2}},
\end{align*}
}
which implies that $$\zeta = \left( \omega_1 \ln \frac{\overline{v}_1}{\omega_1} + \omega_1 + \underline{v}_2 \ln \frac{\overline{v}_2}{\underline{v}_2} + \underline{v}_2 \right)^{-1}.$$
Therefore, the seller's optimal performance ratio obtained under distribution $\F_{\boldsymbol{\omega}}$ is calculated as 
\begin{align*}
\bigof{\omega_1 \ln \frac{\overline{v}_1}{\omega_1} +  {\omega_1} +\underline{v}_2 \ln \frac{\overline{v}_2}{\underline{v}_2} +  {\underline{v}_2}}^{-1}   \cdot (\omega_1+\underline{v}_2)
\end{align*}
\QED
\end{proof}

\begin{proof}{Proof of \Cref{prop:2item-opt}.}
For a given nature's strategy $\F_{\boldsymbol{\omega}}$, by \Cref{prop:2item}, the optimal selling mechanism achieves an approximation ratio $ \bigof{\omega_1 \ln \frac{\overline{v}_1}{\omega_1} +  {\omega_1} +\underline{v}_2 \ln \frac{\overline{v}_2}{\underline{v}_2} +  {\underline{v}_2}}^{-1}   \cdot (\omega_1+\underline{v}_2)$. Now we hope to find the optimal $\omega_1$ for nature to minimize the performance ratio.
    The derivative of  $ \bigof{\omega_1 \ln \frac{\overline{v}_1}{\omega_1} +  {\omega_1} +\underline{v}_2 \ln \frac{\overline{v}_2}{\underline{v}_2} +  {\underline{v}_2}}^{-1}   \cdot (\omega_1+\underline{v}_2)$ with respect to $\omega_1$ is $\frac{\omega_1+\underline{v}_2\cdot\bigof{\ln\frac{\overline{v}_2\omega_1}{\underline{v}_2\overline{v}_1}+1}}{\bigof{\omega_1 \ln \frac{\overline{v}_1}{\omega_1} +  {\omega_1} +\underline{v}_2 \ln \frac{\overline{v}_2}{\underline{v}_2} +  {\underline{v}_2}}^2}$. Notice the denominator is always positive.
The numerator is increasing in $\omega_1$ and is always positive when $\omega_1\ge \frac{\overline{v}_1\underline{v}_2}{\overline{v}_2}$, so the minimizer $\omega_1$ should be within $[\underline{v}_1, \frac{\overline{v}_1\underline{v}_2}{\overline{v}_2}]$. 
Thus, consider the following two cases.
\begin{enumerate}
    \item If $\omega_1+\underline{v}_2\cdot\bigof{\ln\frac{\overline{v}_2\omega_1}{\underline{v}_2\overline{v}_1}+1}< 0$ at $\omega_1=\underline{v}_1$, the derivative is first negative and then positive for $\omega_1 \ge \underline{v}_1$.  By first-order condition, i.e., $\omega_1+\underline{v}_2\cdot\bigof{\ln\frac{\overline{v}_2\omega_1}{\underline{v}_2\overline{v}_1}+1}= 0$, the approximation ratio is minimized at the $\omega_1^*=\underline{v}_2W(\frac{\overline{v}_1}{e\overline{v}_2})$. Hence, the approximation ratio is simplified as $\left(W(\frac{\overline{v}_1}{e\overline{v}_2}) + \ln \frac{\overline{v}_2}{\underline{v}_2}+1\right)^{-1}$. 
    \item If $\omega_1+\underline{v}_2\cdot\bigof{\ln\frac{\overline{v}_2\omega_1}{\underline{v}_2\overline{v}_1}+1}\ge 0$ at $\omega_1=\underline{v}_1$, then the derivative is always nonnegative for $\omega_1 \ge \underline{v}_1$, so the approximation ratio is increasing in $\omega_1$. The minimum approximation ratio is achieved at $\omega_1^*=\underline{v}_1$, which can be expressed as $\frac{\sum_{j=1}^2 \underline{v}_j}{\sum_{j=1}^2 \of{ \underline{v}_j \cdot \of{1+\ln(\overline{v}_j/\underline{v}_j)} }}$. 
\end{enumerate}
The analysis above demonstrates that, the highest achievable approximation ratio by the seller under nature's strategy $\F_{\omega_1^*}$, where $\omega_1^*$ is solved based on $\underline{v}_1,\overline{v}_1,\underline{v}_2,\overline{v}_2$, coincides with the performance ratio obtained by the mechanism proposed in \Cref{cor:2item}. Therefore, the mechanism in \Cref{cor:2item} is robustly optimal.
\QED
\end{proof}

\begin{proof}{Proof of \Cref{prop:ratio-upper bound}.} 
First, due to the definition of $\boldsymbol{\omega}$ in \eqref{eq:def-w} and $\tilde{j}(\eta)=\max\{j\in \cJ \mid \underline{v}_j/\overline{v}_j< e^{-1/\eta}\}$, the normalization factor $\zeta$ in the definition of $\G$ in \Cref{prop:nitem} is solved by:
\begin{align*}
    1= &\int_{1}^{\infty} d\G(\xi)
     =\zeta\cdot \sum_{j=1}^J \bigof{\omega_j \ln \frac{\overline{v}_j}{\omega_j} +  {\omega_j} } 
=     \zeta\cdot \biggof{\sum_{j=1}^{\tilde{j}(\eta)} \bigof{\overline{v}_j e^{-\frac{1}{\eta}} (1+\frac{1}{\eta}) }+
\sum_{j=\tilde{j}(\eta)+1}^{J} \bigof{\underline{v}_j \ln \frac{\overline{v}_j}{\underline{v}_j} +  {\underline{v}_j} } 
}
\end{align*}
Therefore, $\zeta = \left(\sum_{j=1}^{\tilde{j}(\eta)} \bigof{\overline{v}_j e^{-\frac{1}{\eta}} (1+\frac{1}{\eta}) }+
\sum_{j=\tilde{j}(\eta)+1}^{J} \bigof{\underline{v}_j \ln \frac{\overline{v}_j}{\underline{v}_j} +  {\underline{v}_j} } 
\right)^{-1}$. Denote $f(\eta)$ the highest performance ratio the seller can obtain under nature's strategy $\F_{\eta}$. By \Cref{prop:nitem}, the maximum performance ratio is
{\small
\begin{align*}   f(\eta)=&\Bigof{\sum_{j=1}^{\tilde{j}(\eta)} \bigof{\overline{v}_j e^{-\frac{1}{\eta}} (1+\frac{1}{\eta}) }+
\sum_{j=\tilde{j}(\eta)+1}^{J} \bigof{\underline{v}_j \ln \frac{\overline{v}_j}{\underline{v}_j} +  {\underline{v}_j} } 
}^{-1} \cdot \sum_{j\in \cJ} \omega_j\\
=&\Bigof{\sum_{j=1}^{\tilde{j}(\eta)} \bigof{\overline{v}_j e^{-\frac{1}{\eta}} (1+\frac{1}{\eta}) }+
\sum_{j=\tilde{j}(\eta)+1}^{J} \bigof{\underline{v}_j \ln \frac{\overline{v}_j}{\underline{v}_j} +  {\underline{v}_j} } 
}^{-1} \cdot \bigof{\sum_{j=1}^{\tilde{j}(\eta)}\overline{v}_j e^{-\frac{1}{\eta}} +  \sum_{j=\tilde{j}(\eta)+1}^{J} \underline{v}_j}
\end{align*}
}
Now nature aims to optimize $\eta$ to minimize the approximation ratio $f(\eta)$.
The derivative of $f$ with respect to $\eta$ has the same sign as $\phi(\eta) = \eta\cdot e^{-1/\eta} \cdot \sum_{j=1}^{\tilde{j}(\eta)} \overline{v}_j - \sum_{j=\tilde{j}(\eta)+1}^{J} \bigof{\underline{v}_j\cdot\bigof{\eta\ln(\underline{v}_j/\overline{v}_j)-\eta+1}} $. 
By \Cref{increasing-g}, function $\phi(\eta) = \eta\cdot e^{-1/\eta} \cdot \sum_{j=1}^{\tilde{j}(\eta)} \overline{v}_j - \sum_{j=\tilde{j}(\eta)+1}^{J} \bigof{\underline{v}_j\cdot\bigof{\eta\ln(\underline{v}_j/\overline{v}_j)-\eta+1}}$ is increasing in $\eta$ and there exists a unique solution to $\phi(\eta)= 0$. 
Hence, $f(\eta)$ is first decreasing and then increasing in $\eta$, and the lowest approximation ratio is achieved at the unique solution $\eta^*$ such that $\phi(\eta^*)=0$.
Then embedding  $e^{-1/\eta} \cdot \sum_{j=1}^{\tilde{j}(\eta)} \overline{v}_j = \sum_{j=\tilde{j}(\eta)+1}^{J} \bigof{\underline{v}_j\cdot\bigof{\ln(\underline{v}_j/\overline{v}_j)-1+\frac{1}{\eta}}} $ into the expression of approximation ratio $f(\eta)$, we have that
{\small
\begin{align*}
        f(\eta^*)=
&\Bigof{ \sum_{j=1}^{\tilde{j}(\eta^*)} \bigof{\overline{v}_j e^{-\frac{1}{\eta^*}} (1+\frac{1}{\eta^*}) }+
\sum_{j=\tilde{j}(\eta^*)+1}^{J} \bigof{\underline{v}_j \ln \frac{\overline{v}_j}{\underline{v}_j} +  {\underline{v}_j} } 
}^{-1} \cdot \bigof{\sum_{j=1}^{\tilde{j}(\eta^*)}\overline{v}_j e^{-\frac{1}{\eta^*}} +  \sum_{j=\tilde{j}(\eta^*)+1}^{J} \underline{v}_j} \\
=& \Bigof{  \frac{\eta^*+1}{\eta^*}\sum_{j=\tilde{j}(\eta^*)+1}^{J} \bigof{\underline{v}_j(\ln\frac{\underline{v}_j}{\overline{v}_j}-1+\frac{1}{\eta^*})} +\sum_{j=\tilde{j}(\eta^*)+1}^{J} \bigof{\underline{v}_j \ln \frac{\overline{v}_j}{\underline{v}_j} +  {\underline{v}_j} } }^{-1} \cdot \Bigof{\sum_{j=\tilde{j}(\eta^*)+1}^{J} \bigof{\underline{v}_j(\ln\frac{\underline{v}_j}{\overline{v}_j}-1+\frac{1}{\eta^*})} +  \sum_{j=\tilde{j}(\eta^*)+1}^{J} \underline{v}_j} \\ 
=& \biggof{  \frac{1}{\eta^*}\sum_{j=\tilde{j}(\eta^*)+1}^{J} \Bigof{\underline{v}_j\cdot\bigof{\ln(\underline{v}_j/\overline{v}_j)+\frac{1}{\eta^*}}} 
}^{-1} \cdot \biggof{\sum_{j=\tilde{j}(\eta^*)+1}^{J} \Bigof{\underline{v}_j\cdot\bigof{\ln(\underline{v}_j/\overline{v}_j)+\frac{1}{\eta^*}}} } \\
=& \eta^*
\end{align*}
}
which completes our proof.
\QED
\end{proof}

\begin{proof}{Proof of \Cref{thm:bundle-symmetric}.}
First, we show that mechanism \eqref{eq:bundle-symmetric} is incentive-compatible and individually rational. For a buyer with valuation $\bv$, their utility when reporting $\bv'$ is 
\begin{align*}
    u(\bv,\bv') =  \sum_{j\in\cJ}[\bq(\bv')]_j\cdot v_j -t(\bv') =\frac{1+\ln\bigof{\frac{\sum_{j\in\cJ}v_j'}{\underline{v}}}}{1+\ln(\overline{v}/\underline{v})}\cdot(\sum_{j\in\cJ}v_j)-\bigof{\frac{\sum_{j\in\cJ}v_j'}{1+\ln(\overline{v}/\underline{v})} }
\end{align*}
The derivative of $\sum_{j\in\cJ}[\bq(\bv')]_j\cdot v_j -t(\bv')$ w.r.t. $v_j'$ is $\frac{\sum_{i\in\cJ}v_i}{\of{1+\ln(\overline{v}/\underline{v})} \cdot\of{\sum_{i\in\cJ}v_i'}}-\frac{1}{1+\ln(\overline{v}/\underline{v})} =\frac{\sum_{i\in\cJ}v_i - \sum_{i\in\cJ}v_i' }{\of{1+\ln(\overline{v}/\underline{v})} \cdot\of{\sum_{i\in\cJ}v_i'}}$, so $ u(\bv,\bv')$ is increasing in $v_j'$ when $\sum_{i\in\cJ}v_i'$ is less than $\sum_{i\in\cJ}v_i$ and decreasing  in $v_j'$ when $\sum_{i\in\cJ}v_i'$ is greater than $\sum_{i\in\cJ}v_i$. Hence, $ u(\bv,\bv')$ is maximized when $\sum_{i\in\cJ}v_i' = \sum_{i\in\cJ}v_i$, which implies incentive compatibility. 
Moreover, since $ u(\bv,\bv) =\frac{1+\ln\bigof{\frac{\sum_{j\in\cJ}v_j}{\underline{v}}}}{1+\ln(\overline{v}/\underline{v})}\cdot(\sum_{j\in\cJ}v_j)-\frac{\sum_{j\in\cJ}v_j}{1+\ln(\overline{v}/\underline{v})} =\frac{\ln\bigof{\frac{\sum_{j\in\cJ}v_j}{\underline{v}}}}{1+\ln(\overline{v}/\underline{v})}\cdot(\sum_{i\in\cJ}v_i) \ge 0$, the mechanism also satisfies individual rationality.
Second, by \Cref{lemma:single}, the mechanism \eqref{eq:bundle-symmetric} achieves a performance ratio of  $\min\limits_{\bv\in\cV} \frac{t(\bv)}{\allone^\top \bv} = \frac{1}{1+\ln(\overline{v}/\underline{v})} $.

Now we aim to prove the optimality of mechanism \eqref{eq:bundle-symmetric}. By \Cref{assum:ambiguityset}, for any $\xi\in [\underline{v},\overline{v}]$, there exists $\bv\in \cV$ such that $\sum_{j\in \cJ}v_j=\xi$. Since $\cV$ is $\boldsymbol{\rho}$-scaled invariant, then $\xi\cdot\boldsymbol{\rho} \in \cV$ for all $\xi\in [\underline{v},\overline{v}]$.
Let us consider nature's strategy $\F^*(\bv)$ such that  $\bv(\xi) = \of{\xi\rho_1,\xi \rho_2, \dots, \xi\rho_J}$ where $\xi\sim\G$ with $G(\xi)= \begin{cases}
\frac{\ln\xi-\ln \underline{v}}{1+\ln(\overline{v}/\underline{v})} & \xi\in[\underline{v},\overline{v})\\
1 & \xi= \overline{v}
\end{cases}$.

For any seller's strategy $\of{\bq(\bv(\xi)), t(\bv(\xi))}$, denote $\boldsymbol{\alpha}(\xi)$ and $\tau(\xi)$ the allocation probability and payment at $\bv(\xi)$, respectively, i.e., $(\boldsymbol{\alpha}(\xi),\tau(\xi))= \of{\bq(\bv(\xi)), t(\bv(\xi))}$, where $\bv(\xi) = \of{\xi\rho_1,\xi \rho_2, \dots, \xi\rho_J}$. 
By incentive compatibility and the envelope theorem \citep{milgrom2002envelope}, the payment satisfies
\begin{align*}
    \tau(\xi) = \boldsymbol{\alpha}(\xi)^\top \cdot \bv(\xi) -\int_{\underline{v}}^{\xi}  \boldsymbol{\alpha}(x)^\top \,  d\bv(x) 
 = \sum_{j\in \cJ} \bigof{\int_{\underline{v}}^\xi v_j(x) d\alpha_j(x)}.
\end{align*}
Hence, the performance ratio under $\F^*$ defined above is evaluated as 
{\small
\begin{align*}
    \E_{\bv\sim \F} \Bigoff{\frac{t(\bv)}{\allone^\top \bv}} & = \E_{\xi \sim \G} \Bigoff{\frac{\tau(\xi)}{ \sum_{i\in\cJ}v_i(\xi)}} = \int_{\underline{v}}^{\overline{v}} \Bigoff{\frac{ \sum_{j\in \cJ} \bigof{\int_{\underline{v}}^\xi v_j(x) d\alpha_j(x)}}{ \sum_{i\in\cJ}v_i(\xi)}} d\G(\xi)  =\sum_{j\in \cJ} \Bigoff{\int_{\underline{v}}^{\overline{v}} \Bigof{v_j(\xi) \int_{\xi}^{\overline{v}} \frac{d\G(x)}{ \sum_{i\in\cJ}v_i(x)}}\, d\alpha_j(\xi) } \\
    & =\sum_{j\in \cJ}\int_{\underline{v}}^{\overline{v}} \Bigof{ \rho_j \cdot\xi \int_{\xi}^{\overline{v}} \frac{d\G(x)}{ x}}\, d\alpha_j(\xi) =\sum_{j\in \cJ} \int_{\underline{v}}^{\overline{v}} \Bigof{ \rho_j \cdot \xi\cdot
\frac{1}{\xi\of{1+\ln(\overline{v}/\underline{v})}}  }\, d\alpha_j(\xi) 
= \frac{\sum_{j\in \cJ}\rho_j\of{\alpha_j(\overline{v}) - \alpha_j(\underline{v})} 
  }{1+\ln(\overline{v}/\underline{v})} \\ 
   & \le \frac{1}{1+\ln(\overline{v}/\underline{v})} 
\end{align*}
}
where the last inequality is due to $\alpha_j(\overline{v}),\, \alpha_j(\underline{v})\in [0,1]$.
Hence, under nature's strategy $\F^*(\bv)$, no mechanism can obtain a performance ratio higher than $\frac{1}{1+\ln(\overline{v}/\underline{v})} $, which proves the optimality of mechanism \eqref{eq:bundle-symmetric}.
\QED
\end{proof}
 \end{APPENDICES}
 
\end{document}